\documentclass[prd,aps,twocolumn,a4paper,floatfix,nofootinbib]{revtex4-2}

\usepackage{graphicx}
\usepackage{dcolumn}
\usepackage{bm}
\usepackage{amsmath}
\usepackage{graphicx}
\usepackage{array}
\usepackage{booktabs}
\usepackage {xcolor}
\usepackage{amssymb}
\usepackage{pifont}
\newcommand{\cmark}{\ding{51}}%
\newcommand{\xmark}{\ding{55}}%
\usepackage{multirow}
\usepackage{diagbox}
\usepackage{comment}
\usepackage{xcolor}
\usepackage{enumerate}
\usepackage{booktabs}
\usepackage[normalem]{ulem}
\usepackage[utf8]{inputenc}
\usepackage{hyperref}
\usepackage{units}
\usepackage{enumerate}
\hypersetup{
    colorlinks = true,
    linkcolor = {blue},
    citecolor = {blue},
    urlcolor = {blue},
    linkbordercolor = {white},
    }

\newcommand{\newtext}[1]{{\textcolor{black}
{#1} }}

\begin{document}

\def\nrtidalvthree{{\texttt{NRTidalv3}}}
\def\nrtidal{{\texttt{NRTidal}}}
\def\nrtidalvtwo{{\texttt{NRTidalv2}}}
\def\imrphenomdnrtidaltwo{{\texttt{IMRPhenomD\_NRTidalv2}}}
\def\imrphenomdnrtidalthree{{\texttt{IMRPhenomD\_NRTidalv3}}}
\def\imrphenomxasnrtidaltwo{{\texttt{IMRPhenomXAS\_NRTidalv2}}}
\def\imrphenomxasnrtidalthree{{\texttt{IMRPhenomXAS\_NRTidalv3}}}
\def\imrphenomxpnrtidalthree{{\texttt{IMRPhenomXP\_NRTidalv3}}}
\def\imrphenomxpnrtidaltwo{{\texttt{IMRPhenomXP\_NRTidalv2}}}
\def\imrphenompvtwonrtidaltwo{{\texttt{IMRPhenomPv2\_NRTidalv2}}}
\def\seobnrvfourt{{\texttt{SEOBNRv4T}}}
\def\lalsuite{{\texttt{LALSuite}}}
\def\seobnrvfiverom{{\texttt{SEOBNRv5\_ROM}}}
\def\seobnrvfourromnrtidalvtwo{{\texttt{SEOBNRv4\_ROM\_NRTidalv2}}}
\def\seobnrvfiveromnrtidalvtwo{{\texttt{SEOBNRv5\_ROM\_NRTidalv2}}}
\def\seobnrvfiveromnrtidalvthree{{\texttt{SEOBNRv5\_ROM\_NRTidalv3}}}
\def\teobresums{{\texttt{TEOBResumS}}}

\title{New and Robust Gravitational-Waveform Model for High-Mass-Ratio Binary Neutron Star Systems with Dynamical Tidal Effects}

\author{Adrian \surname{Abac}$^{1,2}$}
\author{Tim \surname{Dietrich}$^{1,2}$}
\author{Alessandra \surname{Buonanno}$^{1,3}$}
\author{Jan \surname{Steinhoff}$^{1}$}
\author{Maximiliano \surname{Ujevic}$^{4}$}

\affiliation{${}^1$Max Planck Institute for Gravitational Physics (Albert Einstein Institute), Am M\"uhlenberg 1, Potsdam 14476, Germany}
\affiliation{${}^2$Institut f\"{u}r Physik und Astronomie, Universit\"{a}t Potsdam, Haus 28, Karl-Liebknecht-Str. 24/25, 14476, Potsdam, Germany}
\affiliation{${}^3$Department of Physics, University of Maryland, College Park, MD 20742, USA}
\affiliation{${}^4$Centro de Ciências Naturais e Humanas, Universidade Federal do ABC, Santo André 09210-170, SP, Brazil}

\date{\today}

\begin{abstract}
    For the analysis of gravitational-wave signals, fast and accurate gravitational-waveform models are required. These enable us to obtain information on the system properties from compact binary mergers. In this article, we introduce the \nrtidalvthree\ model, which contains a closed-form expression that describes tidal effects, focusing on the description of binary neutron star systems. The model improves upon previous versions by employing a larger set of numerical-relativity data for its calibration, by including high-mass ratio systems covering also a wider range of equations of state. 
    It also takes into account dynamical tidal effects and the known post-Newtonian mass-ratio dependence of individual calibration parameters.
    We implemented the model in the publicly available \texttt{LALSuite} software library by augmenting different binary black hole waveform models (\texttt{IMRPhenomD}, \texttt{IMRPhenomX}, and \texttt{SEOBNRv5\_ROM}). 
    We test the validity of \nrtidalvthree\, by comparing it with numerical-relativity waveforms, as well as other tidal models. Finally, we perform parameter estimation for GW170817 and GW190425 with the new tidal approximant and find overall consistent results with respect to previous studies.
\end{abstract}


\maketitle

\section{\label{section: introduction}Introduction}
Two years after the discovery of a binary black hole (BBH) merger~\cite{LIGOScientific:2016aoc}, which initiated the era of gravitational-wave (GW) astronomy, the discovery of the binary neutron star (BNS) merger GW170817 inaugurated a new era in multi-messenger astronomy~\cite{LIGOScientific:2017vwq, LIGOScientific:2017zic, LIGOScientific:2017ync}, in which GWs and electromagnetic signals are combined to unravel these highly energetic events. Indeed, the GW detection of GW170817 came along with electromagnetic signatures covering the whole frequency range from radio to gamma-rays~\cite{LIGOScientific:2017zic, LIGOScientific:2017ync}. 
However, GW170817 was just the beginning. 
Two years later, the LIGO-Virgo-Kagra collaboration detected the BNS GW190425~\cite{LIGOScientific:2020aai} and with the increasing sensitivity of current GW detectors as well as planned new-generation detectors, it is expected that the number of BNS events observed will increase in the future~\cite{KAGRA:2013rdx, Chan:2018csa, Lenon:2021zac, Branchesi:2023mws}. 

One of the key scientific achievements of GW170817 was the improved knowledge of neutron star (NS) interior~\cite{LIGOScientific:2017vwq, Dietrich:2020efo}. The structure of the NS is primarily described by an equation of state (EoS) of neutron-rich supranuclear-dense matter~\cite{Lattimer:2012nd, Ozel:2016oaf}.
In fact, matter in the core of NSs is thought to be at least a few times denser than the nuclear saturation density $\rho \sim 10^{14} \, {\rm g/cm^3}$, and there are various (competing) theoretical and phenomenological models in nuclear physics (e.g.,~\cite{Chin:1974sa, Serot:1997xg, Huth:2021bsp, Alford:2022bpp, Lattimer:2012nd, Lattimer:2021emm, Burgio:2021vgk, Zhu:2023ijx}) describing this state of matter.  

Given that the EoS describing the NS interior influences the tidal deformability, which is a measure of the star's deformation in response to an external tidal field (e.g., during the gravitational interaction with another compact object~\cite{Flanagan:2007ix, Hinderer:2007mb}), BNS merger observations will shed light on the behavior of matter at extreme densities~\cite{Flanagan:2007ix, Hinderer:2009ca, LIGOScientific:2017vwq, LIGOScientific:2018hze, LIGOScientific:2020aai}.

Physical parameters of the system, such as mass, spin, and tidal deformabilities, can be extracted from GW signals by comparing the observed data with theoretical predictions obtained by solving the Einstein field equations (EFEs), usually through Bayesian analysis~\cite{Veitch:2014wba,Thrane:2018qnx}. 
In principle, ab-initio, numerical-relativity (NR) simulations which solve the EFEs on supercomputers would be the method of choice for the description of a BNS waveform. However, such simulations come with high computational costs and typically only cover the late stage of the inspiral and merger~\cite{Hotokezaka:2015xka, Hotokezaka:2016bzh, Kawaguchi:2018gvj, Kiuchi:2019kzt, Foucart:2018lhe, Dietrich:2018phi,Ujevic:2022qle, Gonzalez:2022mgo}. Furthermore, these simulations are also characterized by an intrinsic uncertainty due to the numerical discretization that is employed to solve the EFEs. 

Over the years, the community has developed different waveform models for BNS systems. One class of models is based on post-Newtonian (PN) theory, which expands the EFEs in powers of $v/c$, where $v$ is the binary characteristic velocity and $c$ is the speed of light. These approximants include tidal interactions that start at the 5th PN order~\cite{Vines:2011ud} ($\propto ({{v}/{c}})^{10}$) up to 7.5PN order~\cite{Damour:2012yf, Henry:2020ski, Narikawa:2023deu}, which is the highest PN order that is currently known for tidal effects.  
However, even high-PN approximants are still not accurate enough in describing the full GW signal, particularly in the late inspiral shortly before the merger, when the velocities of the objects are larger, and the distance between the stars is smaller. 
One approach for improving the accuracy of PN predictions is made by using the effective-one-body (EOB) formalism~\cite{Buonanno:1998gg, Buonanno:2000ef, Damour:2009zoi, Gamba:2022mgx, Gamba:2023mww, Bohe:2016gbl}. 
Over the years, two different `families' of tidal EOB models have emerged: \seobnrvfourt~\cite{Hinderer:2016eia,Steinhoff:2016rfi} and \teobresums~\cite{Bernuzzi:2014owa,Nagar:2018zoe,Akcay:2018yyh}. Generally, tidal EOB models are more accurate than PN predictions and allow, in most cases, a reliable description (within the uncertainty of NR simulations) of the GW signal up to the merger of the stars. To avoid the increase in computational costs, which comes naturally when using EOB (as these models solve ordinary differential equations) instead of PN approximations, people have introduced reduced-order models (ROM)~\cite{Lackey:2016krb,Lackey:2018zvw, Purrer:2014fza, Purrer:2015tud}, post-adiabatic approximations~\cite{Nagar:2018gnk,Mihaylov:2021bpf}, machine learning techniques~\cite{ Tissino:2022thn}, or a combination of the stationary phase and post-adiabatic approximations~\cite{Gamba:2020ljo}. 

Finally, phenomenological GW models~\cite{Ajith:2009bn, Santamaria:2010yb} also exist, which incorporate EOB and/or NR data with the aim to model the tidal phase contribution during the coalescence as accurately as possible while still being fast and efficient (e.g.\ through the use of closed-form analytical expressions for the tidal contribution itself)~\cite{Dietrich:2017aum, Kawaguchi:2018gvj}. One such model is called \nrtidal~\cite{Dietrich:2017aum}, which extracts information from NR, EOB, and PN and constructs an analytical tidal contribution that augments a given BBH waveform baseline, either from EOB models or phenomenological models (such as PhenomD and PhenomX). Two \nrtidal\ versions (\nrtidal~\cite{Dietrich:2017aum} and \nrtidalvtwo~\cite{Dietrich:2019kaq}) have been implemented in the LIGO Algorithm Library (LAL) \cite{lalsuite}. The second iteration, \nrtidalvtwo, builds upon \nrtidal\ by using improved NR data, adding amplitude corrections to the GW signal, as well as incorporating spin effects \cite{Dietrich:2019kaq}. Though computationally efficient, both \nrtidal\ models come with certain caveats. First, both versions only include equal-mass systems in the calibration of their fitting parameters. Second, the models consider the tidal bulge $Q_{ij}$ of the star to be directly related to the tidal field $\mathcal{E}_{ij}$ generated by its companion via a parameter $\lambda$ known as the tidal deformability, i.e., $Q_{ij} = \lambda \mathcal{E}_{ij}$. In this case, the former \nrtidal\ models considered only adiabatic tides ($\lambda = {\rm const.}$) throughout the duration of the inspiral \cite{Dietrich:2017aum, Dietrich:2019kaq}. However, it has been shown that dynamical tides (implying a non-constant, frequency-dependent $\lambda$) can arise from these systems due to the quadrupolar fundamental oscillation mode of the stars \cite{1994PThPh..91..871S, Flanagan:2007ix, Hinderer:2016eia, Steinhoff:2016rfi}, see also Refs.~\cite{Schmidt:2019wrl, Andersson:2019dwg, Gupta:2020lnv, Steinhoff:2021dsn, Pratten:2021pro, Kuan:2022etu, Pnigouras:2022zpx, Mandal:2023lgy, Mandal:2023hqa}. 

Given these limitations, the aim of this work is to extend the existing \nrtidal\ model. In particular, this new model, called \nrtidalvthree, includes the following improvements: 
\begin{enumerate}[(i)]
\itemsep-3pt
\item a larger set of NR BNS waveforms; in total, we utilize 55 waveforms from BAM \cite{Dietrich:2018phi,Gonzalez:2022mgo, Ujevic:2022qle} and SACRA \cite{Kiuchi:2017pte, Kawaguchi:2018gvj} including various total masses, EoSs, mass ratios;
\item inclusion of mass-ratio dependence of the fitting parameters in the calibrated model; and
\item dynamical tides \cite{Flanagan:2007ix, Hinderer:2016eia, Steinhoff:2016rfi, Schmidt:2019wrl}.
\end{enumerate}
The paper is organized as follows. In Sec.~\ref{section: nr data and extraction}, we discuss the employed NR data and the hybridization of the NR data with \seobnrvfourt\ to construct the hybrid waveforms that are used for the calibration. Sec.~\ref{section: time domain nrt3} introduces the time-domain \nrtidalvthree\ phase, while Sec.~\ref{section: freq domain nrt3} discusses the frequency-domain phase. We discuss the implementation and tests for the validation of the model in LAL in Sec.~\ref{section: implementation}. We then conduct a parameter estimation analysis of the model with existing GW observations in Sec.~\ref{section: pe analysis}. Finally, our main conclusion and recommendations are discussed in Sec.~\ref{section: conclusions}. Throughout the paper, we use geometric units, where $G = c = 1$, unless otherwise stated. For the individual components of the BNS, we define $M_A$ to be the primary mass (mass of the heavier companion), $M_B$ to be the mass of the secondary companion, each with tidal deformability $\Lambda_A$ and $\Lambda_B$, respectively. The total mass is given by $M = M_A + M_B$ and the mass ratio is defined as $q = M_A/M_B \ge 1$. The aligned spin components are denoted by $\chi_A$ and $\chi_B$.

\section{\label{section: nr data and extraction}Numerical Relativity Data and Extraction of Tidal Phase Contributions}

The \nrtidal\ model is a closed-form GW model that describes the tidal effects in the inspiral part of BNS coalescences \cite{Dietrich:2017aum, Dietrich:2018uni, Dietrich:2019kaq}. We start from the GW strain \begin{equation} 
    h(t) = A(t) e^{-i\phi(t)},
\end{equation}
where $A(t)$ is the amplitude, and time-domain $\phi(t)$ is the phase. We assume that we can decompose the phase as
\begin{equation} \label{eq:strain_split}
    \phi(\hat{\omega}) = \phi_0(\hat{\omega}) + \phi_{\rm SO}(\hat{\omega}) + \phi_{\rm SS}(\hat{\omega}) + \phi_{\rm S^3}(\hat{\omega}) + \phi_T(\hat{\omega}) +  ...,
\end{equation}
where $\hat\omega = M\omega = Md\phi/dt = M(2\pi f)$ is the rescaled GW frequency, $\phi_0$ denotes the non-spinning point-particle contribution to the total phase, $\phi_{\rm SO}$ denotes the spin-orbit (SO) coupling, $\phi_{\rm SS}$ denotes the spin-spin (SS) interactions (both self-spin and spin interactions), $\phi_{\rm{S}^3}$ denotes contributions cubic in spin (S$^3$), and $\phi_T$ is the tidal phase contribution~\cite{Flanagan:2007ix}. We are neglecting all other (higher-order) spin terms in Eq.~\eqref{eq:strain_split}. 
Similarly to the time domain, we can also write the frequency-domain strain as 
\begin{equation} 
    \tilde{h}(f) = \tilde{A}(f)e^{-i\psi(f)},
\end{equation}
with 
\begin{equation} 
    \psi(f) = \psi_0(f) + \psi_{\text{SO}}(f) + \psi_{\text{SS}}(f) +\psi_{\text{S}^3}(f) +  \psi_T(f) + ... \ .
\end{equation}
The \nrtidal\ model aims at modeling the tidal contributions $\phi_T$ and $\psi_T$, since, unlike the BBH case, tidal deformabilities are present in BNS and BHNS systems and provide valuable information about the internal composition of the individual stars.

\subsection{Numerical relativity data set}

\begin{figure}[t]
    \centering
    \includegraphics[width = \linewidth]{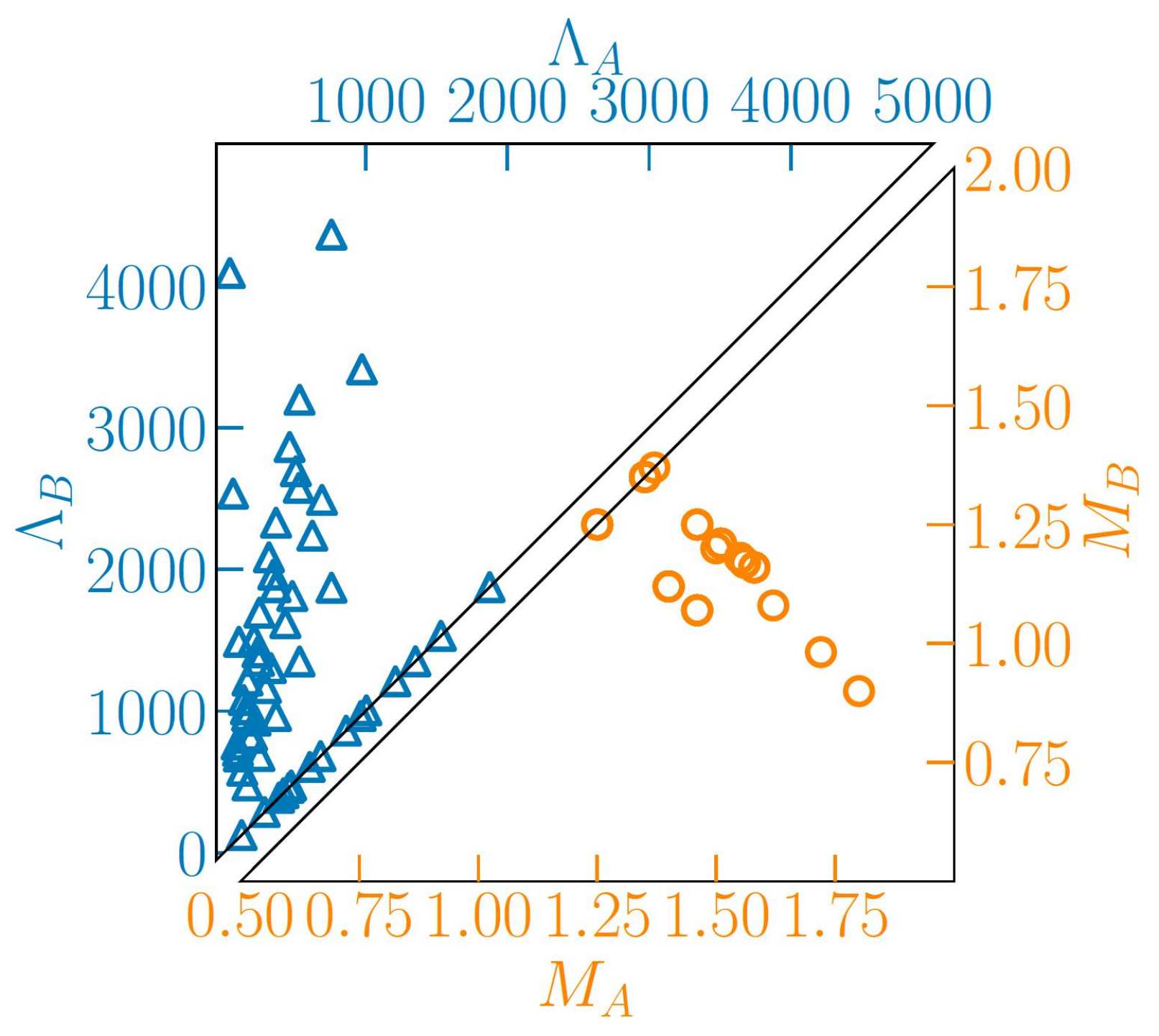}
    \caption{Distribution of the masses (where we set $M_A > M_B$) and corresponding tidal deformabilities $\Lambda_{A,B}$ of the neutron stars used in the NR simulations (see Table~\ref{table: 55nr}). The plot shows overlapping data points for systems with the same masses, but can have different tidal deformabilities due to the various EoSs that were employed.}
    \label{fig: params}
\end{figure}

Previous versions of the \nrtidal\ model start with the tidal part of the GW constructed by combining EOB and NR waveforms~\cite{Dietrich:2017aum, Dietrich:2019kaq}. For the calibration of \nrtidalvthree, 46 NR waveforms simulated by the SACRA code~\cite{Kiuchi:2017pte, Kawaguchi:2018gvj, Kiuchi:2019kzt}, and 9 waveforms simulated with the BAM code~\cite{Dietrich:2018phi, Ujevic:2022qle, Gonzalez:2022mgo} were employed. We provide more detailed information in Appendix~\ref{subsection: appA}. 
Overall, the NR dataset covers ten different EoSs with $q\in [1.0,2.0]$ (see Fig.~\ref{fig: params}).

\subsection{Construction of hybrid waveforms} 

For each of the NR simulations that we employ for calibration, we compute corresponding configurations employing
\seobnrvfourt~\cite{Lackey:2018zvw} from the \lalsuite~\cite{lalsuite} library.
This means that we compute \seobnrvfourt\, waveforms including tidal contributions (which we refer to as EOB-BNS later) and without tidal effects, which we call EOB-BBH later. 
The EOB-BNS waveforms are used in the extraction of the tidal phase contributions, from the early inspiral up to the merger~\cite{Dietrich:2019kaq}. \newtext{The EOB-BBH and EOB-BNS waveforms are computed at a starting frequency of 15 Hz}. These tidal phase contributions will then be used in the calibration of \nrtidalvthree.

For the construction of hybrid waveforms, we first determine the convergence order of the NR data set.  
If we find a clear convergence order, we construct a higher-order Richardson-extrapolated waveform assuming this convergence order~\cite{Bernuzzi:2016pie, Dietrich:2018upm}. In our set of waveforms, this is the case for the BAM waveforms from Ref.~\cite{Dietrich:2019kaq, Ujevic:2022qle}. 
For all other data, we are simply employing the highest resolution for the hybrid construction. 

In the next step, we align and hybridize the EOB-BNS with the NR waveforms. This is required since our NR data only cover the last few orbits before the merger, but we are planning to construct a closed-form approximant that is valid for the entire frequency range. 

For the alignment, we minimize the following integral~\cite{Hotokezaka:2015xka, Hotokezaka:2016bzh, Lackey:2013axa}:
\begin{equation} \label{eq: alignment}
    \mathcal{I}(\delta t, \delta \phi) = \int_{t_1}^{t_2} dt |\phi_{\rm NR}(t) - \phi_{\rm EOB}(t + \delta t) + \delta \phi|, \end{equation}
over the chosen frequency interval (or hybridization window, typically near the beginning of the NR waveform) $[\hat{\omega}_1, \hat{\omega}_2]$ corresponding to times $[t_1, t_2]$. 

\begin{figure}[t]
    \centering
    \includegraphics[width = \linewidth]{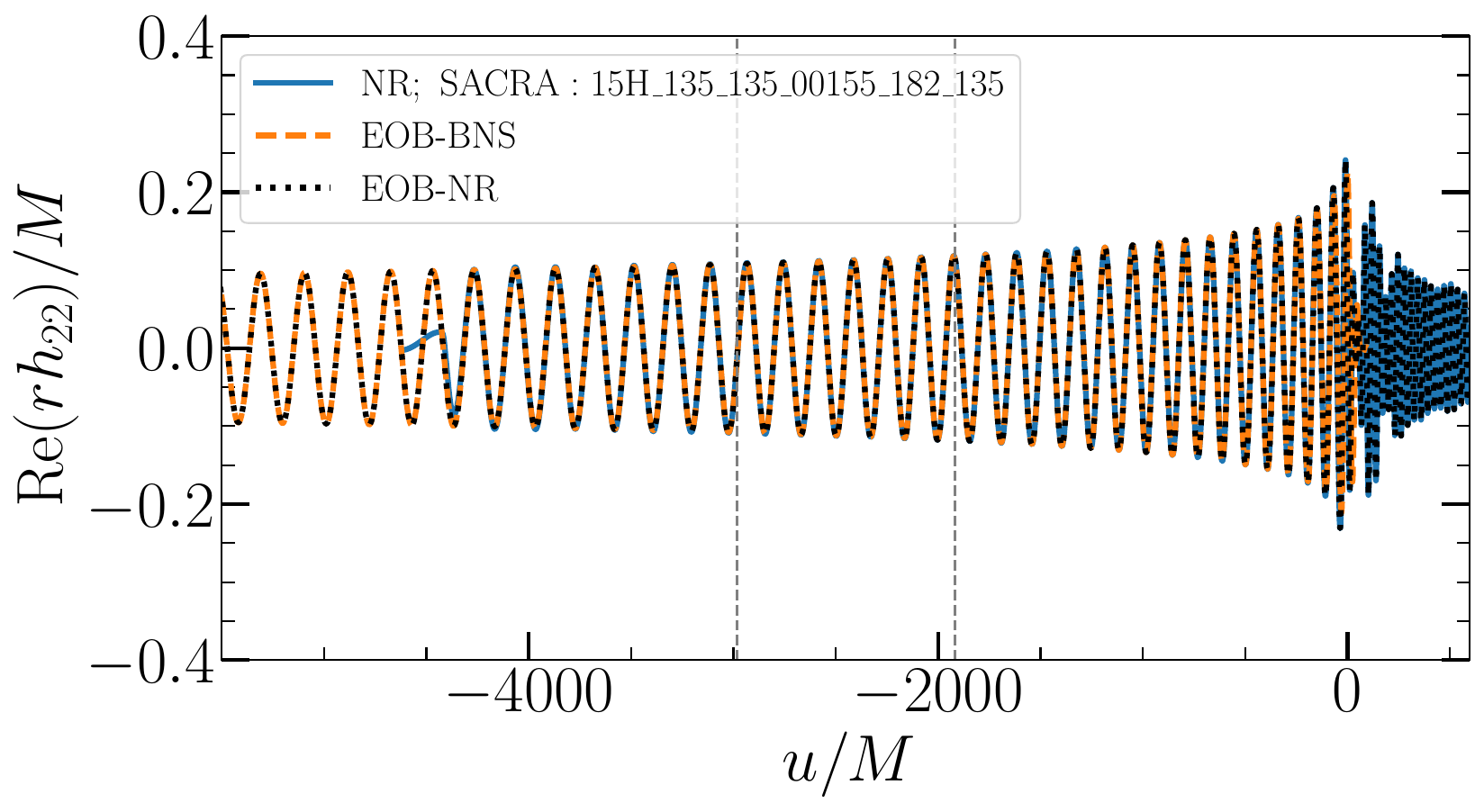}
    \caption{An example of a hybrid Waveform (SACRA:15H\_135\_135\_00155\_182\_135). Here, we plot the real part of the (2,2) mode $rh_{22}$ of the GW strain (where $r$ is the extraction radius) as a function of the retarded time $u$. The blue curve is the NR waveform, the orange dashed curve is the EOB-BNS waveform from \seobnrvfourt, and the black, dotted curve is the hybridized waveform EOB-NR. The gray, dashed, vertical lines denote the boundaries of the hybridization window, $[t_1,t_2]$ corresponding to $\hat{\omega} \in [0.035, 0.04]$. The peak amplitude of the NR and EOB-NR hybrid, indicating merger, is set at $u/M = 0$.  
    }
    \label{fig: hybrid waveform}
\end{figure}

Once the EOB-BNS and the NR waveforms are aligned, we create the hybridized waveform through~\cite{Hotokezaka:2015xka, Hotokezaka:2016bzh, Lackey:2013axa}:
\begin{equation} \label{eq: hybridization}
    h_{\text{EOB-NR}} =  \begin{cases}
        h_{\text{EOB-BNS}}, & t \le t_1\\
        h_{\text{NR}} H(t) + h_{\text{EOB-BNS}}[1 - H(t)], & t_1 \le t \le t_2\\
        h_{\text{NR}}, & t \ge t_2,
    \end{cases}
\end{equation}
where 
\begin{equation} 
    H(t) \equiv \frac{1}{2}\left[1 - \cos\left(\pi \frac{t - t_1}{t_2 - t_1} \right) \right],
\end{equation}
is the Hann window function. This ensures a smooth transition between the EOB and NR waveforms. The result of this hybridization is shown in Fig.~\ref{fig: hybrid waveform} for one example. 
As indicated above, the final waveforms are labeled EOB-NR.

From the expression of the hybrid waveform above in Eq.~\eqref{eq: hybridization}, we expect the tidal phase contributions of the hybrid to be identical to the one of the EOB-BNS waveform right up to the start of the hybridization window. \newtext{The start of the hybridized waveform is chosen to be $\hat{\omega} = 0.0015$ (corresponding to $f \in [17.7,19.4] {\rm Hz}$).} After the window, the tidal contribution would then be identical to that of the NR waveform. For the purposes of constructing \nrtidalvthree, we only consider the EOB-NR tidal phase contributions up to merger. The post-merger parts of the EOB-NR phase are not included. 

\subsection{Extracting the tidal phase}

In the next step, we extract the tidal phase contribution $\phi_T$ up to merger by subtracting the EOB-BBH phase from the EOB-NR phase, which (schematically) means:
\begin{equation} \label{eq: extraction}
    \phi_T = \Delta \phi = \phi_{\text{EOB-NR}} - \phi_{\text{EOB-BBH}}.
\end{equation}

\begin{figure}[t]
    \centering
    \includegraphics[width = \linewidth]{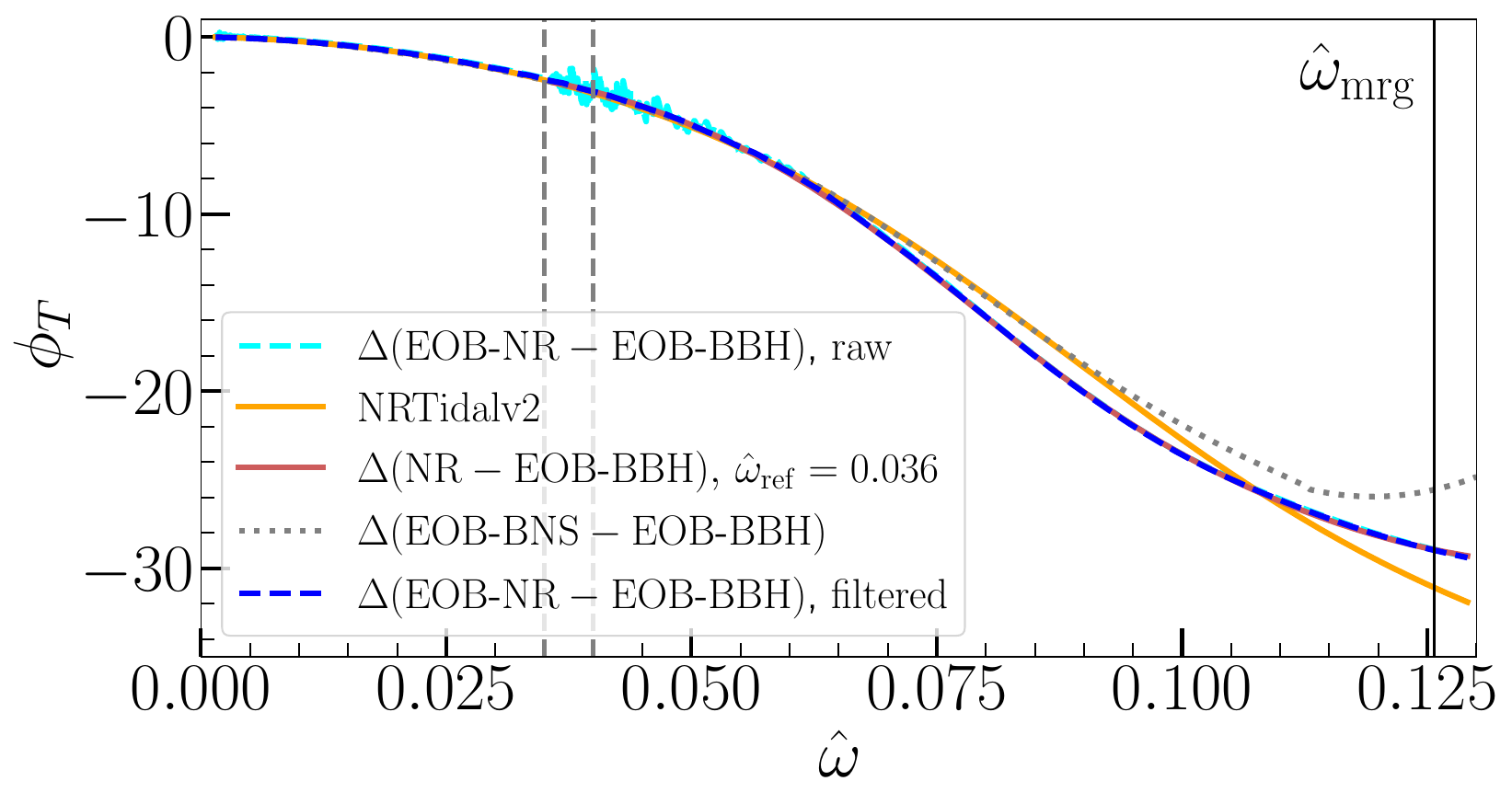}
    \caption{Hybrid tidal phases (EOB -NR) for the configuration SACRA:15H\_135\_135\_00155\_182\_135. The cyan curve represents the EOB-NR data without applying the smoothening operation indicated in the text, the blue dashed line is EOB-NR after filtering. We also show the phase difference between the pure EOB and NR data, as well as for comparison the \nrtidalvtwo\ description. The reference frequency $\hat{\omega}_{\rm ref}$ is used to subtract the EOB-BBH phase from the pure NR phase. \newtext{The hybridization window $\hat{\omega} \in [0.035, 0.04]$ is denoted by the vertical dashed gray lines}. Only the hybrid tidal phase up to the merger frequency $\hat{\omega}_{\rm mrg}$ (indicated by a black vertical line) is used in the construction of \nrtidalvthree.}
    \label{fig: hybridtidal}
\end{figure}

Fig.~\ref{fig: hybridtidal} shows the different tidal phase contributions. We find that the resulting $\phi_T$ calculated from Eq.~\eqref{eq: extraction} has considerable noise, e.g., due to residual eccentricity or remaining density oscillations during the NR simulations. To avoid this, the noise is filtered using a Savitzsky-Golay filter~\cite{Savitzky:1964, Ujevic:2022qle} to smoothen the hybrid waveform.
This leads to the final tidal phase that we will use as an input for the construction of our \nrtidalvthree\ model.

\section{\label{section: time domain nrt3}Dynamical Tides and the Time Domain \nrtidalvthree\ Phase}

\subsection{Employed tidal PN knowledge}

Our construction of the \nrtidalvthree\ approximant begins with the analytical expression of the time-domain tidal phase contribution through 7.5PN order~\cite{Henry:2020ski, Narikawa:2023deu}:
\begin{equation}\label{eq: PNtime}
    \begin{split}
        \phi_T^{\rm PN} =& -\kappa_A c_{\text{Newt}}^A x^{5/2} \Big(1 + c_1^A x + c_{3/2}^A x^{3/2} \\
        & +c_2^A x ^2 + c_{5/2}^Ax^{5/2}\Big) + [A \leftrightarrow B],
    \end{split}
\end{equation}
where $x = (\hat{\omega}/2)^{2/3}$, and the $c_i$'s are
\begin{equation}
    \begin{split}
        c_{\text{Newt}}^A =& \frac{(12 - 11 X_A)(X_A + X_B)^2}{8X_A X_B^2},\\
        c_1^A =&-5 \frac{260X_A^3 - 2286X_A^2 - 919X_A + 3179}{336(11X_A - 12)},\\
        c_{3/2}^A =& -\frac{5}{2}\pi, \nonumber\\
        c_2^A =& \Big[5\big(4572288 X_A^5 -20427120 X_A^4+158378220 X_A^3+\\
        &174965616 X_A^2+43246839 X_A-387973870\big) \Big]\\
        &/[9144576 (11X_A-12)],\\
        c_{5/2}^A =&  - \pi\frac{10520 X_A^3-7598 X_A^2+22415 X_A-27719}{192 (11X_A-12)},
    \end{split}
\end{equation}
where we also have $A \leftrightarrow B$ and $X_{A,B} = M_{A,B}/M$. Note that the coefficients are different from Ref.~\cite{Dietrich:2019kaq} in \nrtidalvtwo, which employed the PN expression derived in Ref.~\cite{Damour:2012yf}\footnote{The PN coefficients in Ref.~\cite{Damour:2012yf} were corrected in Ref.~\cite{Henry:2020ski, Narikawa:2023deu}, though the actual differences in the computed $\phi_T^{\rm PN}$ values are negligible.}. The expression employed for \nrtidalvthree\ uses the updated PN expression introduced in Ref~\cite{Henry:2020ski, Narikawa:2023deu}. We also note that the tidal parameters are given by
\begin{equation}
    \kappa_A = 3X_BX_A^4 \Lambda_A, \quad \Lambda_A = \frac{k_2^A}{C_A^5},
\end{equation}
where $C_A = M_A/R_A$ is the compactness of  star $A$ in isolation, $R_A$ is the radius, and $k_2^A$ is the tidal Love number of the star~\cite{Hinderer:2007mb}. 

For \nrtidalvthree, we will incorporate dynamical tides, i.e., $\Lambda_{A,B}$ will be a function of the orbital frequency $\omega_{\text{orb}} = \omega/2$ of the system~\cite{Steinhoff:2016rfi, Steinhoff:2021dsn}. This stems from the quadrupolar oscillations of the star due to f-mode excitations, which can be represented by a dynamical quadrupole moment obeying the equation of motion of a tidally driven harmonic oscillator plus relativistic corrections such as redshift, frame-dragging, and spin \cite{Steinhoff:2016rfi, Gupta:2020lnv, Steinhoff:2021dsn}. The dynamical Love number for nonspinning systems can be approximately expressed in terms of an enhancement factor $k_\ell^{\rm eff}$ as $k_2 
\rightarrow k_2({\omega}) = k_2 k_2^{\rm eff}({\omega})$ (for $\ell = 2$), where
\begin{equation}\label{eq: k2eff}
    \begin{split}
        k_2^{\text{eff}} =& u_2 + v_2 \Bigg[\frac{\omega_{02}^2}{\omega_{02}^2 + (2\omega_{\text{orb}})^2}
        + \frac{\omega_{02}^2}{2\sqrt{\epsilon_2}\hat{t}|\tilde{\Omega}'|(2\omega_{\text{orb}})^2}\\
        +& \frac{\omega_{02}^2}{\sqrt{\epsilon_2}(2\omega_{\text{orb}})^2}Q_{22}(\hat{t}) \Bigg],
    \end{split}
\end{equation}
where $\omega_{0\ell}$ is the fundamental mode frequency, $\tilde{\Omega}'$ is the derivative (with respect to the gravitational radiation reaction) of the ratio of the mode and tidal forcing frequencies (which in the nonspinning case is $\tilde{\Omega}' = -3/8$),
\begin{equation}
    \begin{split}
        Q_{22}(\hat{t}) =& \cos(|\tilde{\Omega}'|\hat{t}^2)\int_{-\infty}^{\hat{t}}\sin(|\tilde{\Omega}'|s^2)ds \\ &-\sin(|\tilde{\Omega}'|\hat{t}^2)\int_{-\infty}^{\hat{t}}\cos(|\tilde{\Omega}'|s^2)ds,
    \end{split}
\end{equation}	
and 
\begin{equation}
    \hat{t} = \frac{8}{5\sqrt{\epsilon_2}} \left[1 - \left(\frac{\omega_{02}}{2\omega_{\text{orb}}}\right)^{5/3}\right], \quad \epsilon_2 = \frac{256\mu M^{2/3} \omega_{02}^{5/3}}{5\cdot 2^{5/3}},
\end{equation}
where $\mu = M_A M_B/M$ and the quadrupole coefficients are $u_2 = 1/4$ and $v_2 = 3/4$.
The enhancement factor Eq.~\eqref{eq: k2eff} results from a two-timescale approximation for the dynamical tidal quadrupole with a Newtonian estimate for the orbital evolution~\cite{Steinhoff:2016rfi, Steinhoff:2021dsn}, which should be extended to higher PN orders in future work.

\begin{figure}[t]
    \centering
    \includegraphics[width = \linewidth]{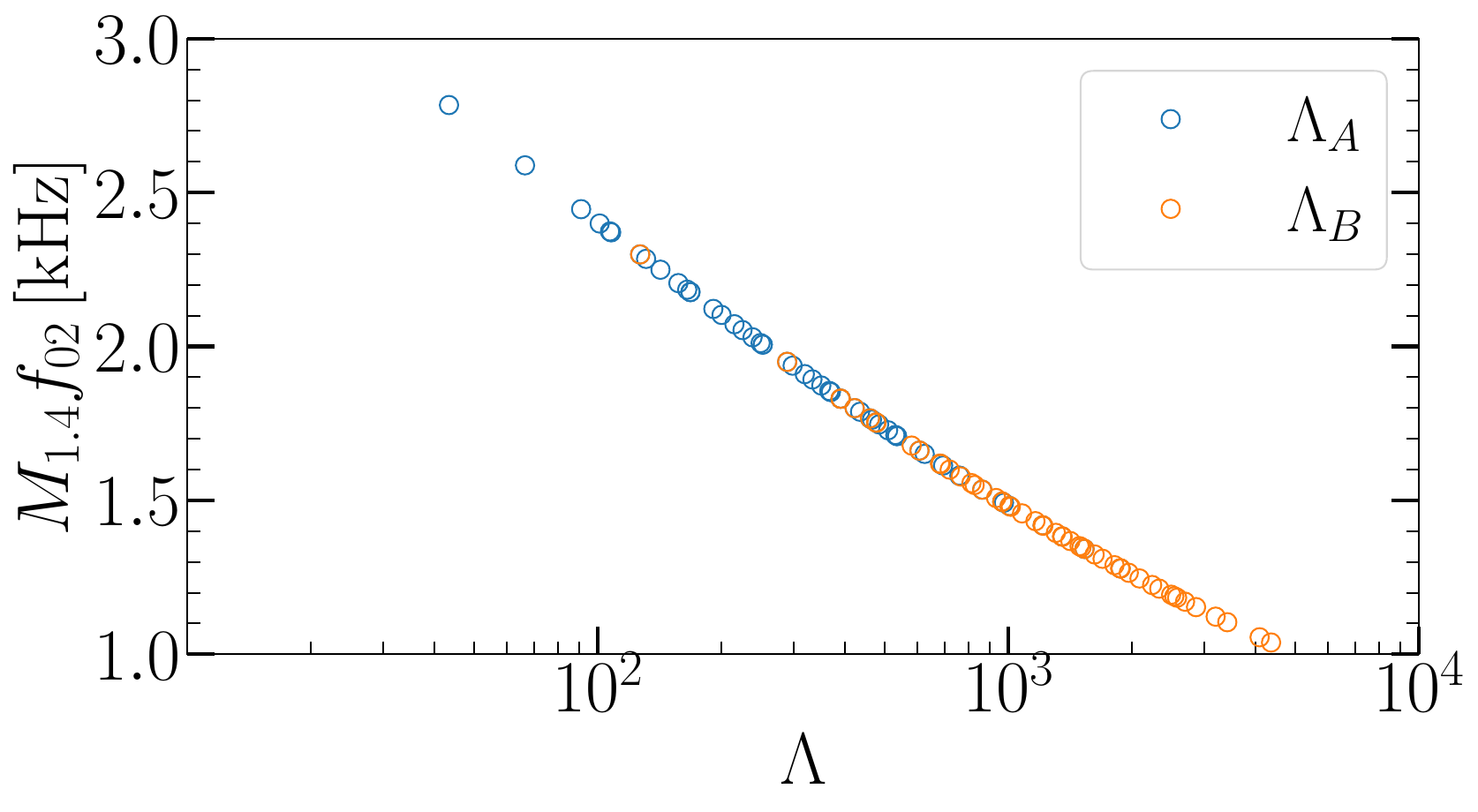}
    \caption{The universal relation between the $f$-mode frequency (in kHz) and $\Lambda$ used for stars A and B.} 
    \label{fig: univrels}
\end{figure}

Since $k_2^{\text{eff}}$ is a tidal enhancement factor, $k_2^{\text{eff}} \rightarrow 1.0$ at $\omega_{\text{orb}} \rightarrow 0$, which means that the tidal deformability of the NS is at its adiabatic value at large distances (i.e., $r \rightarrow \infty)$ from its partner. The fundamental frequency (in kHz) $M_{1.4}f_{02} = (M/1.4M_{\odot})\omega_{02}/(2\pi)$ rescaled to a $1.4M_{\odot}$ NS can be obtained using the quasi-universal relation found in~\cite{Sotani:2021kiw}
\begin{equation}\label{eq: univrel}
    M_{1.4}f_{02} \, [{\rm kHz}] = \sum_{i=0}^5 g_i(\log_{10} \Lambda_2)^i,
\end{equation}
where 
\begin{equation}
    \begin{split}
        g_i \in& [4.2590, -0.47874, -0.45353,\\
        &0.14439, -0.016194, 0.00064163].
    \end{split}
\end{equation}
This relation is shown in Fig.~\ref{fig: univrels} for various values of $\Lambda$, for stars A and B.
Then, for $\ell = 2$, our tidal parameters for \nrtidalvthree\ contains the modification
\begin{equation}
    \kappa_{A,B} \rightarrow \kappa_{A,B}(\hat{\omega}) = \kappa_{A,B} k_{2, A,B}^{\text{eff}}(\hat{\omega}).
\end{equation}

\begin{figure}[t]
    \centering
    \includegraphics[width = \linewidth]{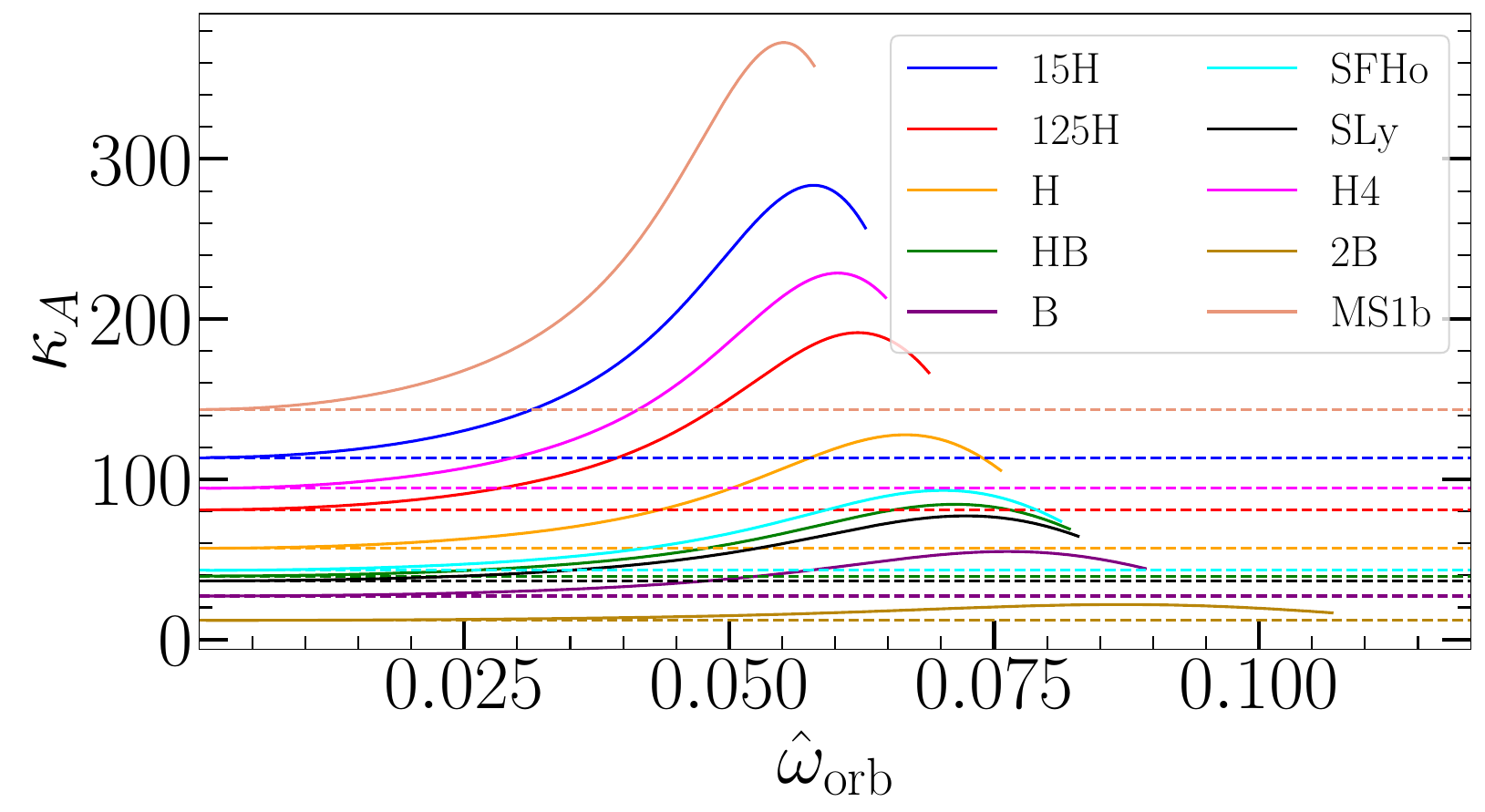}
    \caption{Dynamical tidal parameter in the time-domain as a function of orbital frequency ($\hat{\omega}_{\text{orb}} = \hat{\omega}/2$) for different EoSs, given $q = 1.0$. Note that each curve starts at the given constant value of the tidal parameter, which is provided by the dashed lines. The curves terminate at the merger frequency $\hat{\omega}_{\rm mrg} = 2\hat{\omega}_{\rm orb}^{\rm mrg}$.} 
    \label{fig: kappaA}
\end{figure}
Note that, as constructed, $\kappa$ recovers at large distances (very small frequency) the original (constant) adiabatic value (see Fig.~\ref{fig: kappaA}). 

\subsection{Calibrating the time-domain, tidal phase}

We then write the effective representation of the time-domain tidal phase in \nrtidalvthree\ as:
\begin{equation}\label{eq: Phi_TNRT3}
    \phi_T^{\text{NRT3}} = -\kappa_A(\hat{\omega}) c_{\text{Newt}}^Ax^{5/2} P_{\text{NRT3}}^A(x) + [A\leftrightarrow B],
\end{equation}
where we use the following functional form
\begin{equation}\label{eq: Padetime}
\begin{split}
        &P_{\text{NRT3}}^A(x)\\
        &=\frac{1 + n_1^A x + n_{3/2}^Ax^{3/2} + n_2^Ax^2 + n_{5/2}^Ax^{5/2} + n_3^Ax^3}{1 + d_1^Ax + d_{3/2}^A x^{3/2}},
\end{split}
\end{equation}
and the same with $A \leftrightarrow B$.
The exact functional form employed in this work is based on numerous tests for various different fitting functions and provided overall the best performance. However, it is clearly an assumption, and also other forms could have been used. We also note that a polynomial form of the fitting function could describe the 55 EOB-NR tidal phase hybrids well, but it is not guaranteed to work for extreme cases, i.e., large masses and/or tidal deformabilities. For some of these configurations, the polynomial can become very large for some mass- and tidal-deformability-dependent combination of coefficients at low frequencies.

Taking the Taylor series expansion of Eq.~\eqref{eq: Phi_TNRT3} and comparing it with Eq.~\eqref{eq: PNtime} allows us to enforce the following constraints to ensure consistency with the PN expression:
\begin{equation}
    \begin{split}
        n_1^A =& c_1^A + d_1^A,\\
        n_{3/2}^A =& \frac{c_1^Ac_{3/2}^A - c_{5/2}^A - c_{3/2}^Ad_1^A + n_{5/2}^A}{c_1^A},\\
        n_2^A =& c_2^A + c_1^A d_1^A, \\
        d_{3/2}^A =& -\frac{c_{5/2}^A + c_{3/2}^Ad_1^A - n_{5/2}^A}{c_1^A},
    \end{split}
\end{equation}
and similar constraints for $A \rightarrow B$.\footnote{This is a choice as to what coefficients will be constrained by the PN tidal phase. In principle, we can constrain different combinations of coefficients, e.g. including $n_{5/2}^A$, and $d_1^A$, by constructing a linear system (like the above), as long as there exists a solution in that system. However, we find the above choice is robust and allows for reasonable results also outside the calibration region.} This means that we are left with six unknown parameters, $(n_{5/2}^{A,B}, n_3^{A,B}, d_1^{A,B})$, which can be fitted to the data. However, to make sure that the parameters remain symmetric with respect to stars $A$ and $B$, we can impose additional constraints on these parameters. We find the following functional form to be sufficient for our model:
\begin{equation}\label{eq: parametrization_time}
    \begin{split}
        p_i^{A,B}(\hat{\omega}) =& a_{i,0} + a_{i,1} X_{A,B} \\
        &+ a_{i,2} (\kappa_{A,B}+1)^{\alpha}\\
        &+ a_{i,3} X_{A,B}^{\beta} , \, \text{for} \, p_i \in [n_{5/2}, n_3],\\
        d_1^{A,B} =& d_{1,0} + d_{1,1} X_{A,B} + d_{1,2} X_{A,B}^{\beta}.
    \end{split}
\end{equation}

\begin{table}[t] 
\renewcommand*{\arraystretch}{1.4}
\caption{\label{table: nrt3 paramstime} Parameters for the \nrtidalvthree\ timed-domain approximant. The table lists the fitting parameters for the Padé approximant, given in Eqs.~\eqref{eq: Padetime} and \eqref{eq: parametrization_time}.}
\begin{ruledtabular}
    \begin{tabular}{c|cccc}
    Tides & \multicolumn{4}{c}{$k_2^{\text{eff}}(\hat{\omega})$ (Eq.~\eqref{eq: k2eff})} \\
    \hline
    $\alpha$ & \multicolumn{4}{c}{0.762130731}\\
    $\beta$ & \multicolumn{4}{c}{-0.577611983}\\
    \hline
    & $a_0$ & $a_1$ & $a_2$ & $a_3$ \\
    \hline
    $n_{5/2}$& 10973.4227& -7775.85588& -0.113688274& -4483.08830\\
    $n_{3}$ &-11424.2843& 8026.17700& -0.379126345& 4665.72647\\
    \hline
    & $d_{1,0}$ & $d_{1,1}$ & $d_{1,2}$ & -\\
    \hline
    $d_{1}$& -546.799216& 379.280986& 223.018238& -\\
        \end{tabular}
\end{ruledtabular}
\end{table}

This leaves us with the 13 parameters $a_i$'s, $d_{1,i}$'s, $\alpha$, and $\beta$ that we now determine by fitting the function to the EOB-NR hybrid tidal phase data. We note that this parametrization, Eq.~\eqref{eq: parametrization_time}, should ensure that the coefficients of the Padé approximants themselves are functions of the mass-ratio and tidal deformability, thus making itself applicable to a wide range of EoSs. This is unlike the attempt that was made in Ref.~\cite{ODell:2023hbm} where each set of coefficients ($n$'s and $d$'s) of the Padé approximant in Eq.~\eqref{eq: Padetime} was specifically determined for two equations of state describing two SpEC waveforms. The parameters for the time-domain NRtidal phase are shown in Table~\ref{table: nrt3 paramstime}.  

\begin{figure}[hbt!]
    \centering
    \includegraphics[width = \linewidth]{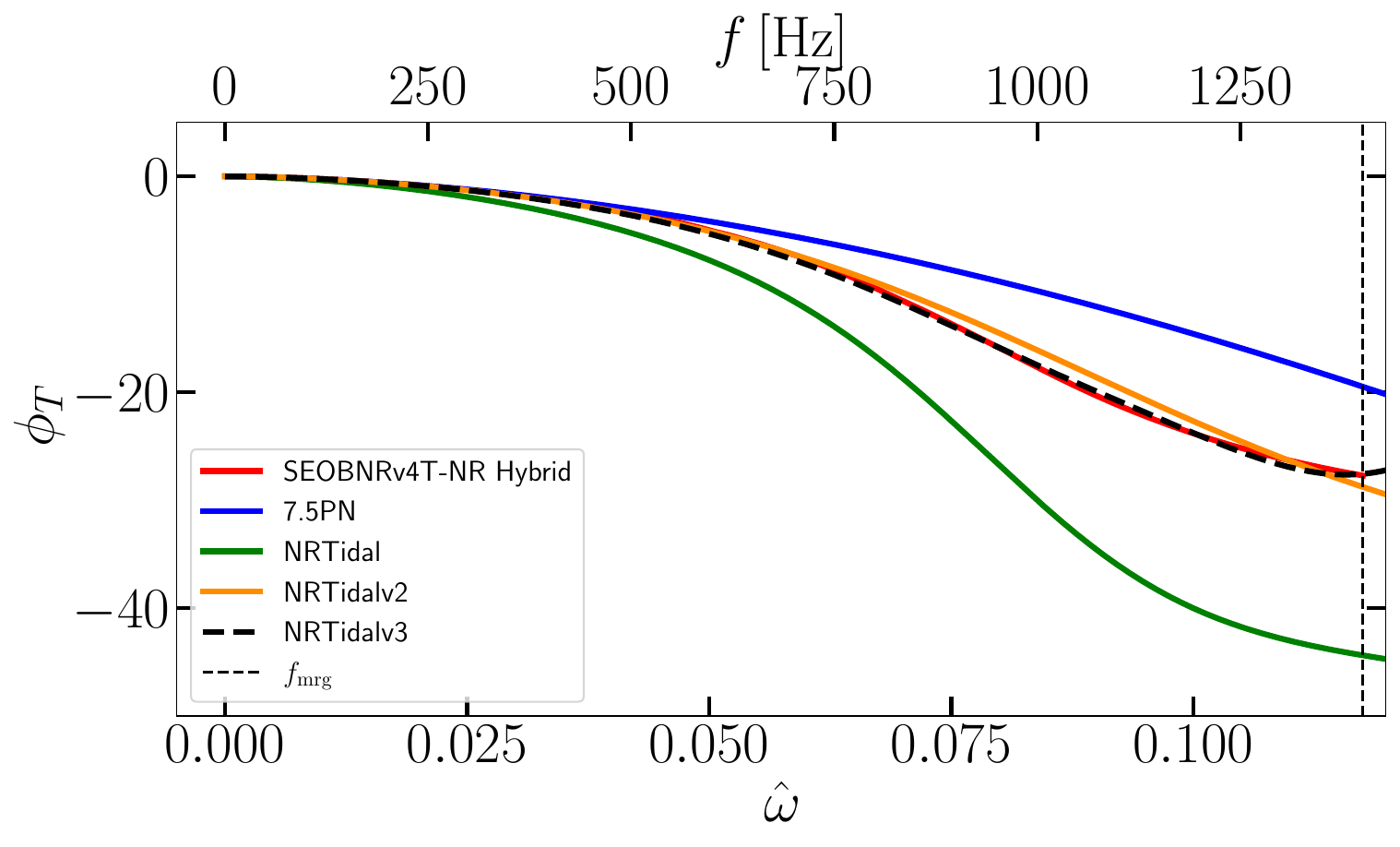}
    \caption{Time-domain tidal phase contributions calculated from different Models for the configuration of the NR waveform SACRA:15H\_125\_146\_00155\_182\_135. We also show the EOB-NR phase for comparison. The vertical dashed line marks the merger frequency.} 
    \label{fig: time-domain ex}
\end{figure}

We show in Fig.~\ref{fig: time-domain ex} the comparison between $\phi_T^{\text{NRT3}}$ and the previous versions of the NRTidal model, as well as the 7.5PN approximation of the tidal phase, for the SACRA:15H\_125\_146\_00155\_182\_135 configuration.

\begin{figure}[hbt!]
    \centering
    \includegraphics[width = \linewidth]{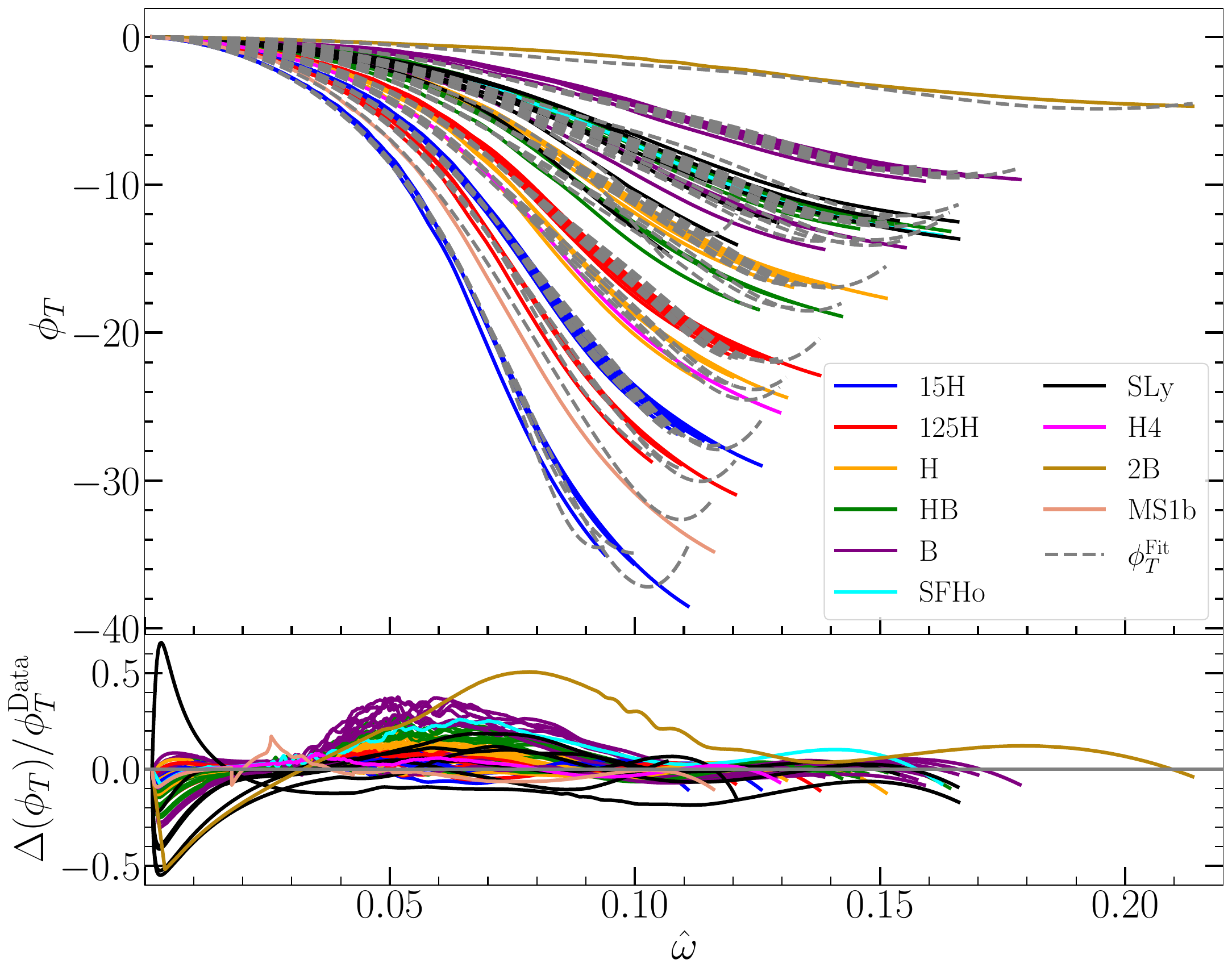}
    \caption{\textit{Top}: The fits (dashed lines) plotted on top of the time-domain hybridized tidal phase data (solid lines). \textit{Bottom}: Fractional differences between the fits and the data. The EOB-NR hybrid tidal phases are color-indicated by their EoS.}
    \label{fig: time domain fits}
\end{figure}

The results of the fitting in the time-domain are shown in Fig. \ref{fig: time domain fits}, together with the fractional differences from the hybridized data. We note the ``curving up" of the fits near the merger part of the waveforms; this is primarily due to the dynamical nature of the tides that were incorporated here. For the fits, significant fractional differences $\Delta \phi_T/\phi_T^{\rm Data} = (\phi_T^{\rm NRT3} - \phi_T^{\rm Data})/\phi_T^{\rm Data}$ are found within the earlier frequencies (where very small phase magnitudes can correspond to large fractional differences) and in the vicinity of the hybridization window, where most of the noise from the NR data can persist even after filtering. 

\section{\label{section: freq domain nrt3} Frequency Domain Representation}

\subsection{PN knowledge and SPA}
We now construct the frequency-domain phase using Eq.~\eqref{eq: PNtime} in the stationary phase approximation (SPA) \cite{Droz:1999qx, Damour:2012yf}
\begin{equation}\label{eq: SPA}
    \frac{d^2\psi_{T}(\omega)}{d\omega^2} = \frac{1}{\omega}\frac{d\phi_T(\omega)}{d\omega}.
\end{equation}
The 7.5PN order analytical expression~\cite{Henry:2020ski, Narikawa:2023deu} of the frequency-domain tidal phase with constant tidal deformability is then
\begin{equation}\label{eq: PNfreq}
    \begin{split}
        \psi_T^{\rm PN} =& -\kappa_A \bar{c}_{\text{Newt}}^A x^{5/2} \Big(1 + \bar{c}_1^A x + \bar{c}_{3/2}^A x^{3/2} \\
        & +\bar{c}_2^A x ^2 + \bar{c}_{5/2}^Ax^{5/2}\Big) + [A \leftrightarrow B],
    \end{split}
\end{equation}
with
\begin{equation}
    \begin{split}
        \bar{c}_{\text{Newt}}^A =& \frac{3(12 - 11 X_A)(X_A + X_B)^2}{16X_A X_B^2},\\
        \bar{c}_1^A =&-5 \frac{260X_A^3 - 2286X_A^2 - 919X_A + 3179}{672(11X_A - 12)},\\
        \bar{c}_{3/2}^A =& -\pi,\\
        \bar{c}_2^A =& \Big[5\big(4572288 X_A^5 -20427120 X_A^4+158378220 X_A^3\\
        &+174965616 X_A^2+43246839 X_A-387973870\big) \Big]\\
        &/[27433728(11X_A-12)],\\
        \bar{c}_{5/2}^A =&  - \pi\frac{10520 X_A^3-7598 X_A^2+22415 X_A-27719}{672 (11X_A-12)},
    \end{split}
\end{equation} 
and similarly for $A \leftrightarrow B$. We use Eq.~\eqref{eq: PNfreq} in constraining the frequency-domain phase for \nrtidalvthree.

\subsection{Dynamical tides in the frequency domain}

One of the main differences between \nrtidalvtwo\ and \nrtidalvthree\ is the inclusion of dynamical tides in the latter. Therefore, it is not straightforward to define a constant effective $\kappa_{\text{eff}}$ (as was done in \nrtidal~\cite{Dietrich:2017aum} and \nrtidalvtwo~\cite{Dietrich:2019kaq}) that can be used in both the time- and frequency-domain tidal phases, see also Ref.~\cite{Schmidt:2019wrl}. We model the entire frequency-domain dynamical tidal deformability (or Love number) in the same manner as in Eq.~\eqref{eq: k2eff}, serving as a frequency-dependent correction/enhancement factor to the adiabatic Love number. This is unlike in Ref.~\cite{Schmidt:2019wrl}, where the frequency-domain dynamical $f$-mode tidal effects were rewritten as an additive correction to the adiabatic component. 
Hence, we write the PN expression Eq.~\eqref{eq: PNfreq}, with $\kappa \rightarrow \kappa(\omega)$ as (for $\ell = 2$)
\begin{equation}\label{eq: PNfreqdyn}
    \psi_{T,\text{dyn}}^{\rm PN} = \bar{k}_2^{\text{eff}}(\hat{\omega})\psi_T^{\rm PN}(\hat{\omega}),
\end{equation}
where $\bar{k}_2^{\text{eff}}(\hat{\omega})$ is the frequency-domain effective tidal enhancement factor. Note that in this case, the same $\bar{k}_2^{\rm eff}$ applies to both stars, which makes the calculation of the frequency-domain tidal phase more convenient. We can then solve for $\bar{k}_2^{\text{eff}}$ by substituting Eq.~\eqref{eq: PNfreqdyn} into the SPA in Eq.~\eqref{eq: SPA}, yielding the following differential equation:
\begin{equation}\label{eq: dek2}
    \frac{d^2\bar{k}_2^{\text{eff}}}{d\hat{\omega}^2}\psi_T^{\rm PN} + 2\frac{d\bar{k}_2^{\text{eff}}}{d\hat{\omega}}\frac{d\psi_T^{\rm PN}}{d\hat{\omega}} + \bar{k}_2^{\text{eff}}\frac{d^2\psi_T^{\rm PN}}{d\hat{\omega}^2} = \frac{1}{\hat{\omega}}\frac{d\phi_{T,\text{dyn}}^{\rm PN}}{d\hat{\omega}}.
\end{equation}
We can then solve for $\bar{k}_2^{\text{eff}}$ by imposing the following initial conditions:
\begin{equation}\label{eq: init_cond}
    \bar{k}_2^{\text{eff}}(0) = 1, \quad \frac{d\bar{k}_2^{\text{eff}}(0)}{d\hat{\omega}} = 0,
\end{equation}
for it to behave similarly to the enhancement factor in the time-domain. Solving for $\bar{k}_2^{\text{eff}}$, however, for different values of the tidal parameter and mass can be computationally inefficient, especially when implemented in \texttt{lalsuite}. For this reason, we introduce a phenomenological representation of the enhancement factor of the frequency-domain Love number $\bar{k}_{2, \text{rep}}^{\text{eff}}$ as follows:
\begin{equation}
\begin{split}\label{eq: k2bareff}
        \bar{k}_{2, \text{rep}}^{\text{eff}} =& 1 + \frac{s_1-1}{\exp[-s_2(\hat{\omega} - s_3)] + 1} - \frac{s_1 - 1}{\exp(s_2s_3) + 1}\\
        & - \frac{s_2(s_1-1)\exp(s_2s_3)}{[\exp(s_2 s_3) + 1]^2}\hat{\omega},
\end{split}
\end{equation}
where the parameters $s_i,\,\, (i = 1, 2, 3)$ are constrained using $s_{i,j},\,\, (j = 0, 1, 2)$:
\begin{equation}
    s_i = s_{i,0} + s_{i,1} \kappa_{\text{eff}} + s_{i,2}q\kappa_{\text{eff}},
\end{equation}
and $ \kappa_{\text{eff}}$ is the effective tidal parameter
\newtext{
\begin{equation}
    \kappa_{\text{eff}} = \frac{2}{13}\left\{\left[1 + \frac{12X_B}{X_A}\left(\frac{X_A}{C_A}\right)^5k_2^A \right] + [A\leftrightarrow B] \right\}.
\end{equation}
}
The fitting parameters $s_{i,j}$ for $s_i$ are given in Table~\ref{table: k2eff params}. The fitting formula was chosen to ensure the monotonicity of the enhancement factor $\bar{k}_2^{\rm eff}$ up to some maximum at least near (or beyond) the merger frequency, and that it follows the initial conditions laid out in Eq.~\eqref{eq: init_cond}. The fitting formula satisfies the numerically calculated frequency-domain effective enhancement factor, with a maximum (deviation) error of $1.8 \%$ and mean error of $0.2 \%$ \newtext{(see Appendix~\ref{section: k2eff})}.
\begin{table}[t]
    \caption{\label{table: k2eff params} Fitting parameters $s_{i,j}$ for $s_i$ that comprise the representation of the frequency-domain Love number enhancement factor, given in Eq.~\eqref{eq: k2bareff}.}
    \begin{ruledtabular}
        \begin{tabular}{c|c|c|c|c}
            \backslashbox{$i$}{$j$} & 0 & 1 & 2\\
            \hline
            $1$& \newtext{1.273000423}& 0.00364169971& 0.00176144380\\
            $2$& \newtext{27.8793291}& 0.0118175396& -0.00539996790\\
            $3$& 0.142449682& $-1.70505852 \times 10^{-5}$& $3.38040594 \times 10^{-5}$
        \end{tabular}
    \end{ruledtabular}
\end{table}

\subsection{\nrtidalvthree\ in the frequency domain}

The previously derived representation $\bar{k}_{2, \text{rep}}^{\text{eff}}$ in Eq.~\eqref{eq: k2bareff}, which we write from this point onward for convenience as just $\bar{k}_{2}^{\text{eff}}$, can finally be used for building the frequency-domain \nrtidalvthree\ phase:
\begin{equation}\label{eq: Psi_TNRT3}
    \psi_T^{\text{NRT3}} = -\bar{\kappa}_A(\hat{\omega})\bar{ c}_{\text{Newt}}^Ax^{5/2} \bar{P}_{\text{NRT3}}^A(x) + [A\leftrightarrow B],
\end{equation}
where 
\begin{equation}
    \bar{\kappa}_{A,B}(\hat{\omega}) = \kappa_{A,B}\bar{k}_2^{\text{eff}}(\hat{\omega}),
\end{equation}
\begin{equation}\label{eq: Padefreq}
\begin{split}
        &\bar{P}_{\text{NRT3}}^A(x)\\
        &= \frac{1 + \bar{n}_1^A x + \bar{n}_{3/2}^Ax^{3/2} + \bar{n}_2^Ax^2 + \bar{n}_{5/2}^Ax^{5/2} + \bar{n}_3^Ax^3}{1 + \bar{d}_1^Ax + \bar{d}_{3/2}^A x^{3/2}},
\end{split}
\end{equation}
with 
\begin{equation}
    \begin{split}
        \bar{n}_1^A =& \bar{c}_1^A + d_1^A,\\
        \bar{n}_{3/2}^A =& \frac{\bar{c}_1^A\bar{c}_{3/2}^A - \bar{c}_{5/2}^A - \bar{c}_{3/2}^A\bar{d}_1^A + \bar{n}_{5/2}^A}{\bar{c}_1^A},\\
        \bar{n}_2^A =& \newtext{\bar{c}_2^A + \bar{c}_1^A \bar{d}_1^A}\\
        \bar{d}_{3/2}^A =& -\frac{\bar{c}_{5/2}^A + \bar{c}_{3/2}^A\bar{d}_1^A - \bar{n}_{5/2}^A}{\bar{c}_1^A}.
    \end{split}
\end{equation}
The remaining parameters are given by 
\begin{equation}\label{eq: freeconstraint}
\begin{split}
    \bar{p}_i^{A,B}(\hat{\omega}) =&\bar{ a}_{i,0} + \bar{a}_{i,1} X_{A,B} \\
    &+ \bar{a}_{i,2} (\kappa_{A,B}+1)^{\bar{\alpha}}\\
    &+ \bar{a}_{i,3} X_{A,B}^{\bar{\beta}} , \, \text{for} \, \bar{p}_i \in [\bar{n}_{5/2}, \bar{n}_3]\\
    \bar{d}_1^{A,B} =& \bar{d}_{1,0} + \bar{d}_{1,1} X_{A,B} + \bar{d}_{1,2} X_{A,B}^{\bar{\beta}}.
\end{split}
\end{equation}
The values of the fitting parameters are found in Table~\ref{table: nrt3 paramsfreq}. For comparison, recently, a phenomenological tidal approximant (with the Padé function up to the $x^2$ term in the denominator) was also developed in Ref.~\cite{Gamba:2023mww}, where some coefficients of the Padé were constrained to the PN expressions for the tidal and self-spin contributions, while the free coefficients were fitted to the \teobresums\ waveform model; here, two sets of these free coefficents were obtained for $q = 1$ and $q > 1$. However, for \nrtidalvthree, we consider the tides to be dynamical, and the free coefficients are themselves explicit functions of the individual masses and/or tidal deformabilities of the system (as indicated in Eq.~\eqref{eq: freeconstraint}). 

\begin{table}[t] 
    \renewcommand*{\arraystretch}{1.4}
    \caption{\label{table: nrt3 paramsfreq} Parameters for the \nrtidalvthree\ frequency-domain approximant. The table include the fitting parameters for the dynamical love number enhancement, as well as the respective fitting parameters for the Padé approximant given in Eq.~\eqref{eq: Padefreq}.}
    \begin{ruledtabular}
        \begin{tabular}{c|cccc}
        Tides & \multicolumn{4}{c}{$\bar{k}_{2, \text{rep}}^{\text{eff}}$ (Eq.~\eqref{eq: k2bareff})} \\
        \hline
        $\bar{\alpha}$& \multicolumn{4}{c}{-0.00808155404} \\
        $\bar{\beta}$& \multicolumn{4}{c}{-1.13695919} \\
        \hline
        & $\bar{a}_0$ & $\bar{a}_1$ & $\bar{a}_2$ & $\bar{a}_3$\\
        \hline
        $\bar{n}_{5/2}$& -940.654388& 626.517157& 553.629706& 88.4823087\\
        $\bar{n}_{3}$& 405.483848& -425.525054& -192.004957& -51.0967553\\
        \hline
        & $\bar{d}_{1,0}$ & $\bar{d}_{1,1}$ & $\bar{d}_{1,2}$ & -\\
        \hline
        $d_{1}$& 3.80343306& -25.2026996& -3.08054443& -\\
            \end{tabular}
    \end{ruledtabular}
\end{table}

\begin{figure}[hbt!]
    \centering
    \includegraphics[width = \linewidth]{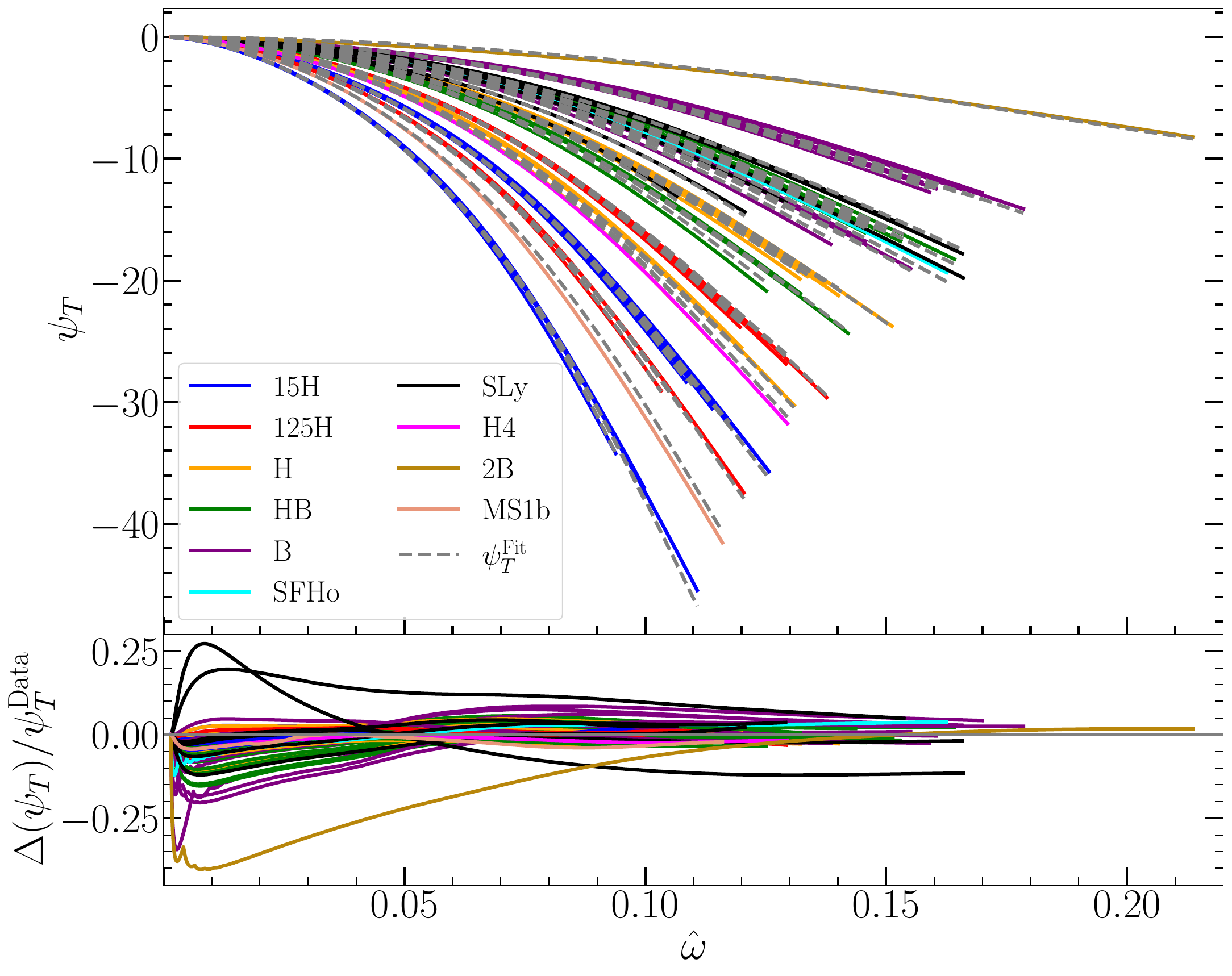}
    \caption{\textit{Top}: The fits plotted on top of the frequency-domain hybridized tidal phase data. \textit{Bottom}: Fractional differences between the fits and the data, where $\Delta \psi_T = \psi_T^{\rm Fit} - \psi_T^{\rm Data}$.} 
    \label{fig: freqdomain fits}
\end{figure}

\begin{figure}[hbt!]
    \centering
    \includegraphics[width = \linewidth]{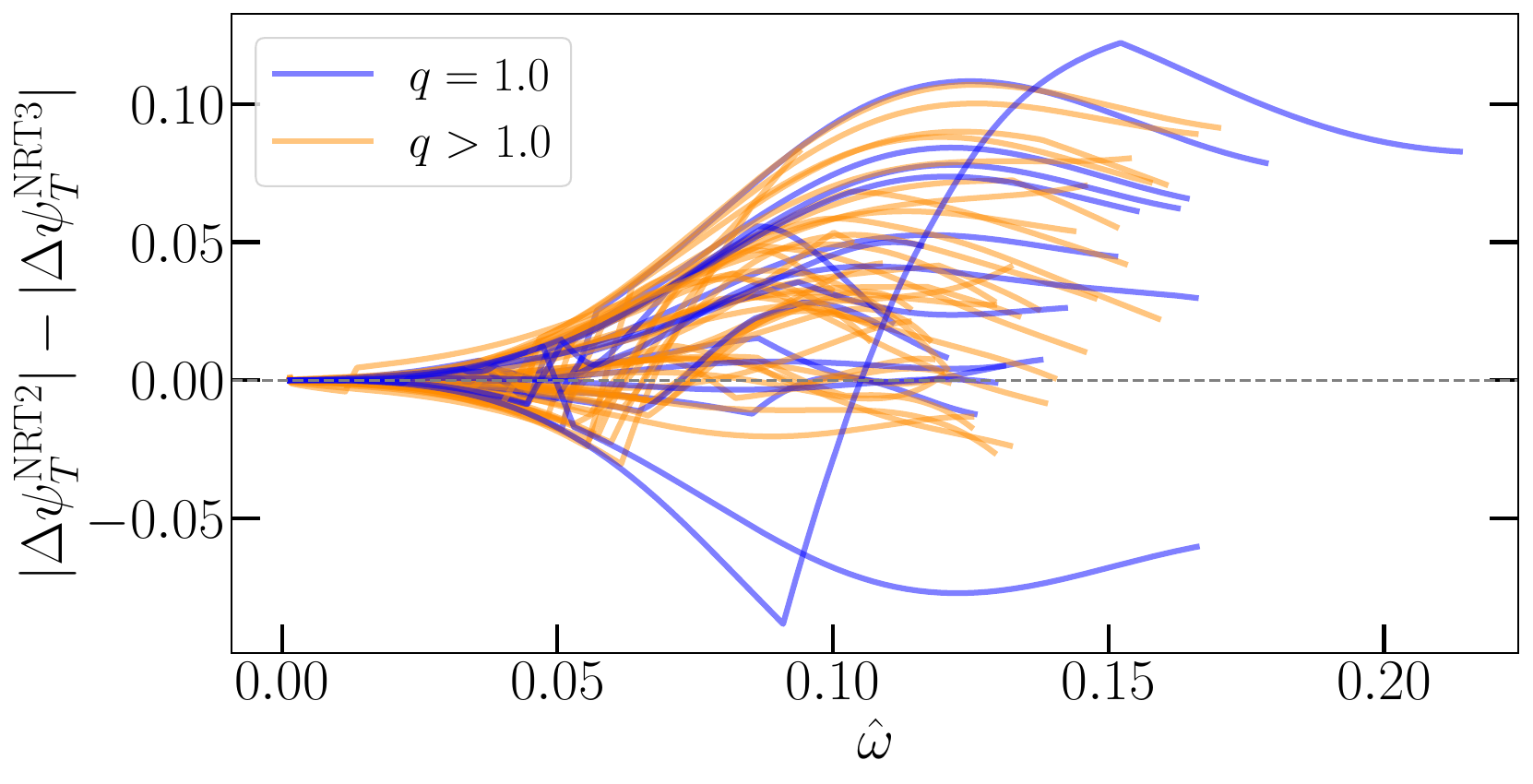}
    \caption{The difference between the absolute fractional differences of \nrtidalvthree\ and \nrtidalvtwo\ with respect to the hybrid data. The curves for $q = 1.0$ and $q>1.0$ are indicated. We observe \nrtidalvthree\ to perform better (i.e., most curves are above the zero line) than \nrtidalvtwo\ for both equal and unequal mass ratios.}
    \label{fig: absdiff}
\end{figure}

We show the results of the fitting in Fig.~\ref{fig: freqdomain fits}. In this case, the significant fractional differences with respect to the EOB-NR hybrid data $\Delta\psi_T^{\rm NRT3} = \Delta \psi_T/\psi_T^{\rm Data} = (\psi_T^{\rm NRT3} - \psi_T^{\rm Data})/\psi_T^{\rm Data}$ are found again early in the frequencies and around the hybridization window. Furthering the investigation of the performance of \nrtidalvthree, we plot the difference between the absolute fractional differences of \nrtidalvtwo\ and \nrtidalvthree\ with respect to the data, that is, $|\Delta\psi_T^{\rm NRT2}| - |\Delta\psi_T^{\rm NRT3}|$, against $\hat{\omega}$, and show the results in Fig.~\ref{fig: absdiff}. We also indicate in the figure $|\Delta\psi_T^{\rm NRT2}| - |\Delta\psi_T^{\rm NRT3}| = 0$, denoted by a black dotted horizontal line. Above this line, \nrtidalvtwo\ has greater error with respect to the data than \nrtidalvthree\, and below this line, \nrtidalvthree\ has a greater error. We observe that \nrtidalvthree\ performs overall better than \nrtidalvtwo\ for most of the configurations, both for $q=1.0$ (yellow curves) and $q>1.0$ (blue curves).

\begin{figure*}[hbt!]
\centering
\includegraphics[width=0.47\linewidth]{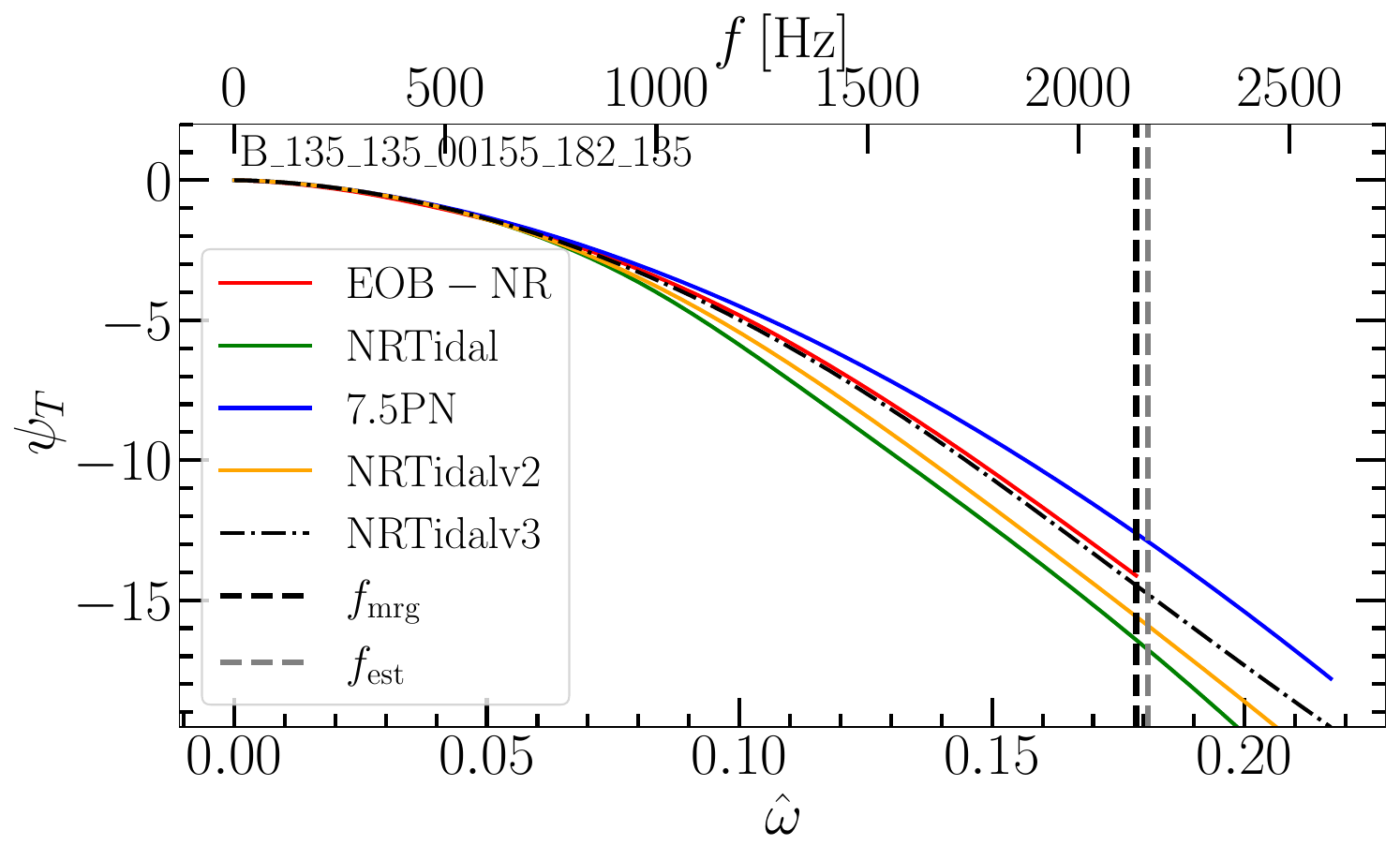}
\includegraphics[width=0.47\linewidth]{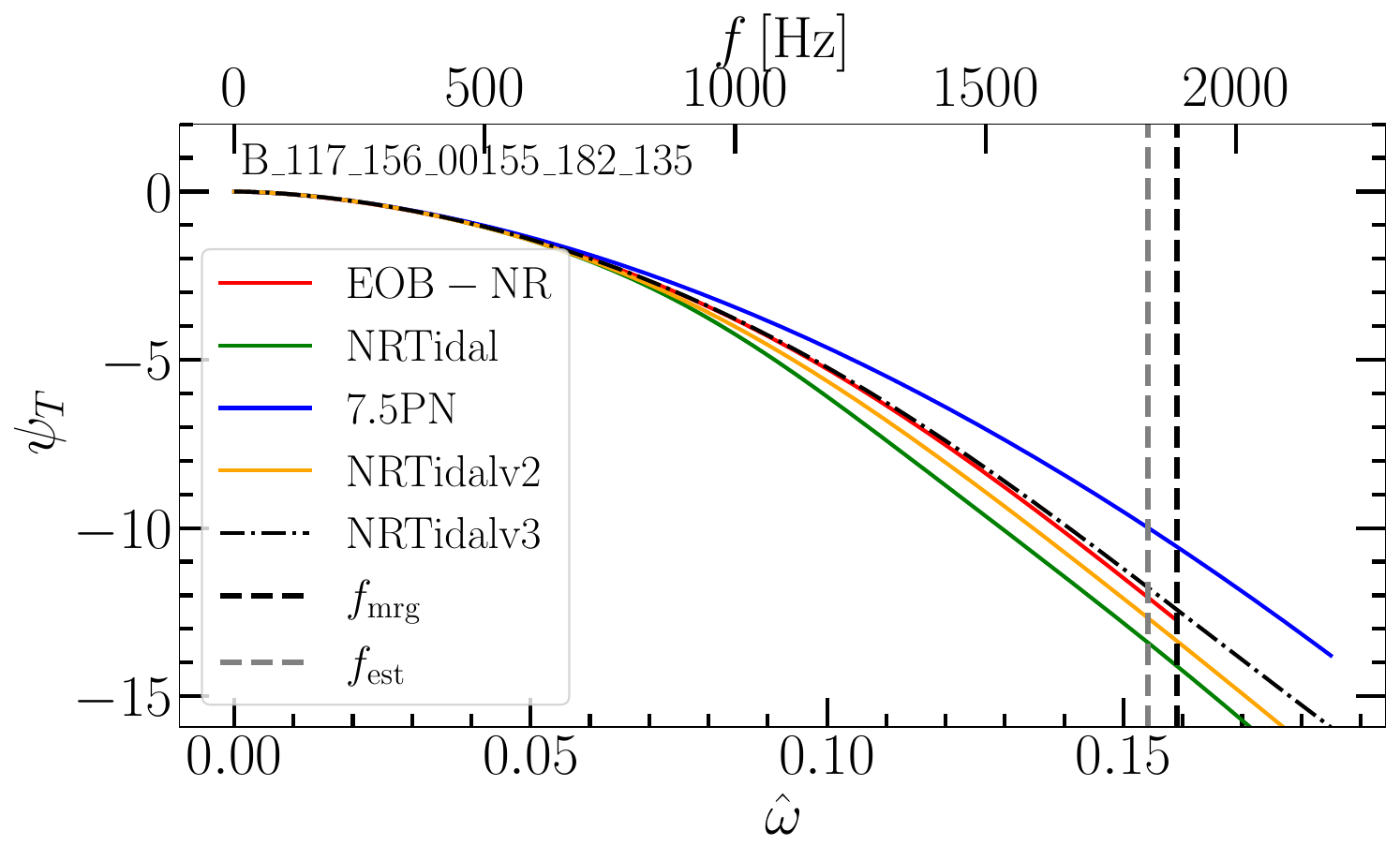}
\includegraphics[width=0.47\linewidth]{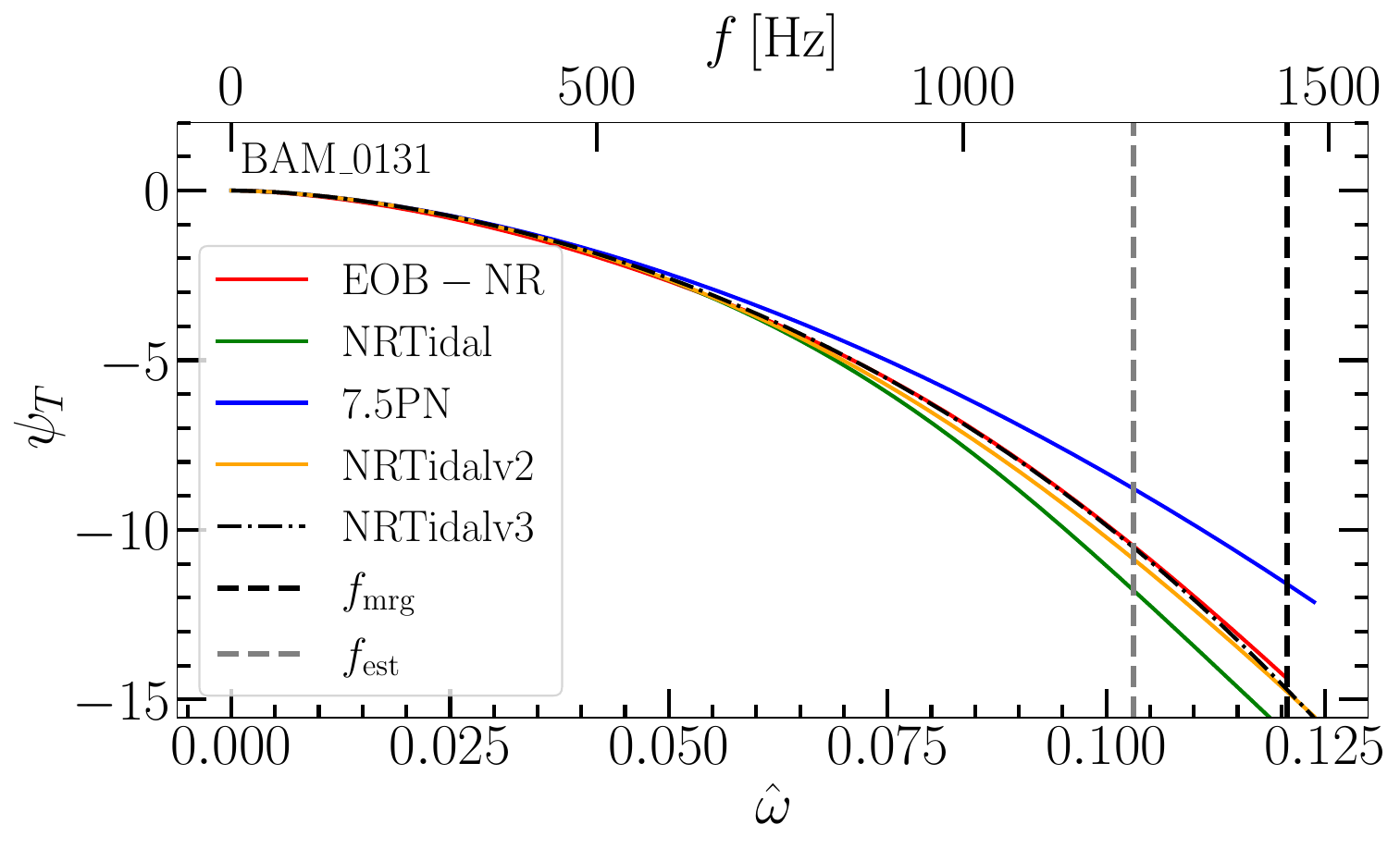}
\includegraphics[width=0.47\linewidth]{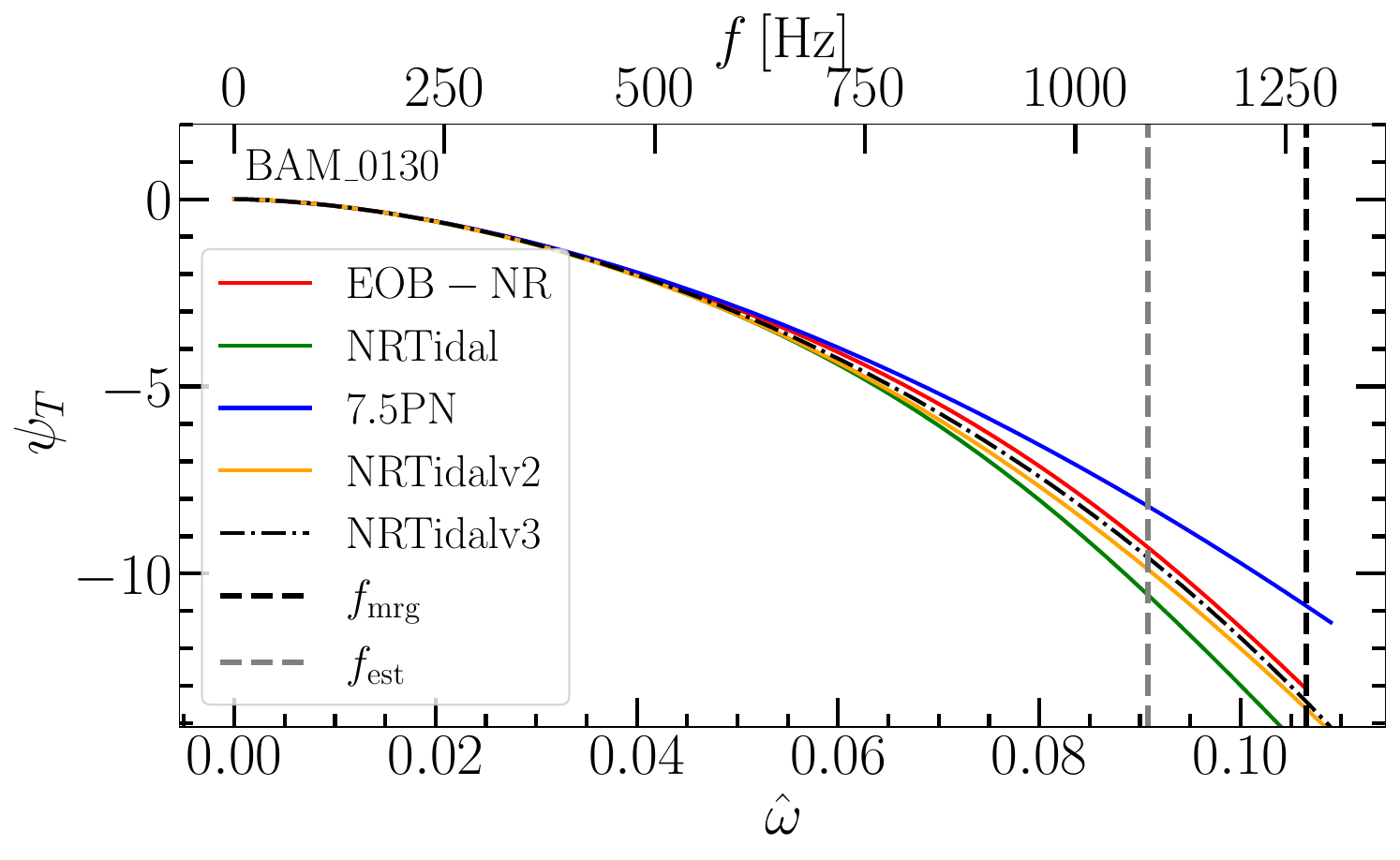}
\caption{Frequency-domain, EOB-NR hybrid tidal phase contributions of SACRA:B\_135\_135\_00155\_182\_135, SACRA:B\_117\_156\_00155\_182\_135, BAM\_0131, and BAM\_0130, representing mass ratios [1.0, 1.33, 1.75, 2.0]. The tidal phase contributions for \nrtidal, \nrtidalvtwo, \nrtidalvthree, and 7.5PN approximants are also shown.  We also indicate the estimated merger frequency $f_{\rm est}$ from the fitting function in Sec.~\ref{subsection: Merger Frequency}, and the merger frequency of the NR simulations $f_{\rm mrg}$. The tidal contributions using other tidal models are also shown for comparison.} 
    \label{fig: freqdomain tidal phases}
\end{figure*}

In addition, we plot the frequency-domain tidal phases of the different \nrtidal\ models for four waveform hybrids in Fig.~\ref{fig: freqdomain tidal phases}, representing $q = [1.0, 1.33, 1.75, 2.0]$. We note that \nrtidalvthree\ describes the EOB-NR tidal phase contributions better than \nrtidalvtwo\ and 7.5PN approximant. We also indicate the merger frequency fit $f_{\rm est}$ from the fitting function shown in Sec.~\ref{subsection: Merger Frequency}.

\subsection{Spin effects}
\label{subsection: spin corrections}

Spinning NS and BH have an infinite series of nonzero multipole moments~\cite{Pappas:2013naa, Yagi:2014bxa} which will depend on the type of object and its internal structure. 
Hence, aside from the corrections in the tidal phase contributions that we discussed above, we now include spin effects in \nrtidalvthree. Here, we follow the same approach as \nrtidalvtwo\, where EoS-dependent spin-squared terms up to 3.5PN were included, as well as leading order spin-cubed terms that enter at 3.5PN order~\cite{Nagar:2018plt, Dietrich:2019kaq}. But note that the dynamical-tidal enhancement factor Eq.~\eqref{eq: k2eff} is specialized to the nonspinning case in this work.

In terms of the spin-induced quadrupolar $C_{\rm Q}^{A,B}$ and octupolar deformabilities $C_{\rm Oc}^{A,B}$ for stars $A$ and $B$, the self-spin terms in the phase that are added to the BBH baseline are given by (for aligned spins)
\begin{equation}
    \begin{split}
        \psi_{\rm SS} =& \frac{3x^{-5/2}}{128\nu}\Bigg(
        \hat{\psi}_{\rm SS, 2PN}^{(A)}x^2 + \hat{\psi}_{\rm SS, 3PN}^{(A)}x^3 + \hat{\psi}_{\rm SS, 3.5PN}^{(A)}x^{7/2} \Bigg)\\
        &+[A\leftrightarrow B],
    \end{split}
\end{equation}
with 
\begin{equation}
    \begin{split}
        \hat{\psi}_{\rm SS, 2PN}^{(A)} =& -50 \hat{C}_{\rm Q}^AX_A^2 \chi_A^2,\\
        \hat{\psi}_{\rm SS, 3PN}^{(A)} =& \frac{5}{84}\left(9407 + 8218X_A - 2016 X_A^2\right)\\
        &\times \hat{C}_{\rm Q}^AX_A^2 \chi_A^2,\\
        \hat{\psi}_{\rm SS, 3.5PN}^{(A)} =& -400\pi \hat{C}_{\rm Q}^AX_A^2 \chi_A^2,
    \end{split}
\end{equation}
where $\hat{C}_{\rm Q}^A  = C_{\rm Q}^A - 1$ and $\hat{C}_{\rm Oc}^A  = C_{\rm Oc}^A - 1$. We subtract one to remove the BH multipole contribution that is already present in the baseline BBH phase. Meanwhile, the spin-cubed term is given by
\begin{equation}
    \begin{split}
            \psi_{\rm S^3, 3.5PN}^{(A)} =& \frac{3x^{-5/2}}{128\nu}\Bigg\{10 \Bigg[\left(X_A^2 + \frac{308}{3}X_A \right)\chi_A\\
        &+ \left(X_B^2 - \frac{89}{3}X_B \right)\chi_B\Bigg]\hat{C}_{\rm Q}^AX_A^2\\
        &-440\hat{C}_{\rm Oc}^AX_A^3\chi_A^3\Bigg\}x^{7/2} + [A\leftrightarrow B].
    \end{split}
\end{equation}
To reduce the number of free parameters, we link $C_{\rm Q}^A$ to $\Lambda_A$, and $C_{\rm Oc}^A$ to $C_{\rm Q}^A$ via the following EoS-independent relations~\cite{Yagi:2016bkt}
\begin{equation}
    \begin{split}
        \log C_{\rm Q}^A =& 0.1940 + 0.09163\log \Lambda_A\\
        &+ 0.04812\log^2\Lambda_A - 0.004286\log^3\Lambda_A\\
        &-0.00012450\log^4\Lambda_A,
    \end{split}
\end{equation}
and
\begin{equation}
    \begin{split}
        \log C_{\rm Oc}^A =& 0.003131 + 2.071\log C_{\rm Q}^A\\
        &-0.7152\log^2C_{\rm Q}^A +0.2458\log^3C_{\rm Q}^A\\
        &-0.03309\log^4C_{\rm Q}^A.
    \end{split}
\end{equation}

\subsection{Tidal amplitude corrections}

We also include amplitude corrections in the \nrtidalvthree\ frequency-domain model following \nrtidalvtwo~\cite{Dietrich:2019kaq}. We incorporate these corrections as an ansatz whose form was taken from the frequency-domain representation of \teobresums-NR hybrids, as discussed in Ref.~\cite{Dietrich:2019kaq}:
\begin{equation}
    \tilde{A}_T^{\rm NRT2} = -\sqrt{\frac{5\pi \nu}{24}}\frac{9M^2}{D}\kappa_{\rm eff}x^{13/4}\frac{1 + \frac{449}{108}x + \frac{22672}{9}x^{2.89}}{1+13477.8x^4},
\end{equation}
where $D$ is the luminosity distance to the source.

\subsection{\label{subsection: Precession} Spin Precession Effects}

As with \nrtidalvtwo, we also consider BNS systems whose individual spins have an intrinsic rotation and where this rotation is not necessarily aligned with the orbital angular momentum, causing the orbital plane to precess. 

We augment \nrtidalvthree\ onto the BBH baseline \texttt{IMRPhenomXP}, as was done for \imrphenomxpnrtidaltwo~\cite{Colleoni:2023}, already available in \lalsuite. The \texttt{IMRPhenomXP} baseline improves on \texttt{IMRPhenomPv2} by incorporating double-spin effects in the twisting-up construction instead of a single-spin approximation. Moreover, the Euler angles are calculated via a precession-averaged treatment of the PN precession dynamics, and there is also an option offered to the user for a more accurate twisting-up prescription based on the solutions of the orbit-averaged SpinTaylorT4 equations implemented in \lalsuite. Note that here, the spin-induced multipole corrections introduced in Section~\ref{subsection: spin corrections} are applied in the co-precessing frame and then twisted up.

\subsection{\label{subsection: Merger Frequency}Merger Frequency}

As a final ingredient for the construction of the full model, we have to define the stopping criterion or the final frequency until which our model is applicable. 
For this purpose, we use the estimated merger frequency (peak in the GW amplitude) $Mf_{\text{est}}/\nu$ (with $\nu = X_AX_B = M_AM_B/M^2$) from Ref.~\cite{Gonzalez:2022mgo}. 
The expression is then given by 
\begin{equation}\label{eq: MergerFit}
    \frac{Mf_{\text{est}}}{\nu}= w_0V^M(X), V^S(\hat{S}, X)V^T(\kappa_2^T,X),
\end{equation}
where the factor $V^M$ depends on the mass ratio, $V^S$ depends on the spin contributions (specifically the aligned-spin components of the binary), and $V^T$ contains the tidal contributions~\cite{Gonzalez:2022mgo}. The factors are
\begin{equation}
    \begin{split}
        V^M =& 1 + a_1^MZ,\\
        V^S =& 1 + p_1^S\hat{S},\\
        V^T =& \frac{1 + p_1^T\kappa_2^T + p_2^T(\kappa_2^T)^2}{1 + p_3^T\kappa_2^T + p_4^T(\kappa_2^T)^2} ,\\
    \end{split}
\end{equation}
where
\begin{equation}
    \begin{split}
        Z =& 1-4\nu,\\
        \kappa_2^T =& 3\nu\left[\left(\frac{M_A}{M} \right)^3\Lambda_A + (A\leftrightarrow B) \right],\\
        \hat{S} =& \left(\frac{M_A}{M} \right)^2\chi_A + (A\leftrightarrow B),\\
        p_1^S =& a_1^S(1 + b_1^SZ),\\
        p_i^T =& a_i^T(1 + b_i^TZ),
    \end{split}
\end{equation}
with coefficients
\begin{equation}
    \begin{split}
        w_0 =& 0.22,\\
        a_1^M =& 0.80,\\
        a_1^S =& 0.25,\,\, b_1^S = -1.99,\\
        a_i^T \in& [0.0485, 5.86 \times 10^{-6}, 0.1, 1.86 \times 10^{-4}],\\
        b_i^T \in& [1.80, 599.99, 7.80, 84.76].
    \end{split}
\end{equation}

Since the Padé approximant is a rational function, asymptotes (corresponding to a zero denominator in the approximant) can appear depending on the specific source parameters. For this reason, we have performed careful checks to verify that no unphysical behavior occurs, i.e., that the estimated merger frequency of the system (as calculated in Eq.~\eqref{eq: MergerFit}) is less than the frequencies at which the asymptotes occur ($f_\text{est} < f_\text{asymp}$). 
Practically speaking, we generated 30,000 random non-spinning configurations with masses $M_A, \, M_B \in [0.5, 3.0]$, (corresponding to $X_A,\, X_B \in [0.14, 0.86]$)  and $\Lambda_A, \, \Lambda_B \in [0, 20000]$. We then performed another 30,000 random spinning configurations with aligned-spin components $\chi_{A,B} = [-0.5,0.5]$. For all configurations, we verify that $f_\text{est} < f_\text{asymp}$ if $f_\text{asymp}$ exists. Since \nrtidalvthree\ only models the inspiral part up to the merger of the binary, the condition $f_\text{est} < f_\text{asymp}$ suffices for the purposes of the model.

However, to ensure that the phase contribution remains smooth even after the merger, i.e., during the time interval when we taper the waveform, and to remove any asymptote, we make the tidal contribution constant if it reaches a local minimum, then connect the tidal phase from inspiral to merger given by $\psi_T^{\text{NRT3}}$ and the post-merger with the PN tidal contribution $\psi_{T}^{\rm PN}$(which is polynomial in nature and is therefore smooth\footnote{In principle, any smooth function can do this, but we choose the usual PN for convenience.}) to obtain the tidal function as a function of the frequency: 
\begin{equation} 
    \psi_T^{\rm NRT3a}(\hat{\omega}) = (1-\sigma(\hat{\omega}))\psi_T^{\text{NRT3}}(\hat{\omega}) + \sigma(\hat{\omega})\psi_{T}^{\rm PN}(\hat{\omega}),
\end{equation}
where $\psi_T^{\rm NRT3a}$ includes the post-merger information from PN, and $\sigma(\hat{\omega})$ is the Planck taper:
\begin{equation} \label{eq: PlanckTaper}
        \sigma(\hat{\omega}) = \begin{cases}
            0,  \quad & \hat{\omega} \le \hat{\omega}_1\\
            \left[1+\exp\left(\frac{\hat{\omega}_2 - \hat{\omega}_1}{\hat{\omega} - \hat{\omega}_1} + \frac{\hat{\omega}_2 - \hat{\omega}_1}{\hat{\omega} - \hat{\omega}_2}\right)\right]^{-1}, \quad & \hat{\omega}_1 \le \hat{\omega} \le \hat{\omega}_2\\
            1, \quad & \hat{\omega} \ge \hat{\omega}_2,
        \end{cases}
\end{equation}
for a frequency window after the estimated merger frequency $[\hat{\omega}_1, \hat{\omega}_2] = [1.15\hat{\omega}_{\text{est}}, 1.35\hat{\omega}_{\text{est}}] > \hat{\omega}_{\text{est}} = M(2\pi f_{\rm est})$. The same Planck taper is used to taper the entire waveform abruptly up to $1.2 \hat{\omega}_{\rm est}$, to minimize the presence of any postmerger signal (since this is not part of the description of \nrtidalvthree). Then, for the rest of the discussion, we will just refer to the $\psi_T^{\rm NRT3a}$ as $\psi_T^{\rm NRT3}$, knowing that this implicitly includes the tapered phase beyond merger (in its \lalsuite\ implementation).

\section{\label{section: implementation} Implementation and Validation}

\begin{table*}[t]
    \caption{\label{table: nrtidal approx lal} Available \nrtidal\ approximants in LAL. The first column contains the BBH baseline model and the second column contains the versions of \nrtidal\ that were added to the BBH model. 
    Full approximant names are given by joining the BBH baseline name with the \nrtidal\ version. We also include information about the corrections in the approximants (i.e., spin-spin, cubic-in-spin, tidal amplitude, and precession), as well as the computational time $\Delta T_{f_{\rm min}}$ of the models at different lower frequencies $f_{\rm min}$. 
    The computational time was obtained using the configuration $M_{A,B} = 1.35 M_{\odot}$, $\Lambda_{A,B} = 400$, with no spins, and were simulated using an Apple M1 Pro processor.}
    \begin{ruledtabular}
    \begin{tabular}{c|c|c|c|c|c|c|cccc}
        BBH baseline & $\psi_T$ & Implemented & spin-spin & cubic-in-spin & tidal amp. & precession & \multicolumn{4}{c}{$\Delta T_{f_{\rm  min}} \, [{\rm s}]$} \\
        & & by this work? & & & & &10 Hz & 20 Hz & 30 Hz & 40 Hz\\
        \hline
        \hline
        \texttt{IMRPhenomD\_} & \nrtidal & No & up to 3PN (BBH) & \xmark & \xmark & \xmark &0.610&0.153&0.038	&0.019\\
        & \nrtidalvtwo & No & up to 3PN & \xmark & \xmark & \xmark &1.020&0.253&0.063&0.031\\
        & \nrtidalvthree & Yes & up to 3PN & \xmark & \xmark & \xmark &2.349&0.591&0.147&0.072\\
        \hline
        \texttt{IMRPhenomPv2\_} & \nrtidal & No & up to 3PN & \xmark & \xmark & \checkmark &0.745&0.184&0.047&0.023\\
        & \nrtidalvtwo & No & up to 3.5PN & up to 3.5PN & \checkmark & \checkmark &0.969&0.243&0.060&0.030\\
        \hline
        \texttt{IMRPhenomXAS\_} & \nrtidalvtwo & No & up to 3.5PN & up to 3.5PN & \checkmark & \xmark &0.084&0.022&0.006&0.004\\
        &\nrtidalvthree & Yes & up to 3.5PN & up to 3.5PN & \checkmark & \xmark &0.089&0.026&0.008&0.005\\
        \hline
        \texttt{IMRPhenomXP\_} & \nrtidalvtwo & No & up to 3.5PN & up to 3.5PN & \checkmark & \checkmark & 0.353&0.091&0.024&0.013\\
        &\nrtidalvthree & Yes& up to 3.5PN & up to 3.5PN & \checkmark & \checkmark &0.358&0.093&0.025&0.014\\
        \hline
        \texttt{SEOBNRv4\_ROM\_} & \nrtidal & No & up to 3PN & \xmark & \xmark & \xmark &0.423&0.110&0.029&0.016\\
        &\nrtidalvtwo & No& up to 3.5PN & up to 3.5PN & \checkmark & \xmark &0.630&0.163&0.043&0.022\\
        \hline
        \texttt{SEOBNRv5\_ROM\_} & \nrtidal & Yes & up to 3PN & \xmark & \xmark & \xmark &0.418&0.102&0.030&0.017\\
        &\nrtidalvtwo & Yes& up to 3.5PN & up to 3.5PN & \checkmark & \xmark &0.639&0.168&0.045&0.024\\
        &\nrtidalvthree & Yes& up to 3.5PN & up to 3.5PN & \checkmark & \xmark &1.193&0.306&0.080&0.041
    \end{tabular}
    \end{ruledtabular}
\end{table*}

To make use of the newly developed \nrtidalvthree\ model, we implement it into LALSuite~\cite{lalsuite} by adding the outlined corrections to several different BBH baseline models: the spinning, non-precessing models \texttt{IMRPhenomD} \cite{Khan:2015jqa}, \texttt{IMRPhenomXAS} \cite{Pratten:2020fqn}, \texttt{SEOBNRv5\_ROM} \cite{Pompili:2023tna}, and the spinning, precessing model \texttt{IMRPhenomXP} \cite{Pratten:2020ceb}. We list all the new approximants in Table~\ref{table: nrtidal approx lal}. The full approximant name is given by combining the BBH baseline name with the NRTidal version (e.g., \imrphenomxasnrtidalthree). Versions of \nrtidal\ which are currently present in \lalsuite\ are shown in Table~\ref{table: nrtidal approx lal}, both previously existing models and the ones newly implemented in this work. We also show in Table~\ref{table: nrtidal approx lal} their computation times employing different values of starting frequencies and using a sampling rate of $2^{15} = 32768$ Hz, for a system $M_A = M_B = 1.35 M_{\odot}$ and $\Lambda_A = \Lambda_B = 400$. Overall, we find that \nrtidalvthree\ has retained both speed and efficiency (particularly for starting frequencies $f_{\rm min} \ge 20\, {\rm Hz}$) despite having a more accurate (and mathematically more complicated) description of the GW signal than \nrtidalvtwo. 

\begin{table*}
    \caption{\label{table: bns nr configs} BNS configurations for the time-domain dephasing comparisons. We show the configurations of NR simulations using BAM, SACRA, and SpEC that are used in the comparison of waveforms in the time domain. The SACRA waveforms, as well as BAM:0001, BAM:0037 and BAM:0064, are also used in the calibration in \nrtidalvthree. For each NR simulation listed here, we indicated the EOS used, the individual masses $M_{A,B}$, spins $\chi_{A,B}$, tidal deformabilities $\Lambda_{A,B}$, effective tidal deformability $\tilde{\Lambda}$ of the system, resolution $h_{\rm fine}$, eccentricity $e$, and whether or not a Richardson extrapolated waveform was constructed from the simulations.}
    \begin{ruledtabular}
        \begin{tabular}{l||cccccccccc|c}
            Name & EOS & $M_A$ [$M_{\odot}$] & $M_B$ [$M_{\odot}$] & $\chi_A$ & $\chi_B$ & $\Lambda_A$ & $\Lambda_B$ & $\tilde{\Lambda}$ & $h_{\text{fine}} $[$M_{\odot}$] & $e\ [10^{-3}]$ & \text{Richardson}\\
            \hline
            CoRe:BAM:0001 & 2B   & 1.35 & 1.35  & 0      & 0      & 126.7  & 126.7  & 126.7 & 0.093  & 7.1  & \xmark \\
            CoRe:BAM:0011 & ALF2 & 1.5  & 1.5   & 0      & 0      & 382.8  & 382.8  & 382.8 & 0.125  & 3.1  & \xmark \\
            CoRe:BAM:0037 & H4   & 1.37 & 1.37  & 0      & 0      & 1006.2 & 1006.2 & 1006.2 & 0.0833 & 0.9 & \cmark  \\
            CoRe:BAM:0039 & H4   & 1.37 & 1.37  & 0.141  & 0.141  & 1001.8 & 1001.8 & 1001.8 & 0.0833 & 0.5 & \cmark  \\
            CoRe:BAM:0062 & MS1b & 1.35 & 1.35  & -0.099 & -0.099 & 1531.5 & 1531.5 & 1531.5 & 0.097  & 1.8  & \cmark  \\
            CoRe:BAM:0068 & MS1b & 1.35 & 1.35  & 0.149  & 0.149  & 1525.2 & 1525.2 & 1525.2 & 0.097  & 2.2  & \cmark  \\
            CoRe:BAM:0081 & MS1b & 1.5  & 1.00     & 0      & 0   & 863.8  & 7022.4 & 2425.5 & 0.125  & 15.5   & \xmark \\
            CoRe:BAM:0094 & MS1b & 1.94 & 0.944 & 0      & 0      & 182.9  & 9279.9 & 1308.2 & 0.125  & 3.3  & \xmark \\
            \hline
            SACRA:15H\_135\_135 & 15H & 1.35 & 1.35 & 0 & 0 & 1211 & 1211 & 1211 & 0.0508 & $\lesssim 1$& \xmark \\
            \quad \quad \_00155\_182\_135 &&&&&&&&&&&\\
            SACRA:HB\_121\_151 & HB & 1.21 & 1.51 & 0 & 0 & 200 & 827 & 422 & 0.0555 & $\lesssim 1$ & \xmark \\
            \quad \quad \_00155\_182\_135 &&&&&&&&&&&\\
            \hline
            SXS:NSNS:001 & $\Gamma 2$ & 1.40 & 1.40 & 0 & 0 & 791 & 791 & 791 & 0.133 & $\sim 2$ &\xmark\\
            SXS:NSNS:002 & MS1b & 1.35 & 1.35 & 0 & 0 & 1540 & 1540 & 1540 & 0.159 & $\sim 2$ &\xmark\\
        \end{tabular}
    \end{ruledtabular}
\end{table*}

\subsection{Time Domain Comparisons}
\label{subsection: time domain comparisons}

\begin{figure*}[t]
 \centering
\includegraphics[width=0.47\linewidth]{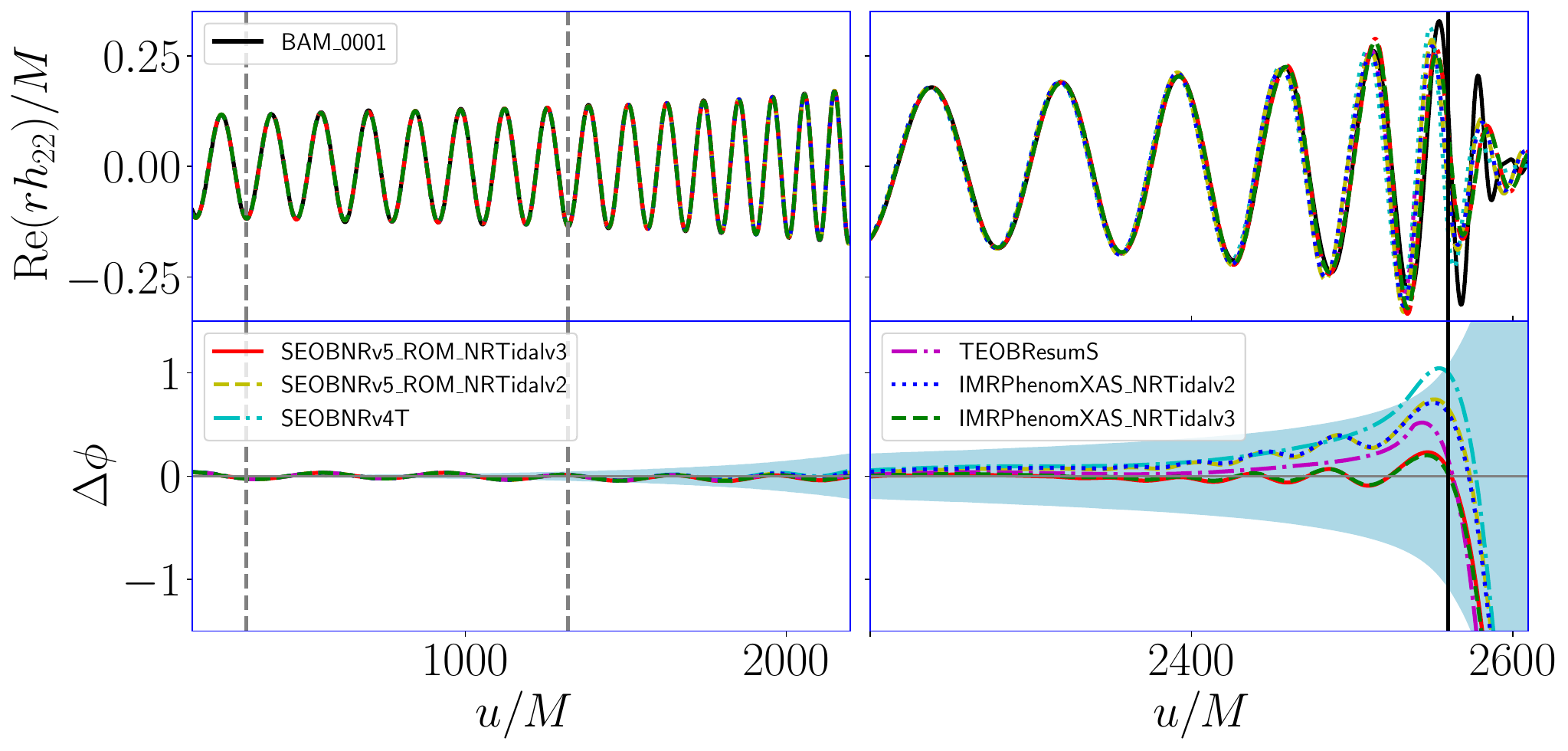}\hfill
\includegraphics[width=0.47\linewidth]{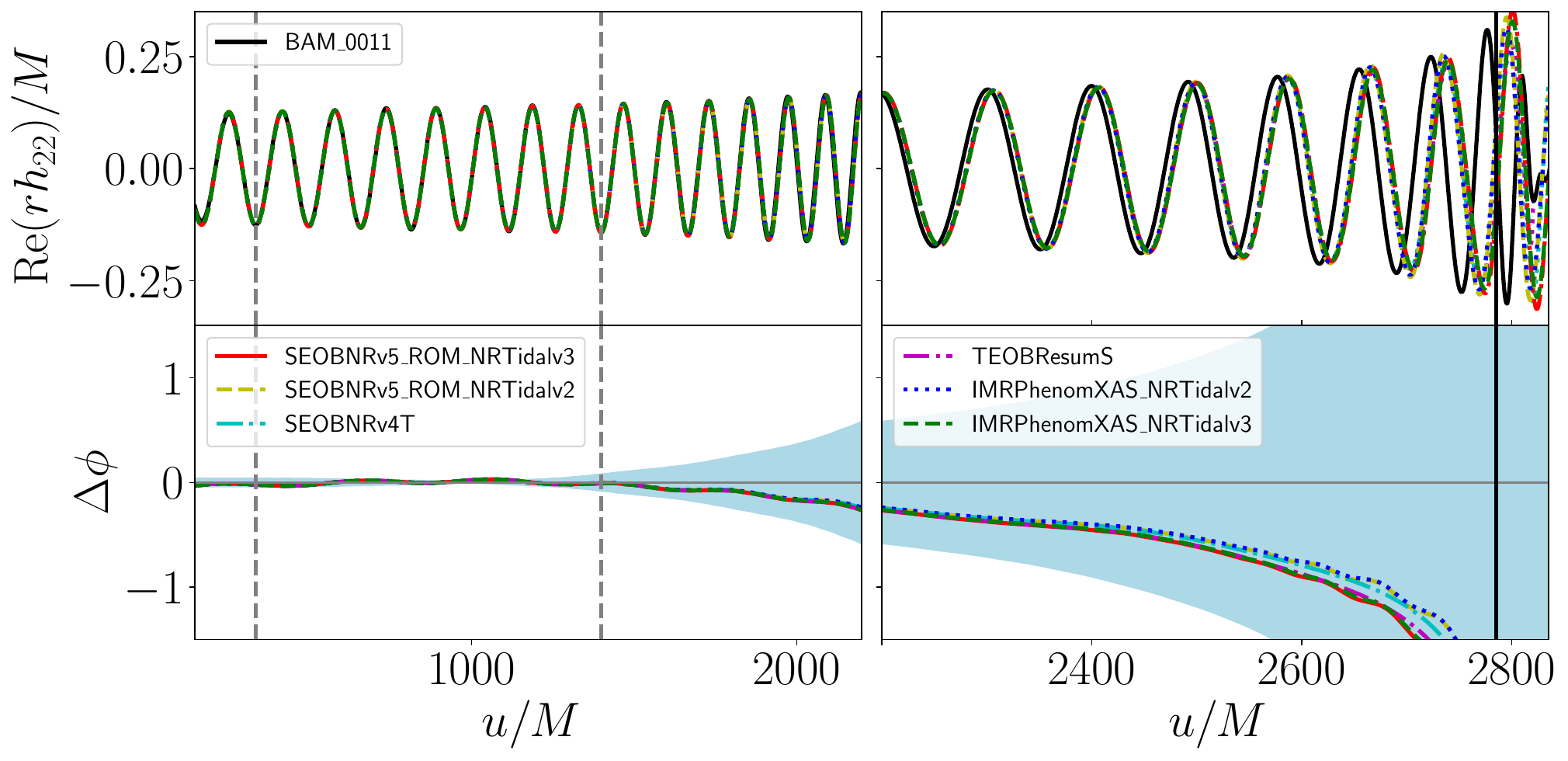}
\includegraphics[width=0.47\linewidth]{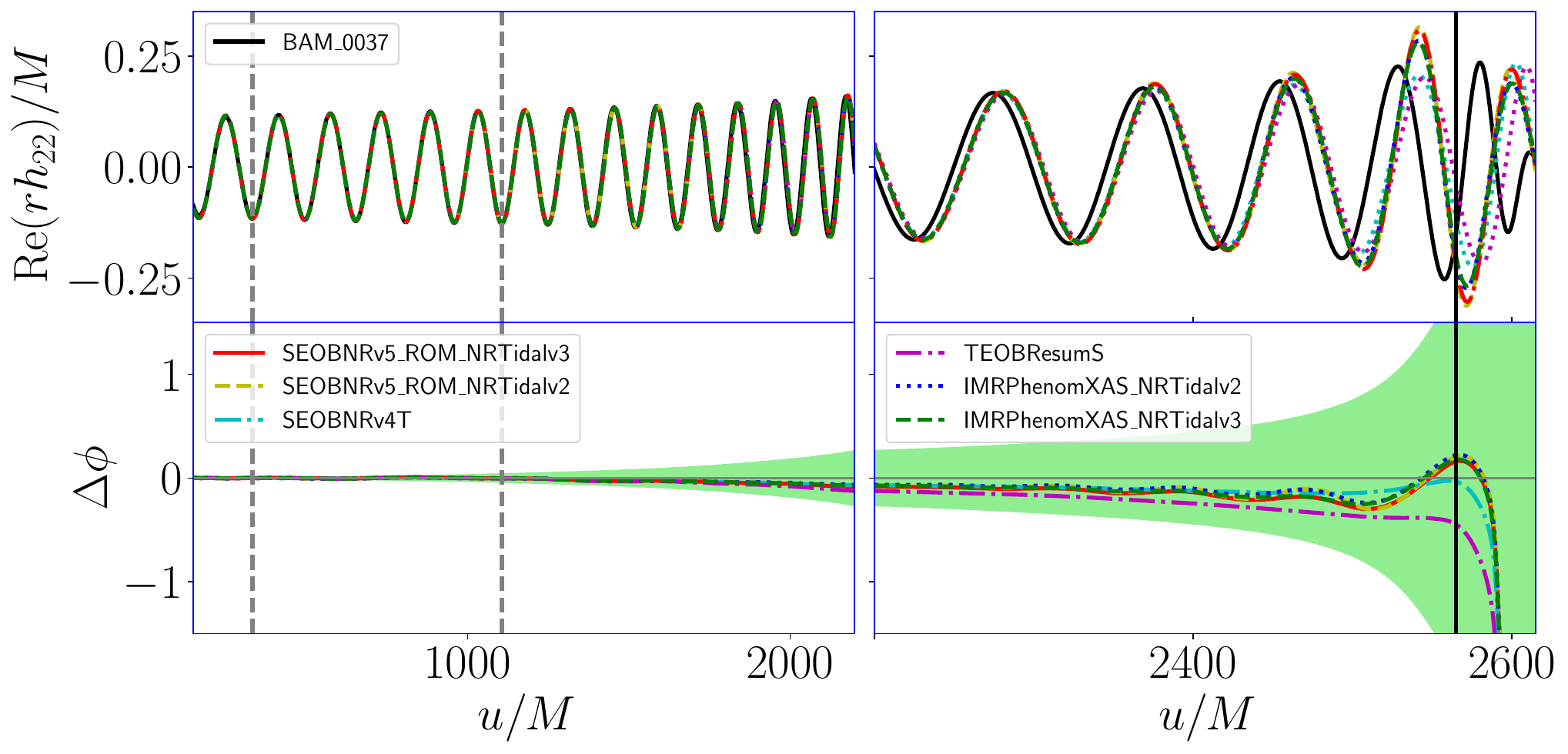}\hfill
\includegraphics[width=0.47\linewidth]{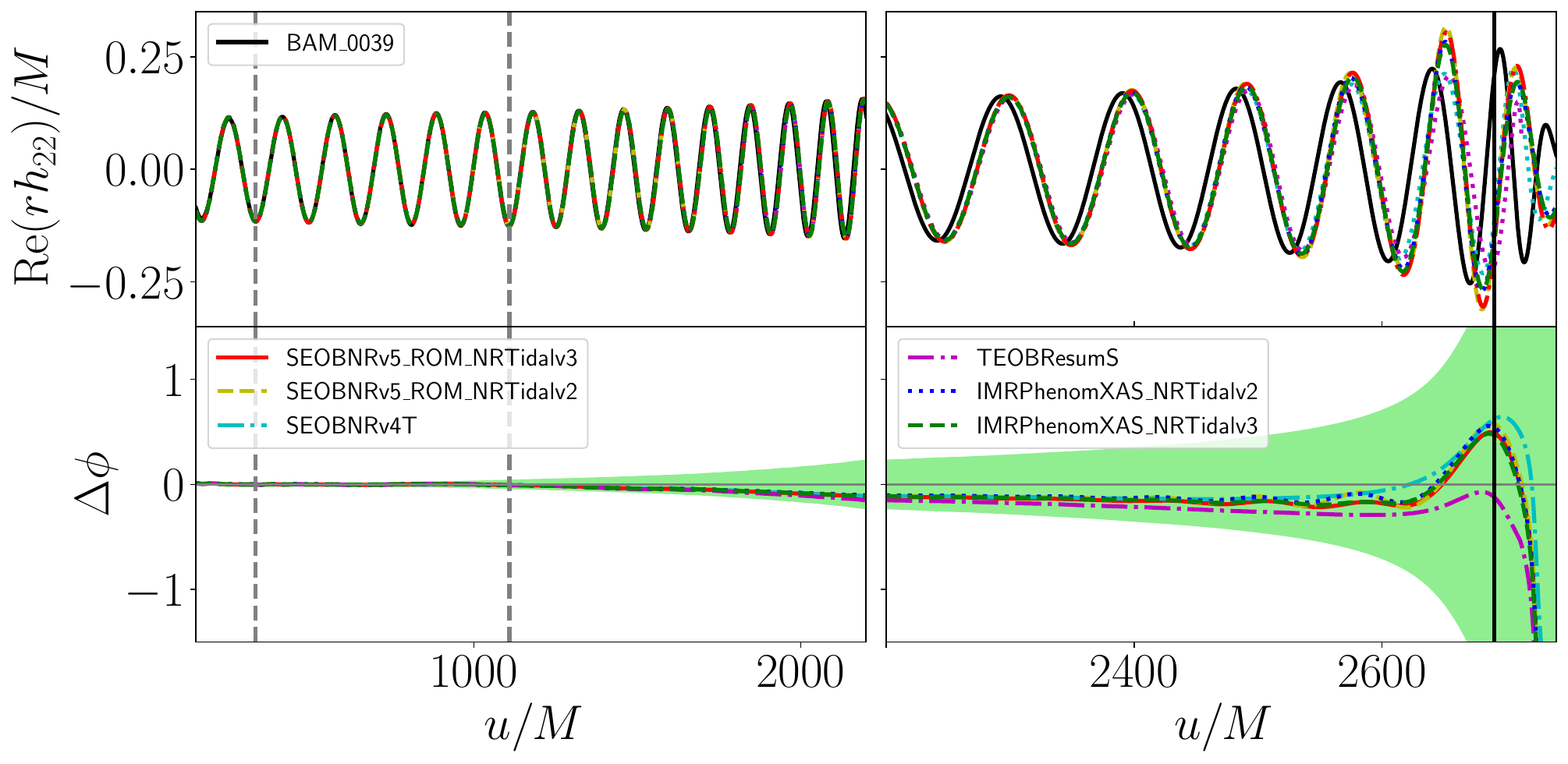}
\includegraphics[width=0.47\linewidth]{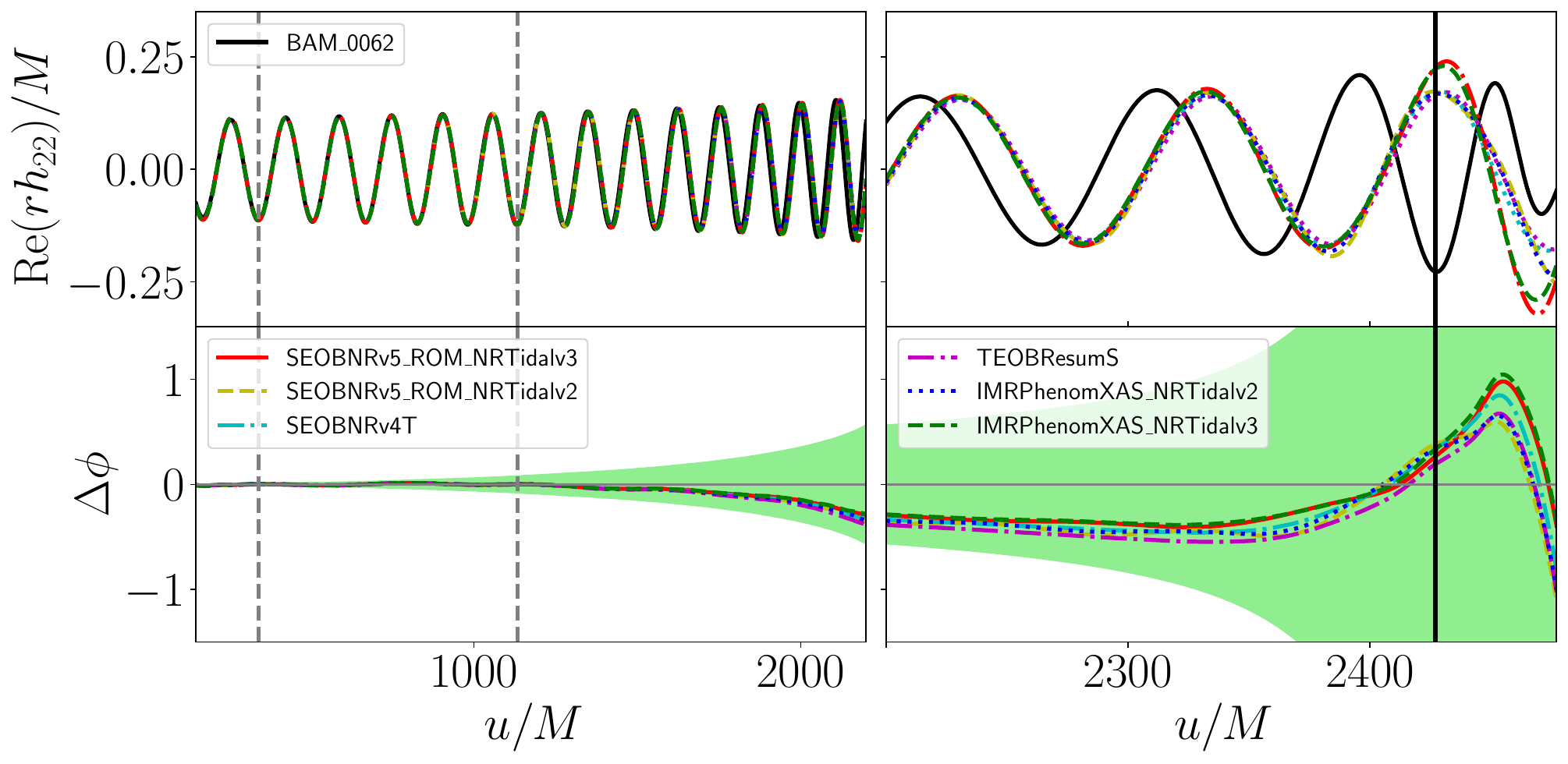}\hfill
\includegraphics[width=0.47\linewidth]{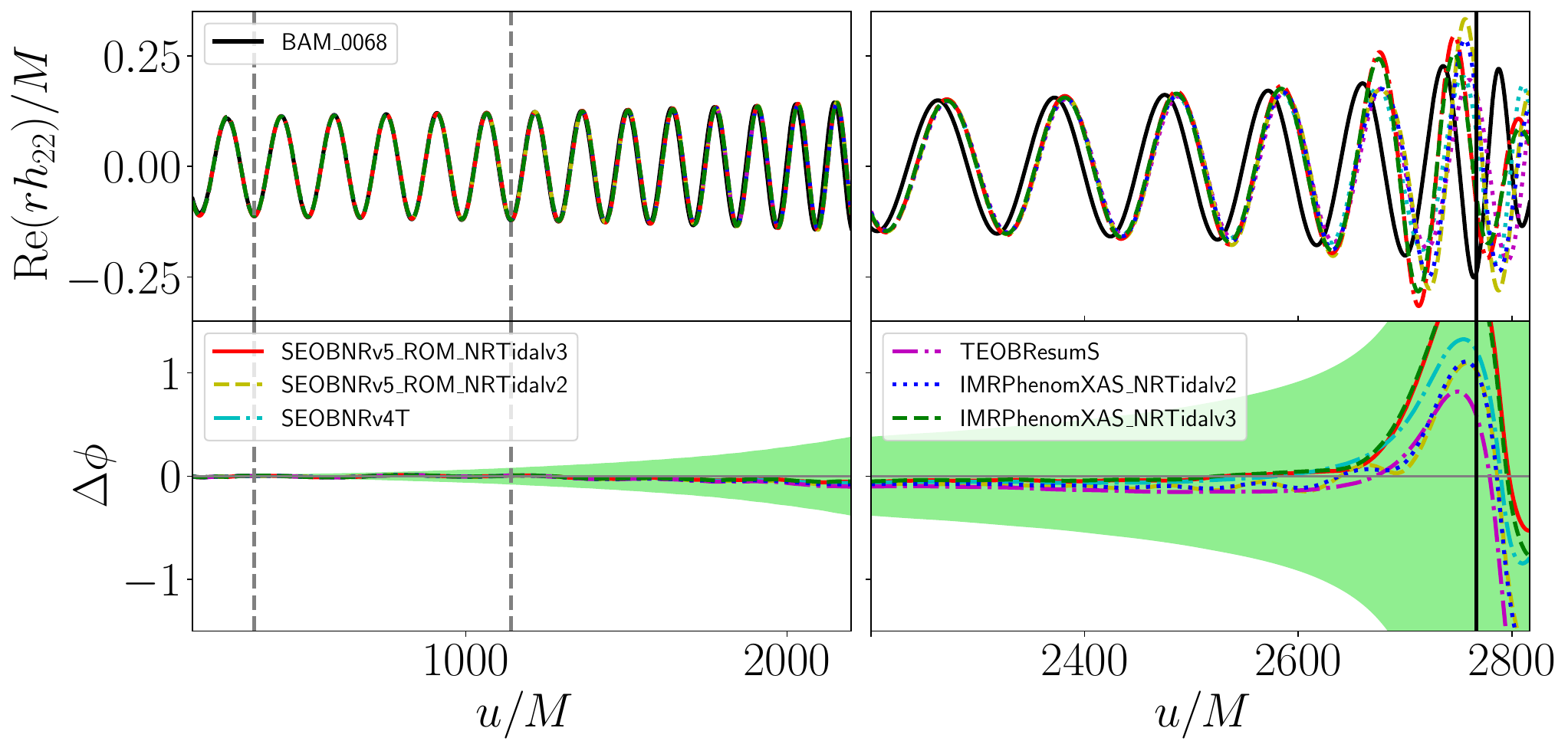}
\includegraphics[width=0.47\linewidth]{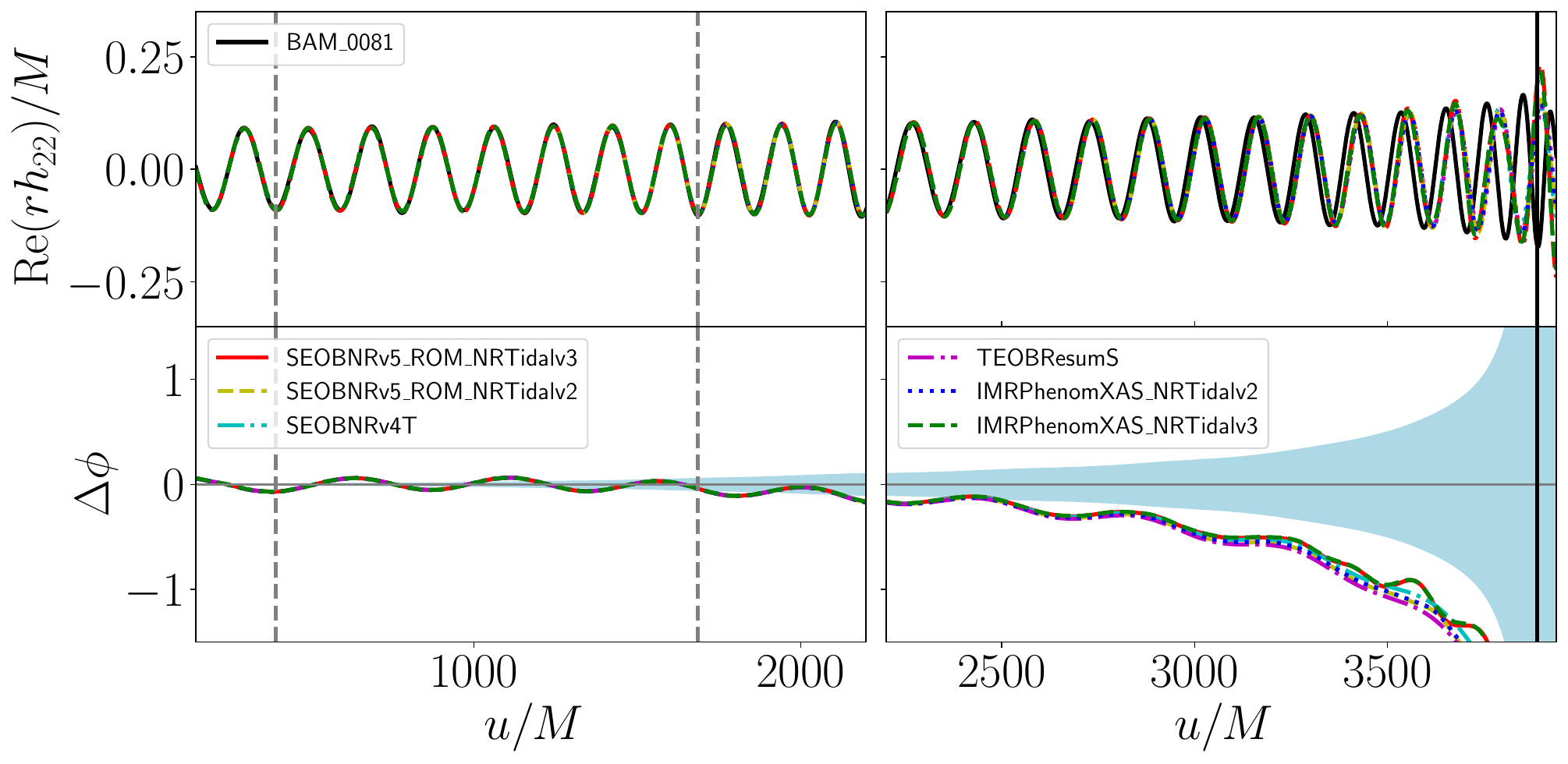}\hfill
\includegraphics[width=0.47\linewidth]{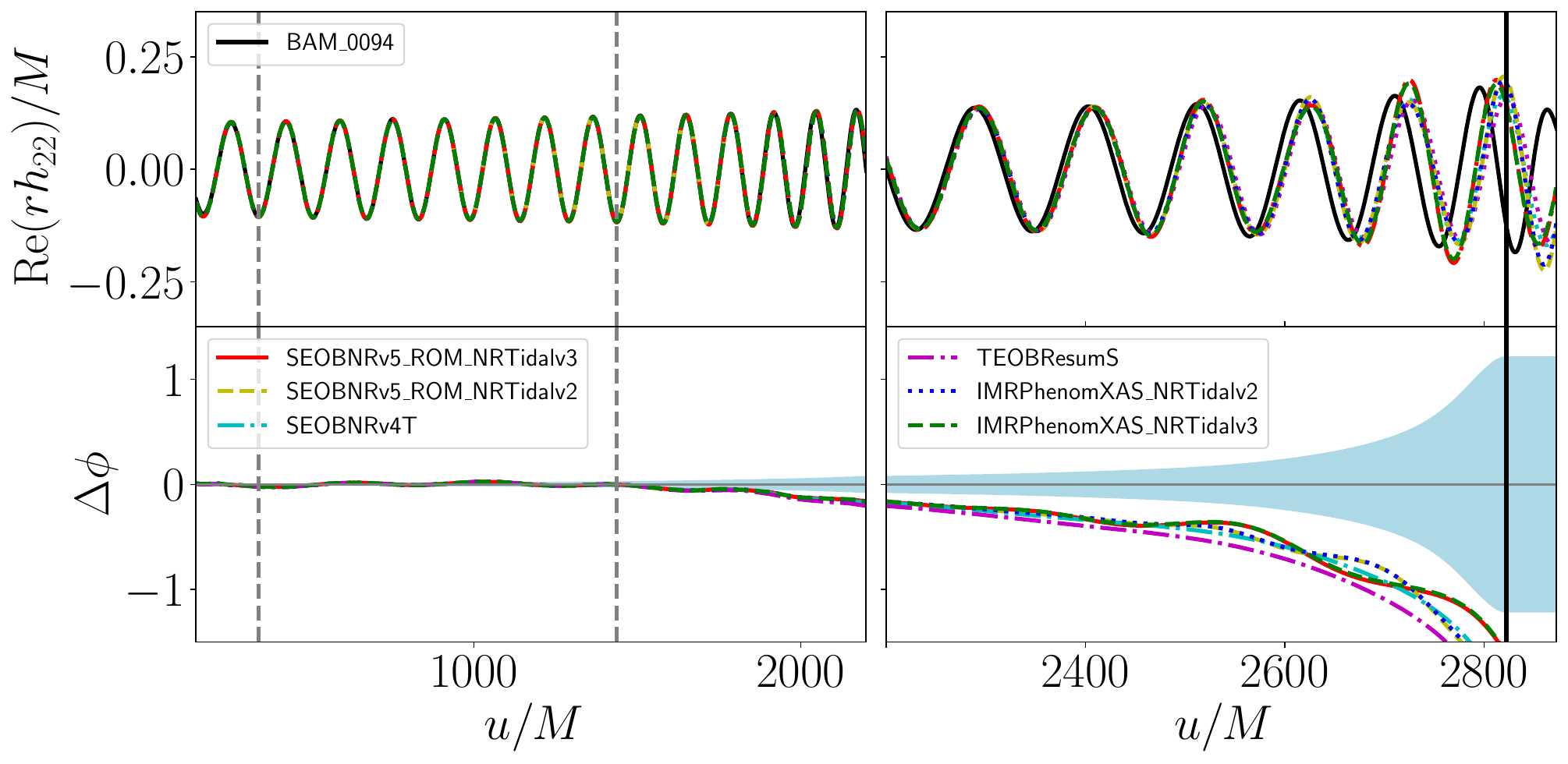}
\includegraphics[width=0.47\linewidth]{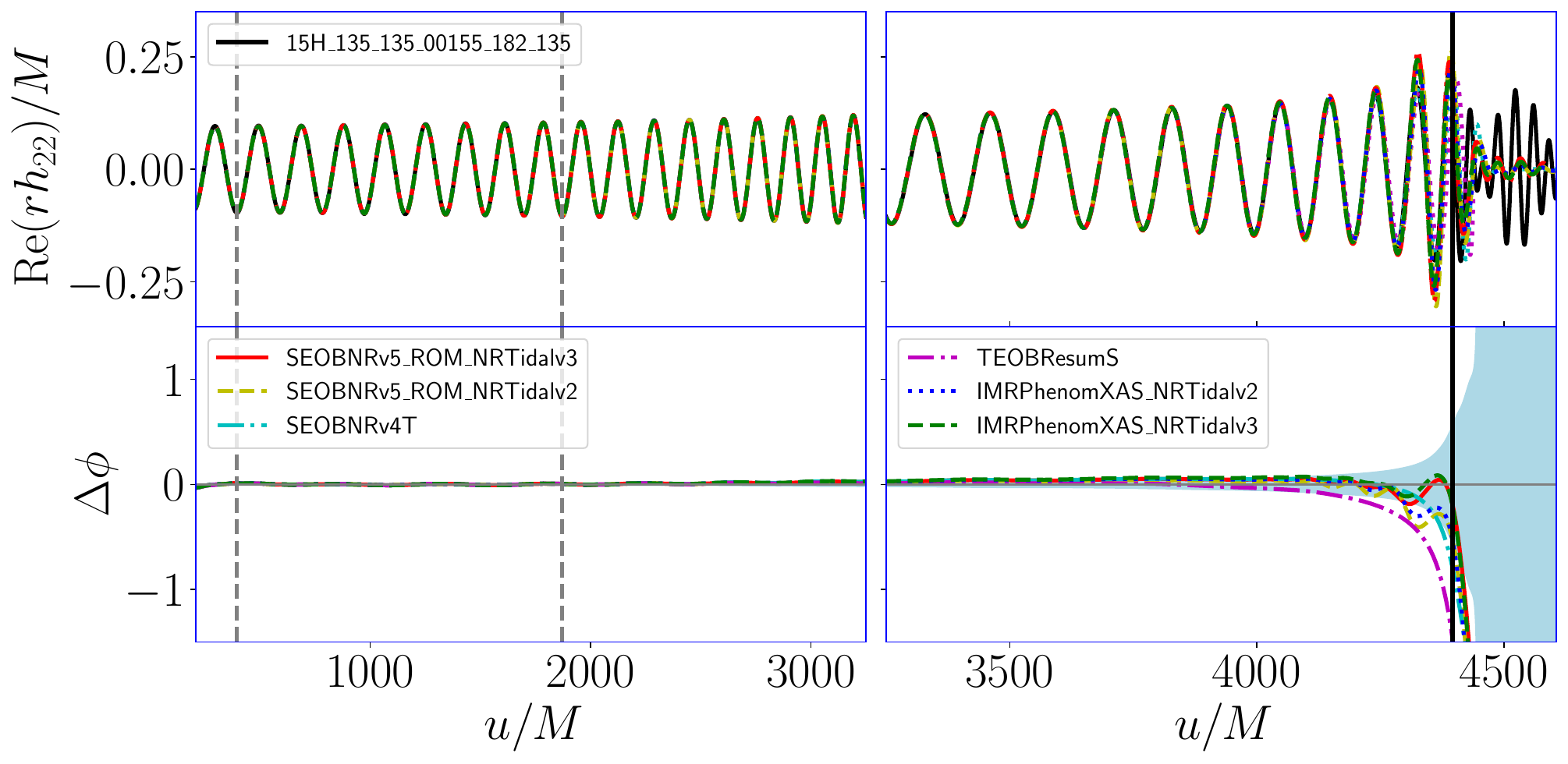}\hfill
\includegraphics[width=0.47\linewidth]{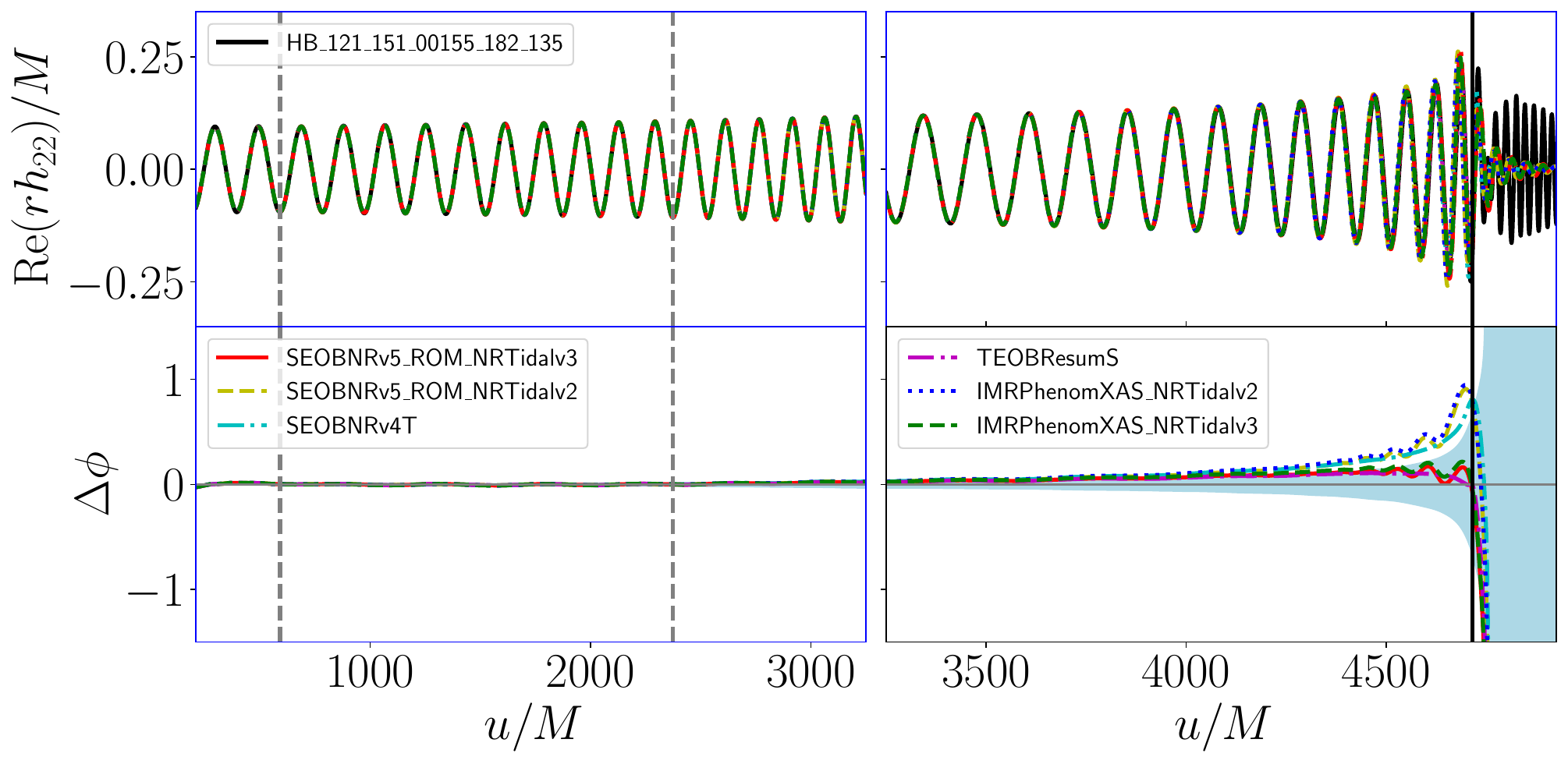}
\includegraphics[width=0.47\linewidth]{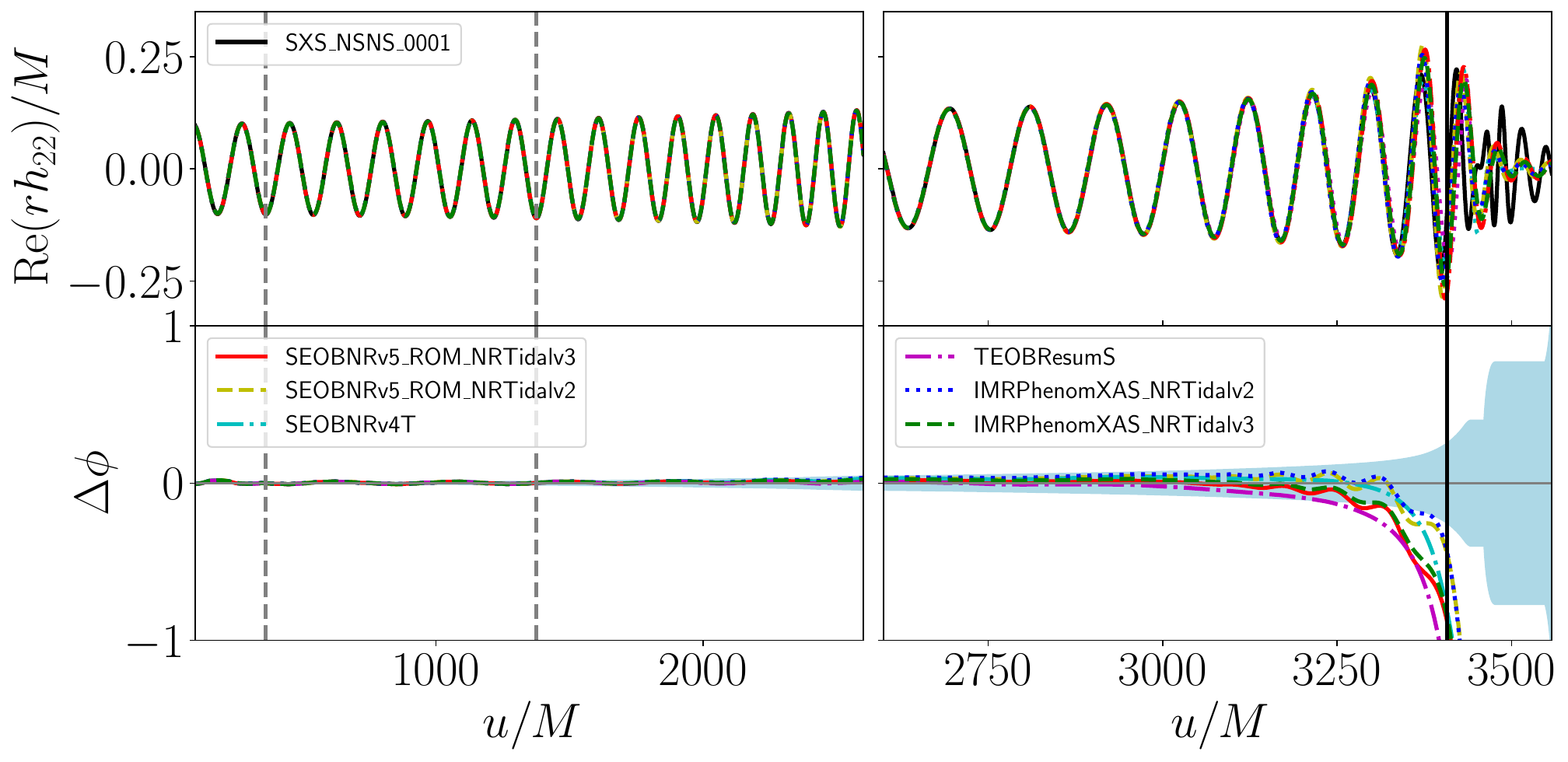}\hfill
\includegraphics[width=0.47\linewidth]{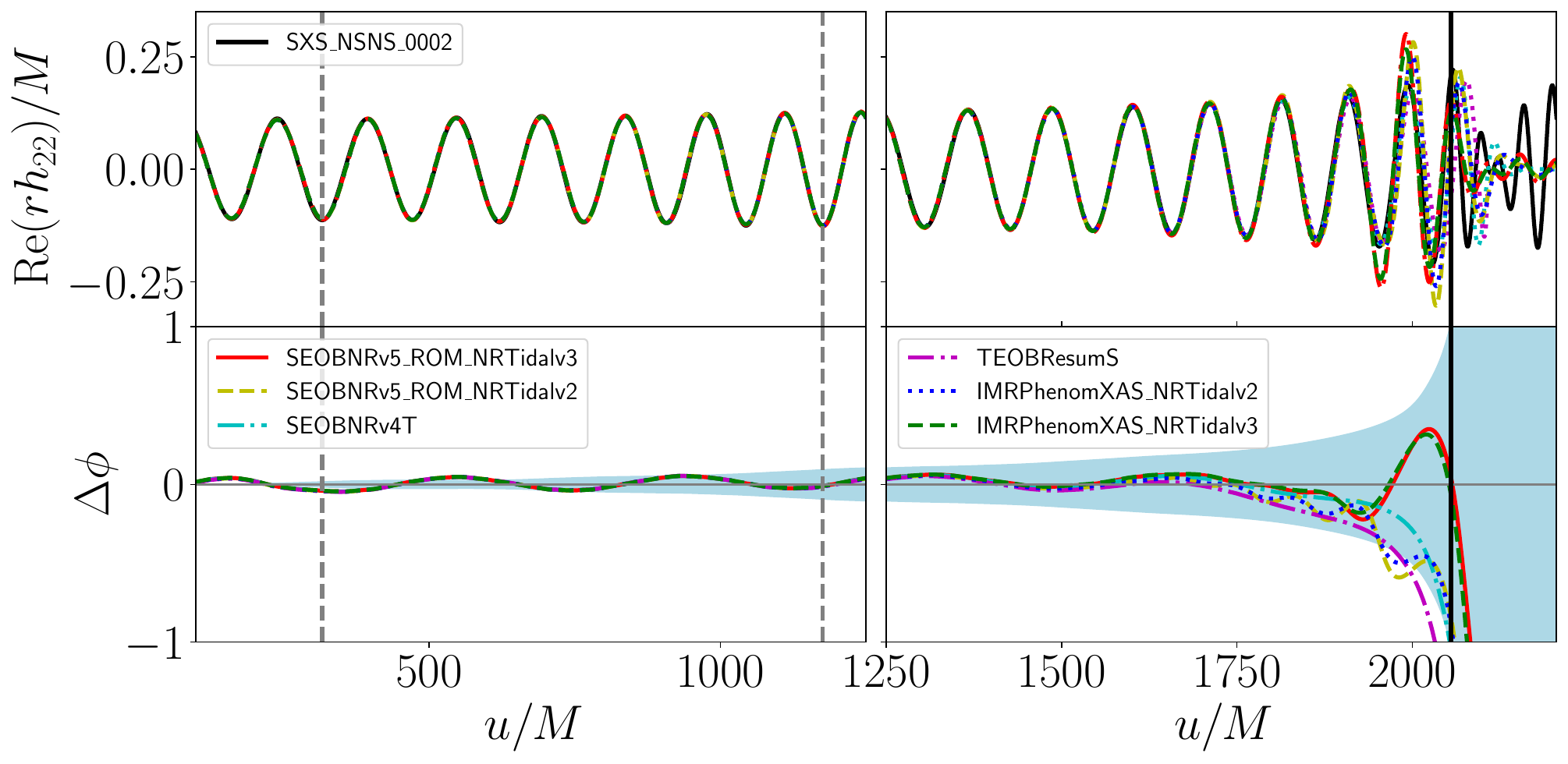}
\caption{Time-domain dephasing comparisons for the BAM, SACRA, and SpEC waveforms. For each NR waveform, the upper panel shows the real part of the gravitational wave strain as a function of the retarded time, while the bottom panel shows the phase difference between the waveform model and the NR waveform. We note that BAM:0001, BAM:0037, and both SACRA waveforms (indicated by blue frames in their plots) were also used in the calibration of \nrtidalvthree.}
\label{fig: timedomaincomparisons}
\end{figure*}

\begin{figure*}[hbt!]
\centering
\includegraphics[width=0.38\linewidth]{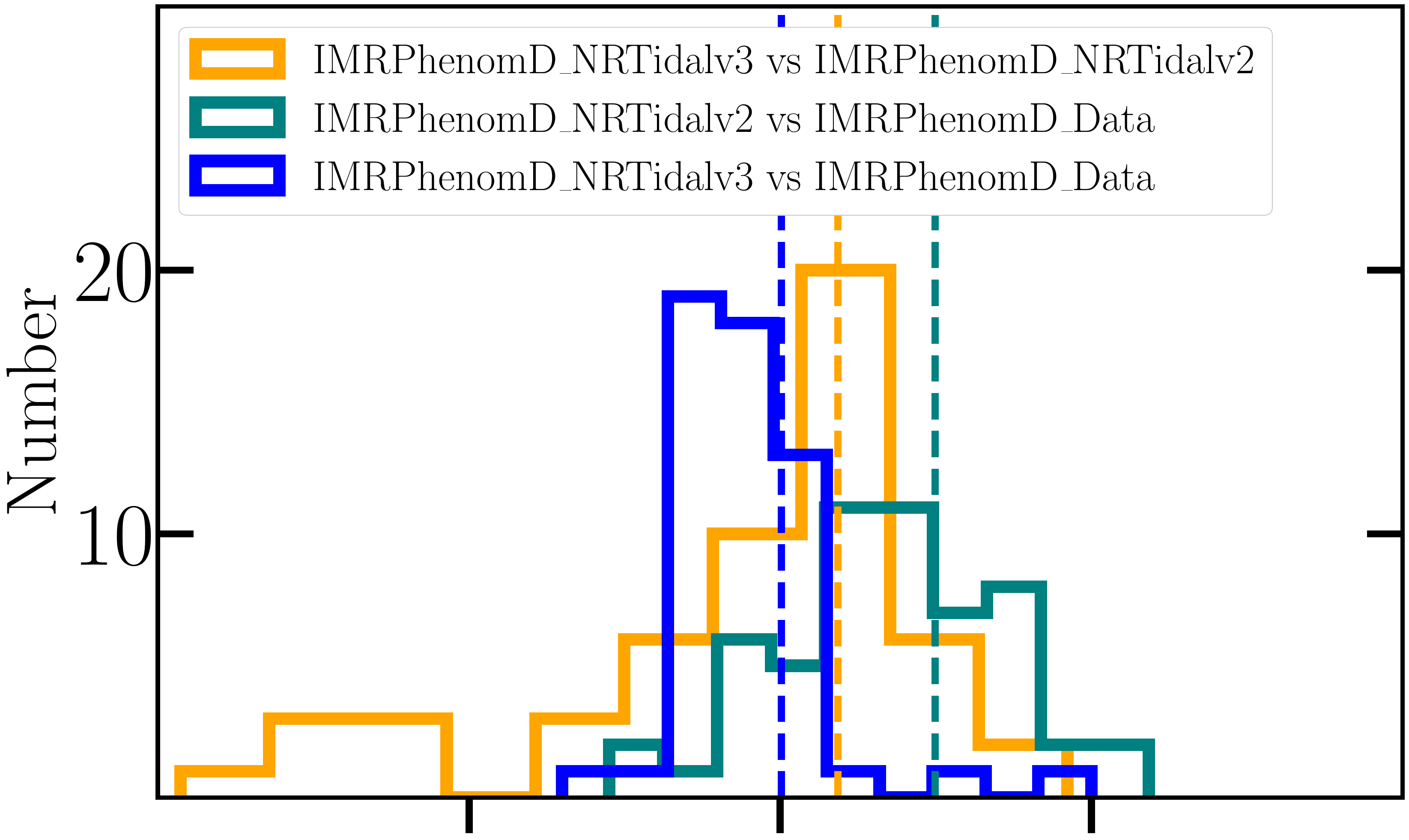}\qquad
\includegraphics[width=0.38\linewidth]
{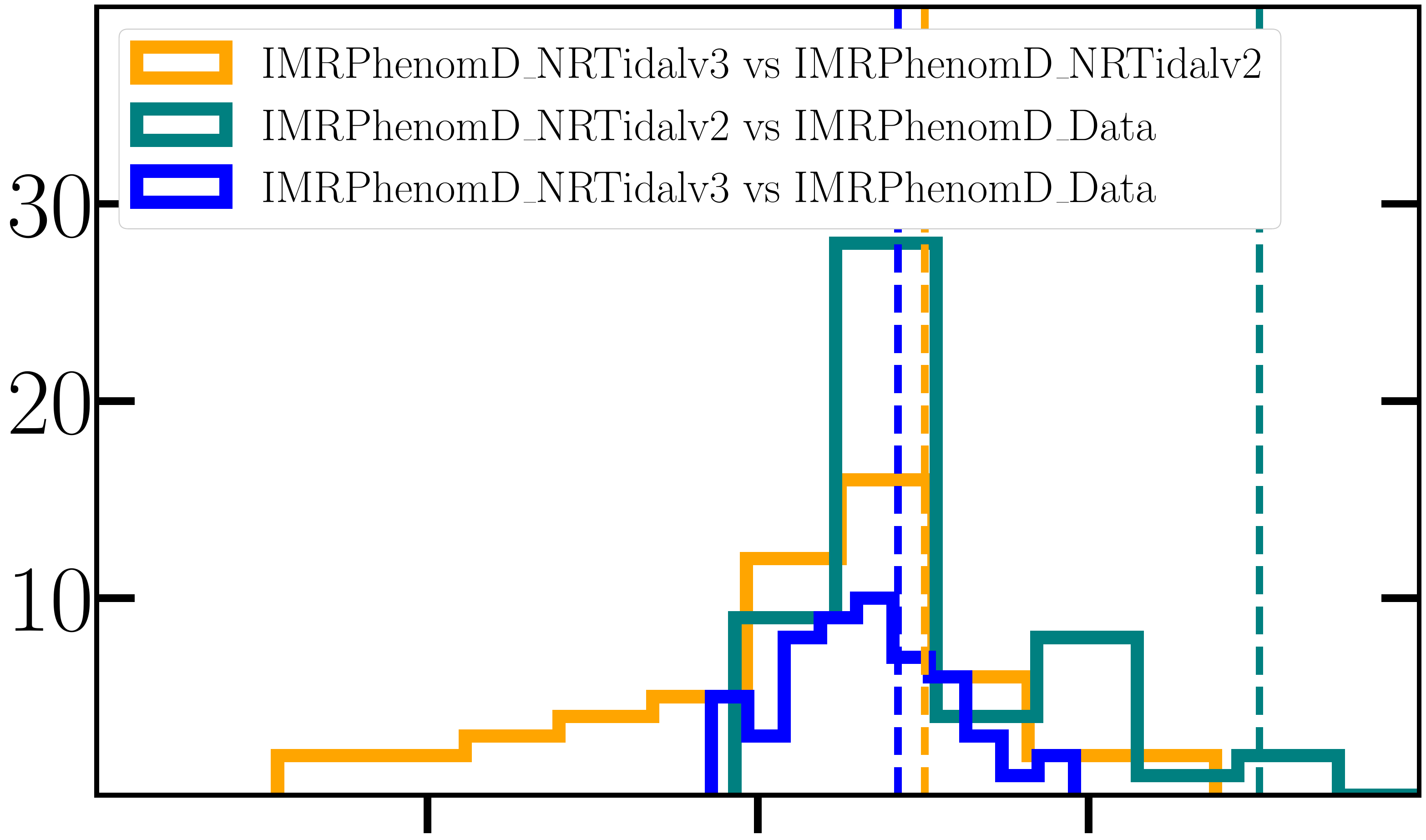}
\vspace{0.00mm}
\includegraphics[width=0.38\linewidth]{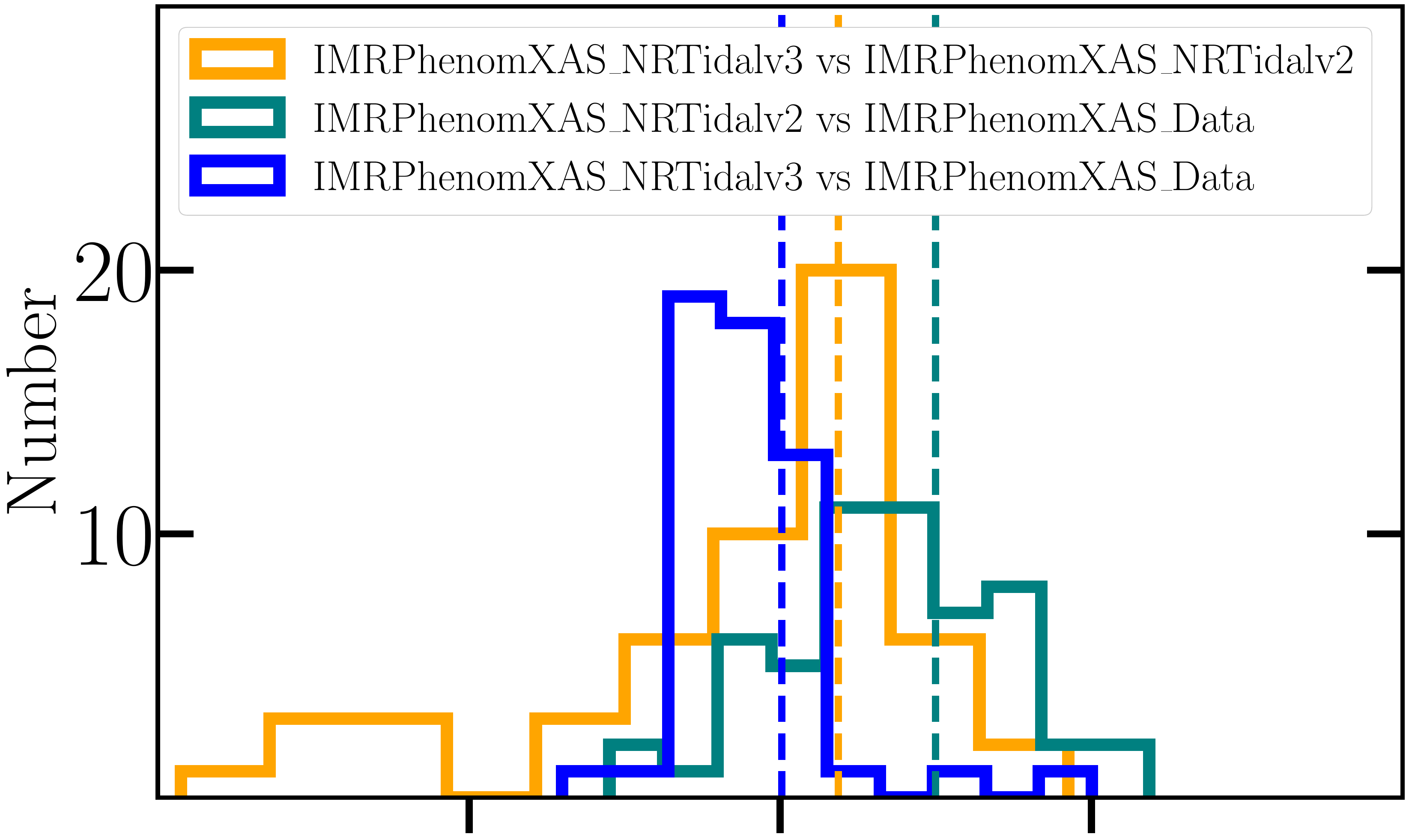}\qquad
\includegraphics[width=0.38\linewidth]{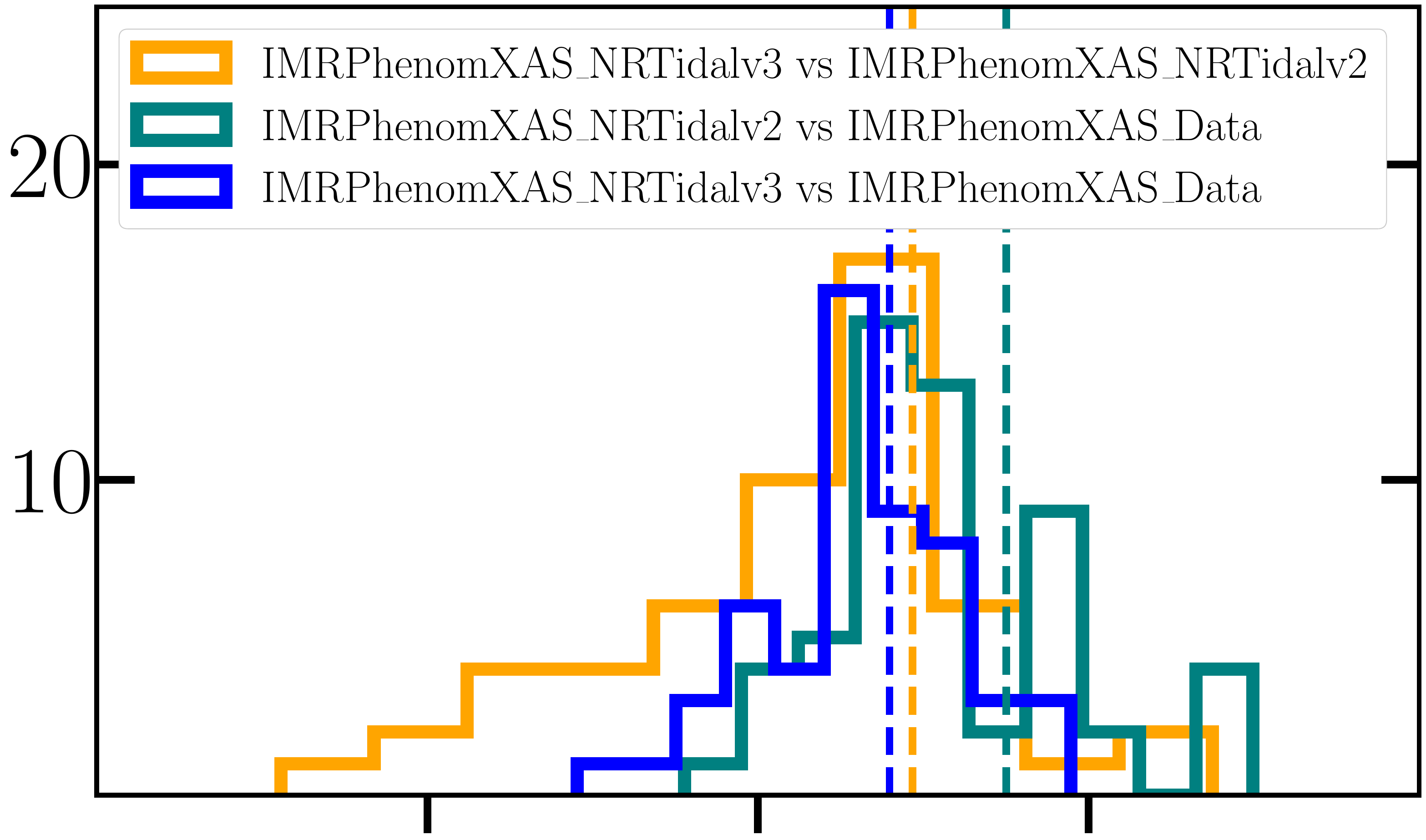}
\vspace{0.00mm}
\includegraphics[width=0.38\linewidth]{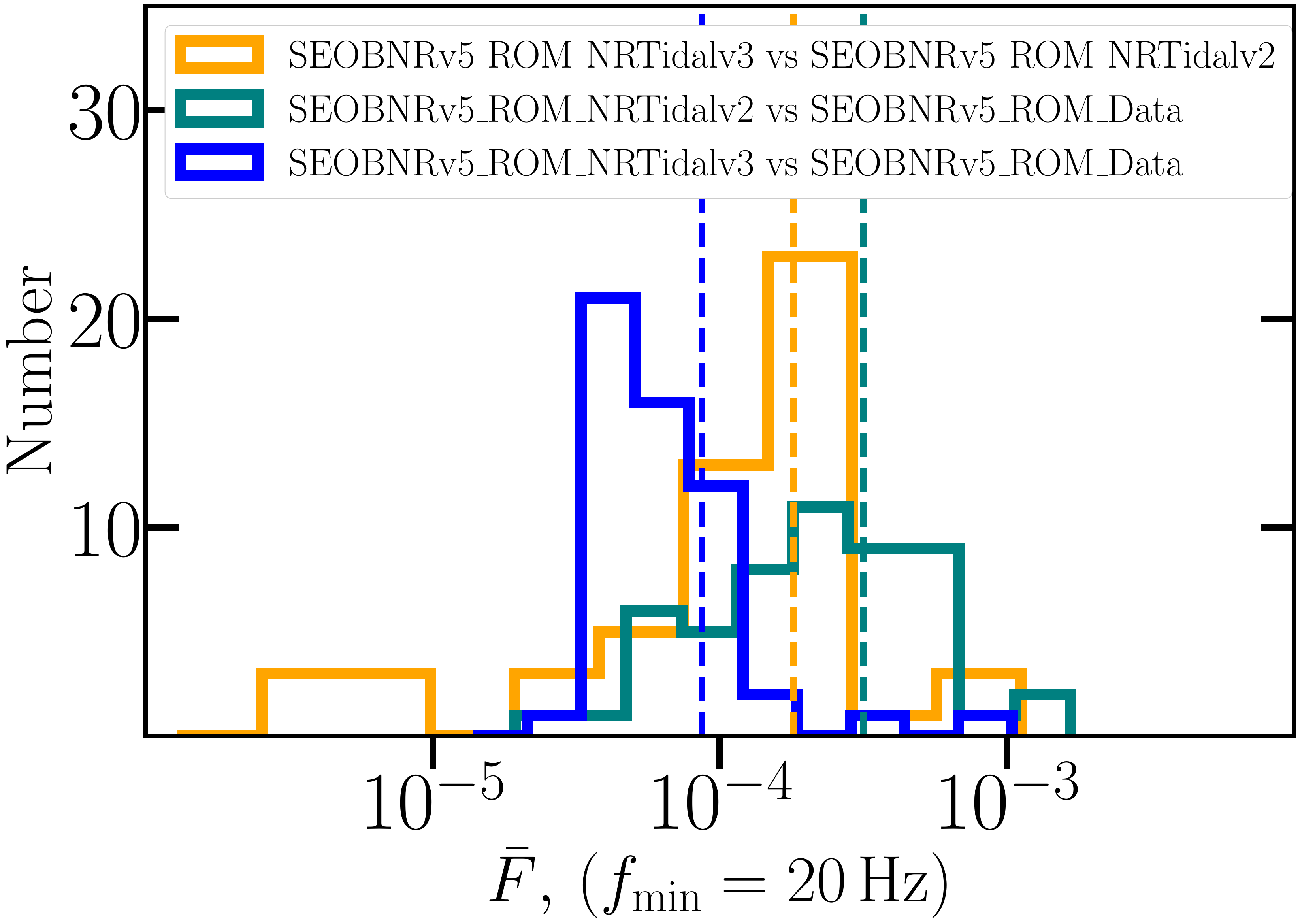}\qquad
\includegraphics[width=0.38\linewidth]{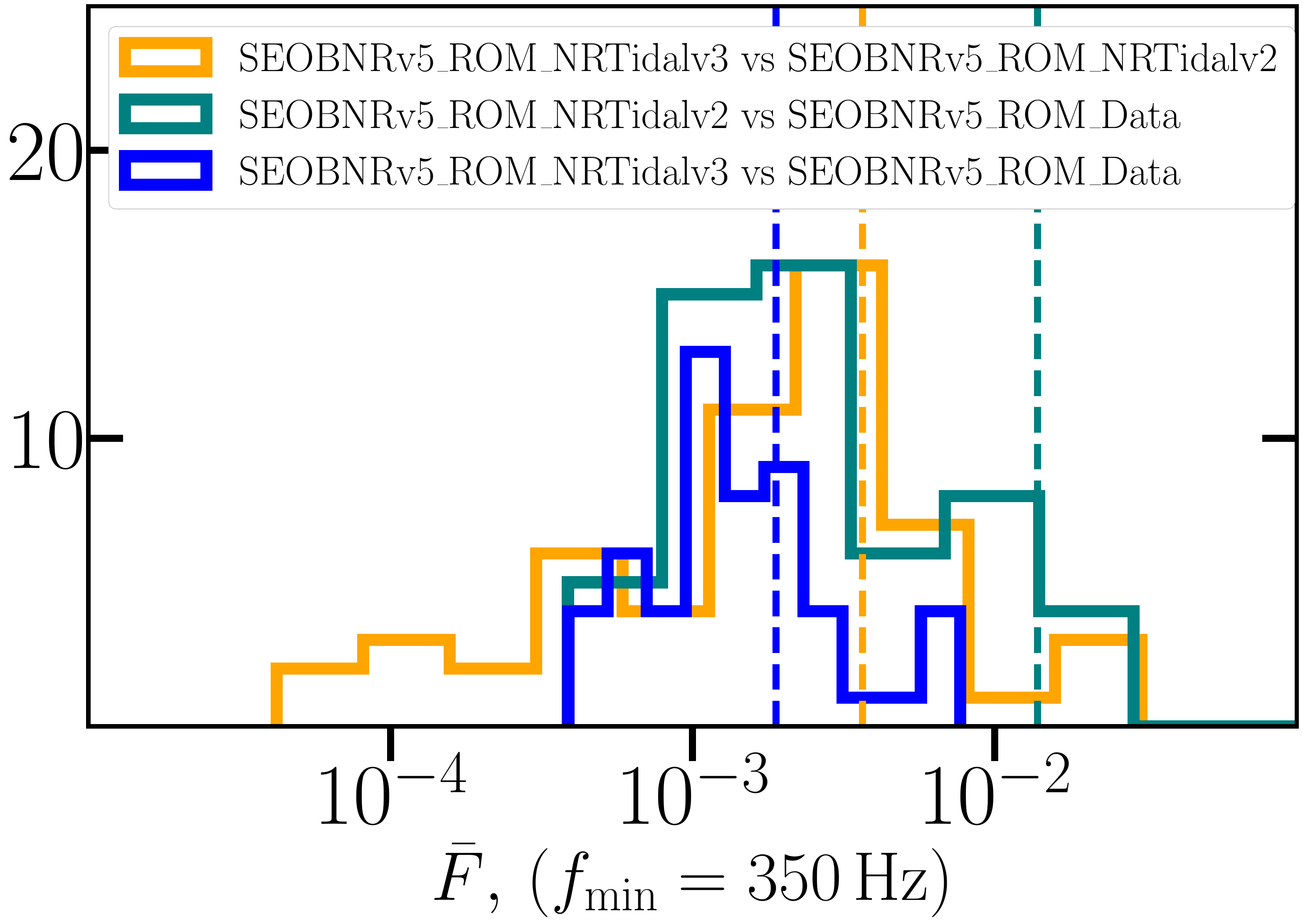}
\caption{Mismatch histograms for different BBH baselines augmented with the \nrtidalvthree\ model, versus the BBH baseline model added with the EOB-NR hybrid tidal phase. Vertical lines denote the mean of the histograms with the same color. The mismatches with \nrtidalvtwo\ are also shown for comparison. For each BBH baseline + \nrtidalvthree\ model, mismatches are computed starting from $f_{\rm min} = 20 \, {\rm Hz}$ in the left panel, indicating the mismatch for the full waveform where EOB contribution is present in the early inspiral. Mismatches are also computed for $f_{\rm min} = 350 \, {\rm Hz}$ in the right panel, where typically the hybridization starts and where the NR contribution becomes dominant up to merger.}
\label{fig: mismatchhistograms}
\end{figure*}

\begin{figure*}[t]
    \centering 
  \includegraphics[width=0.246\linewidth]{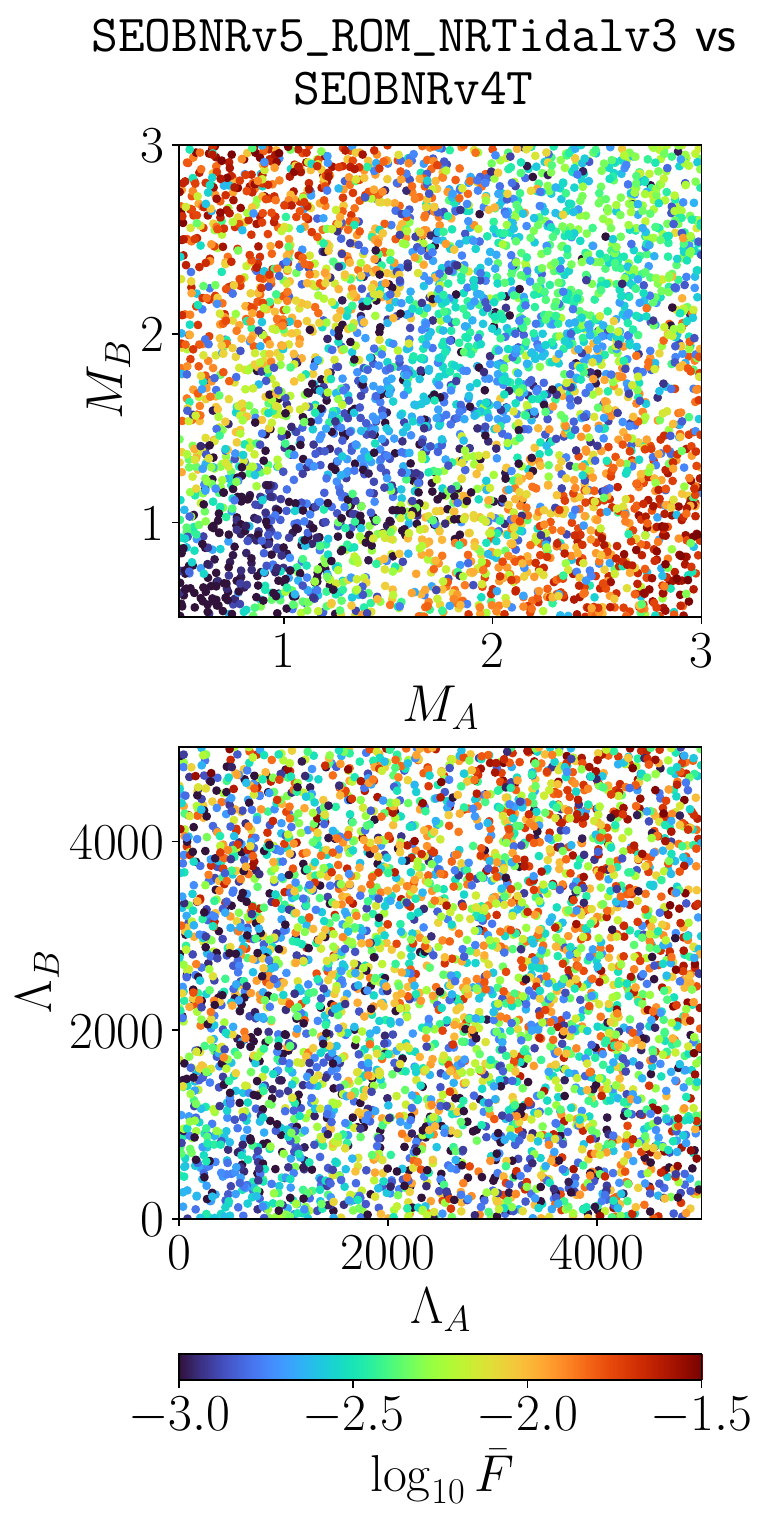}
  \includegraphics[width=0.246\linewidth]{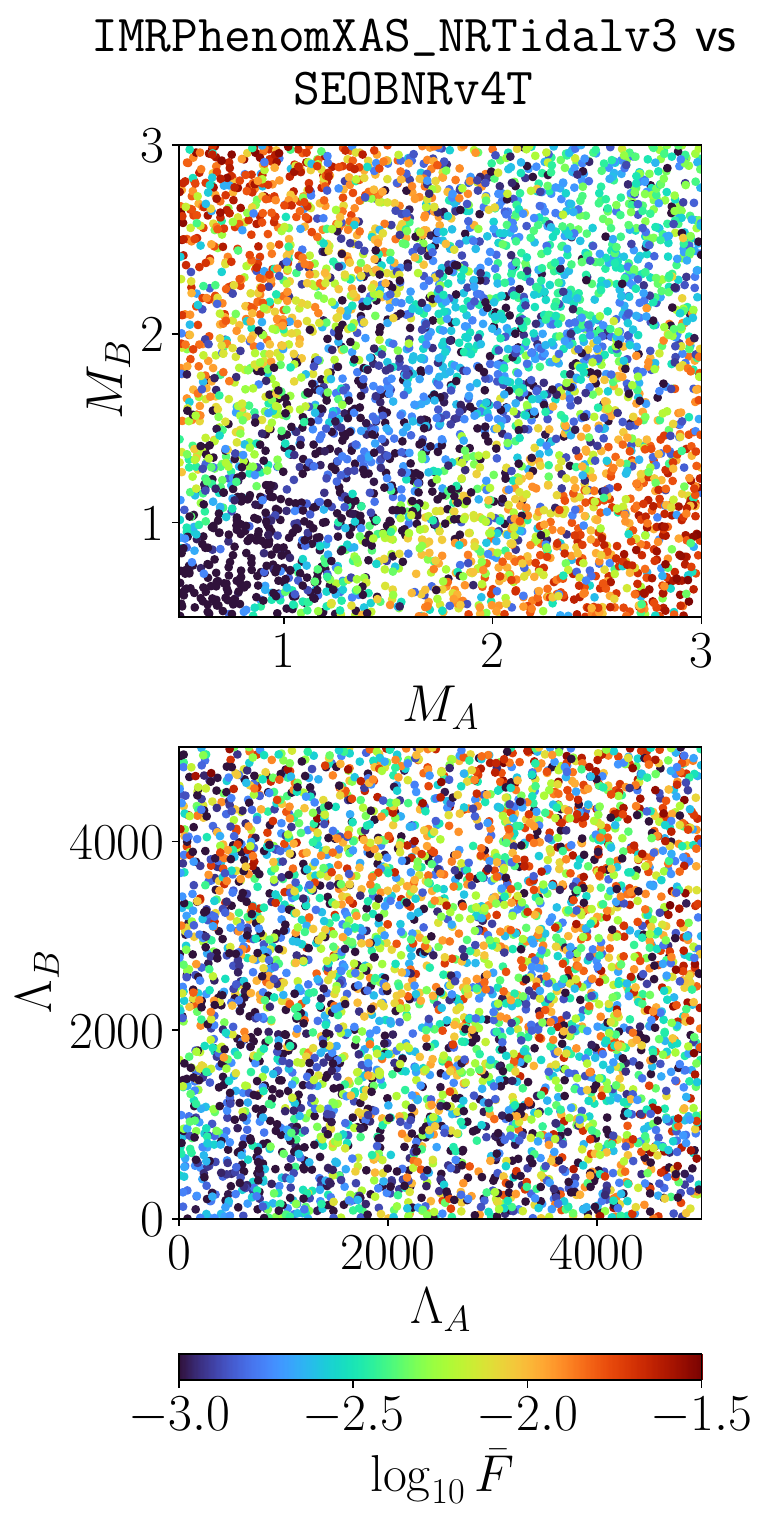}
  \includegraphics[width=0.246\linewidth]{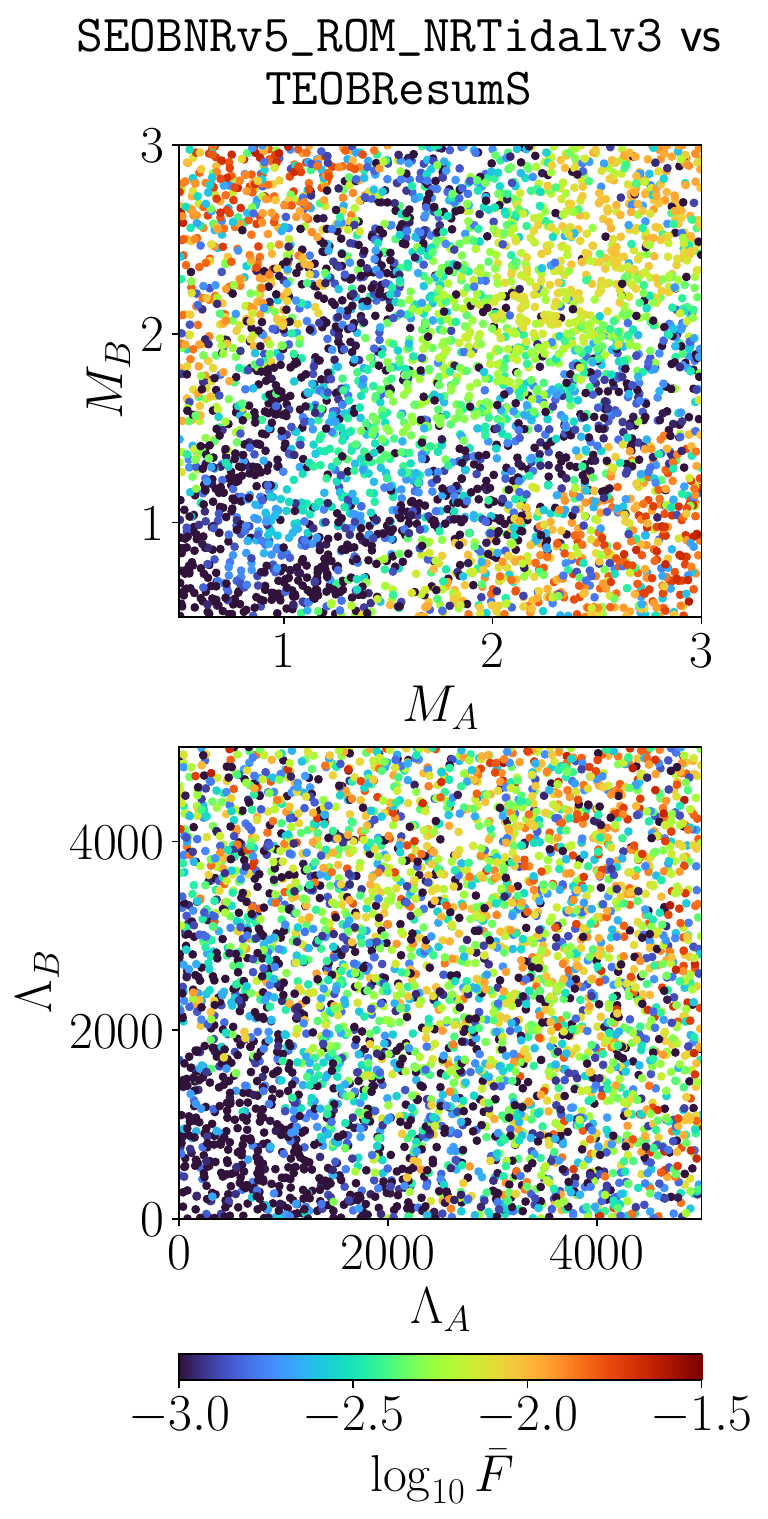}
  \includegraphics[width=0.246\linewidth]{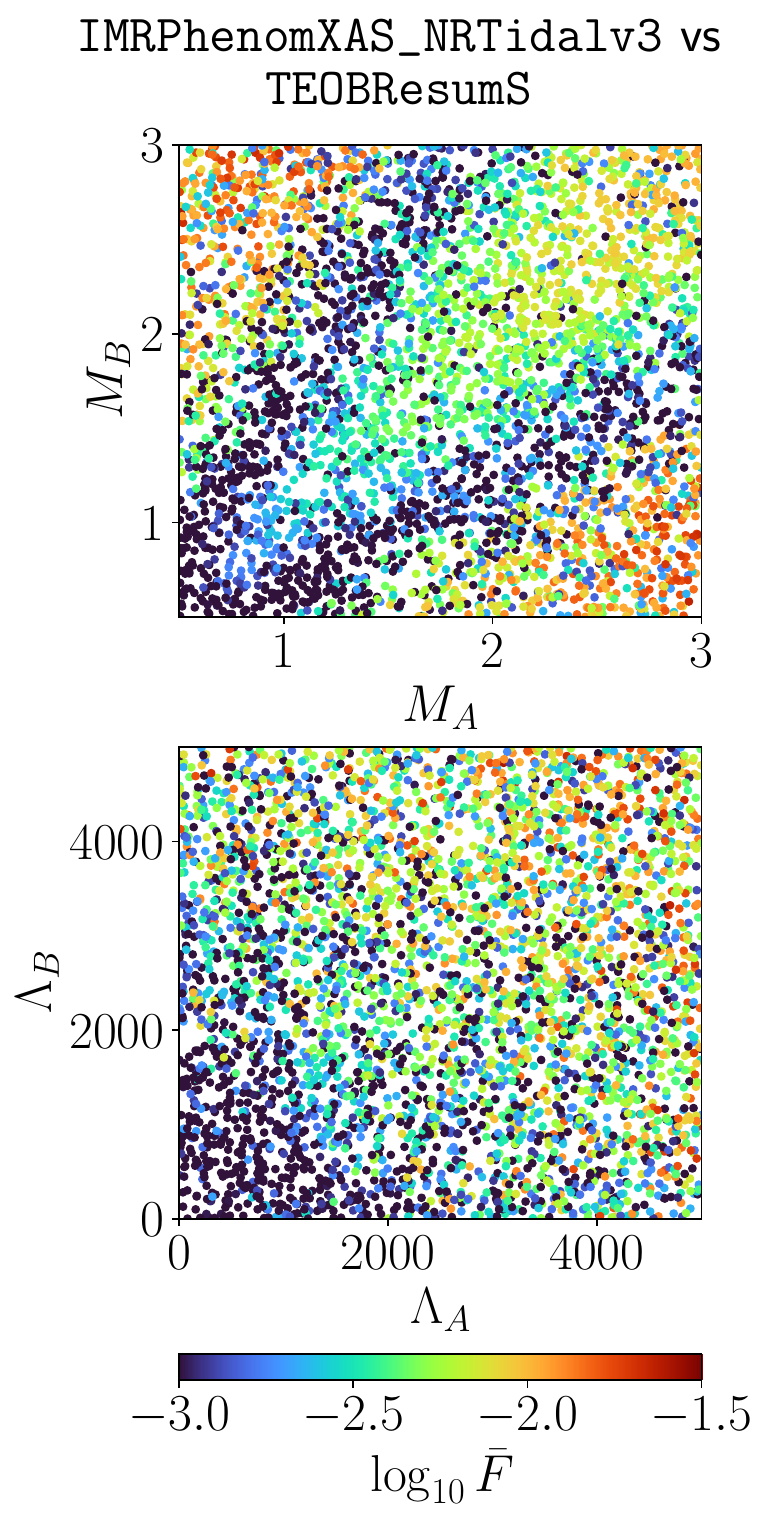}
  \includegraphics[width=0.246\linewidth]{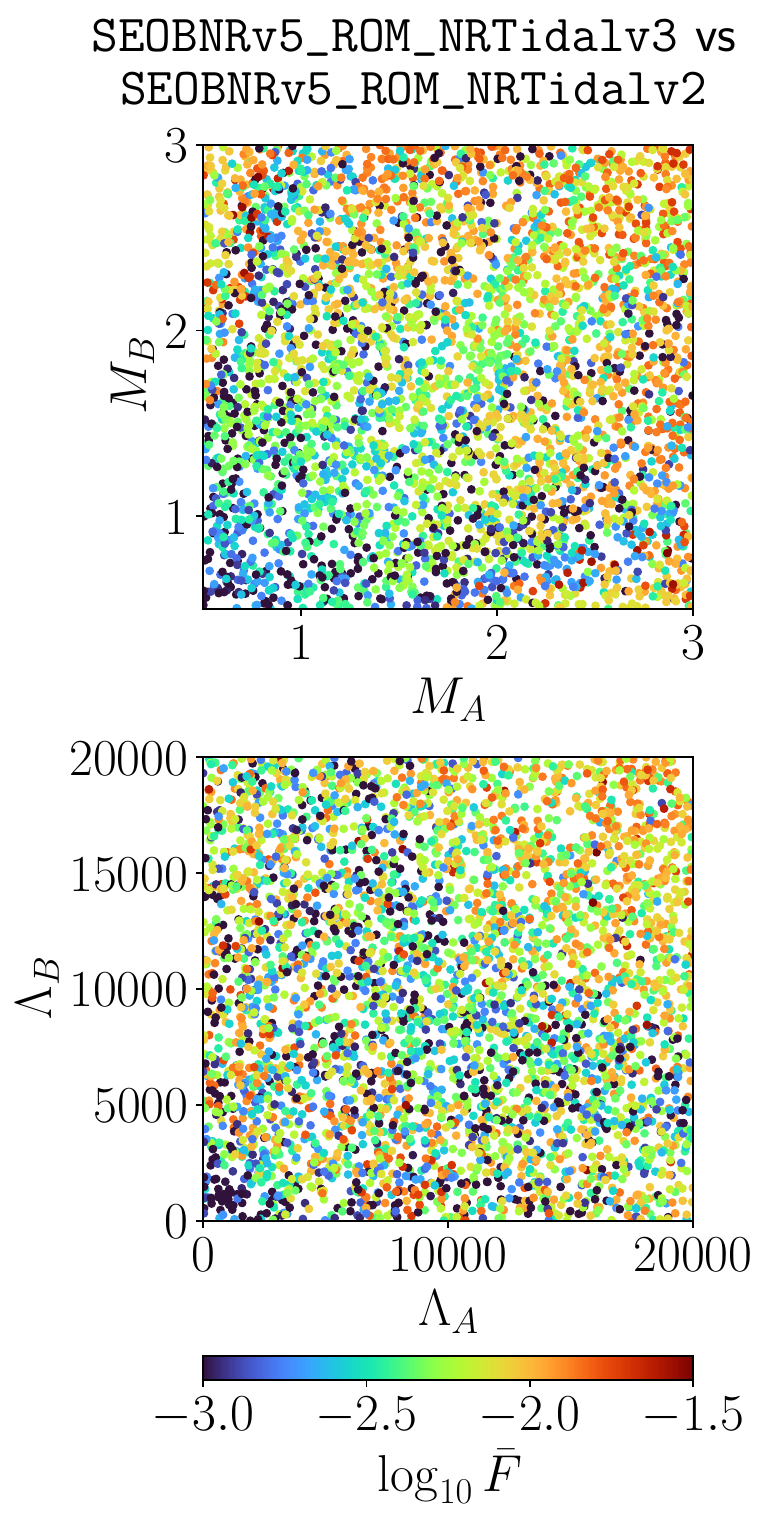}
  \includegraphics[width=0.246\linewidth]{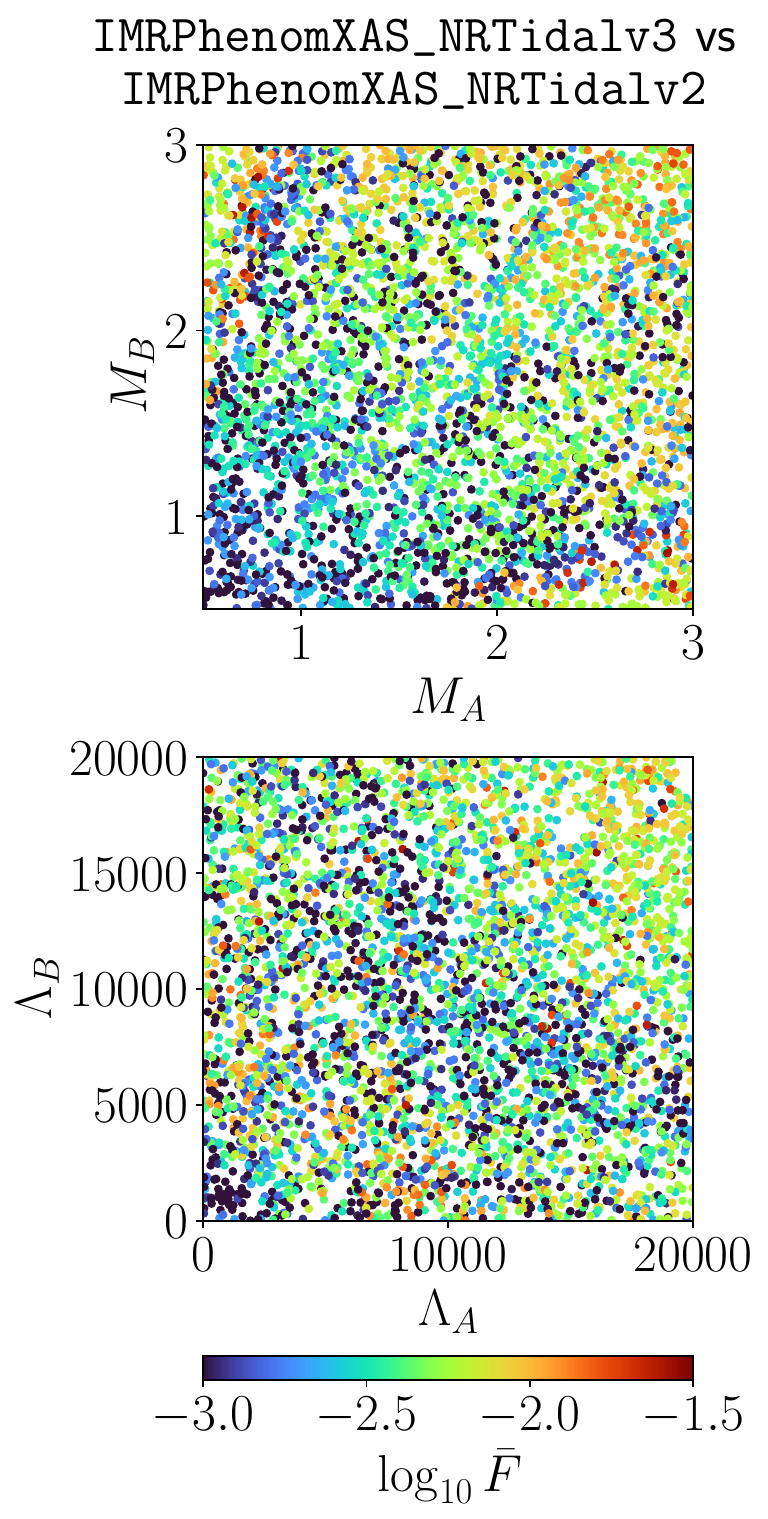}
  \includegraphics[width=0.246\linewidth]{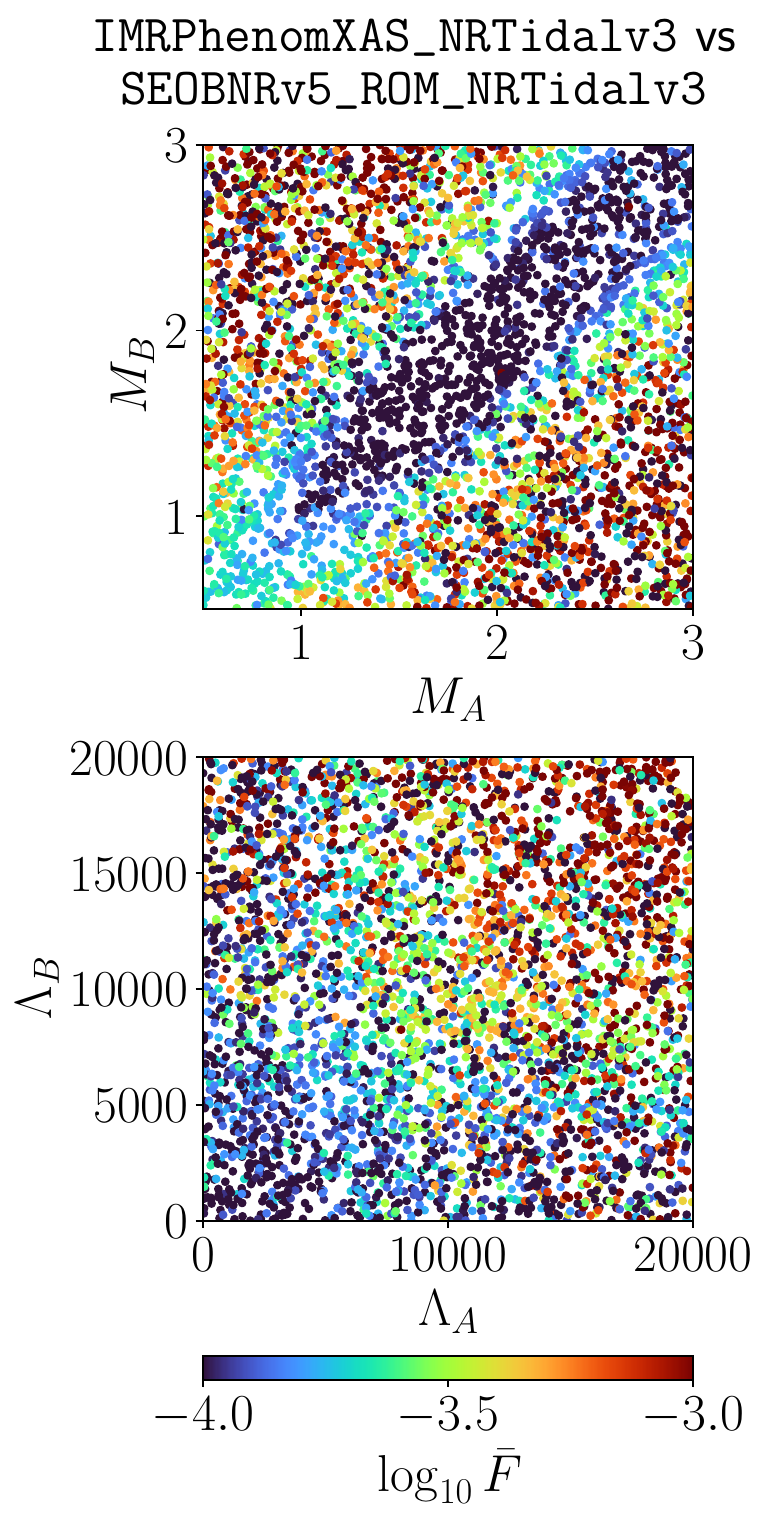}
\caption{Mismatch Comparisons of \nrtidalvthree\ with other tidal waveform models for non-spinning configurations. Each comparison with either \teobresums\ or \seobnrvfourt\, contains 4000 random configurations of mass $M_{A,B} = [0.5, 3.0]$ and tidal deformability $\Lambda_{A,B} = [0, 5000]$. The rest of the comparisons are done with $\Lambda_{A,B} = 20000$. For each subfigure we include a color bar indicating the values of the mismatches. Note that the maximum value in the color bar for \imrphenomxasnrtidalthree\ vs \seobnrvfiveromnrtidalvthree\ is different from the rest of the subfigures due to the \nrtidalvthree\ models yielding very small mismatches with respect to each other.}
\label{fig: mm nospin}
\end{figure*} 

\begin{figure*}
    \centering 
  \includegraphics[width=0.77\linewidth]{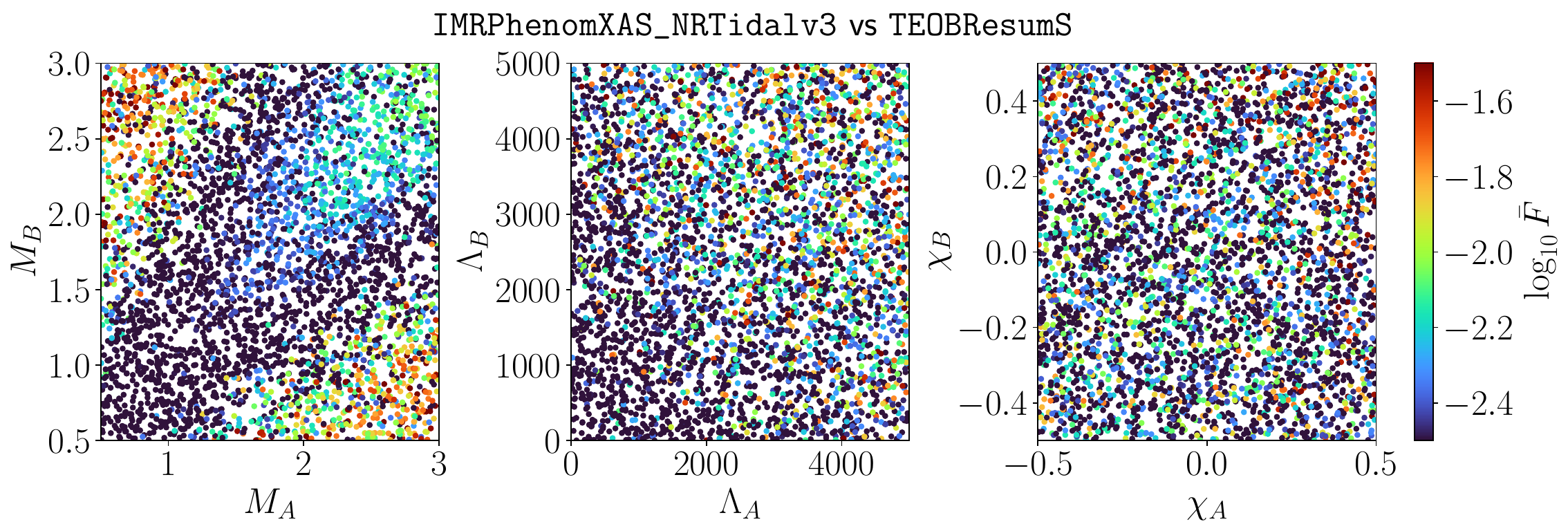}
  \includegraphics[width=0.77\linewidth]{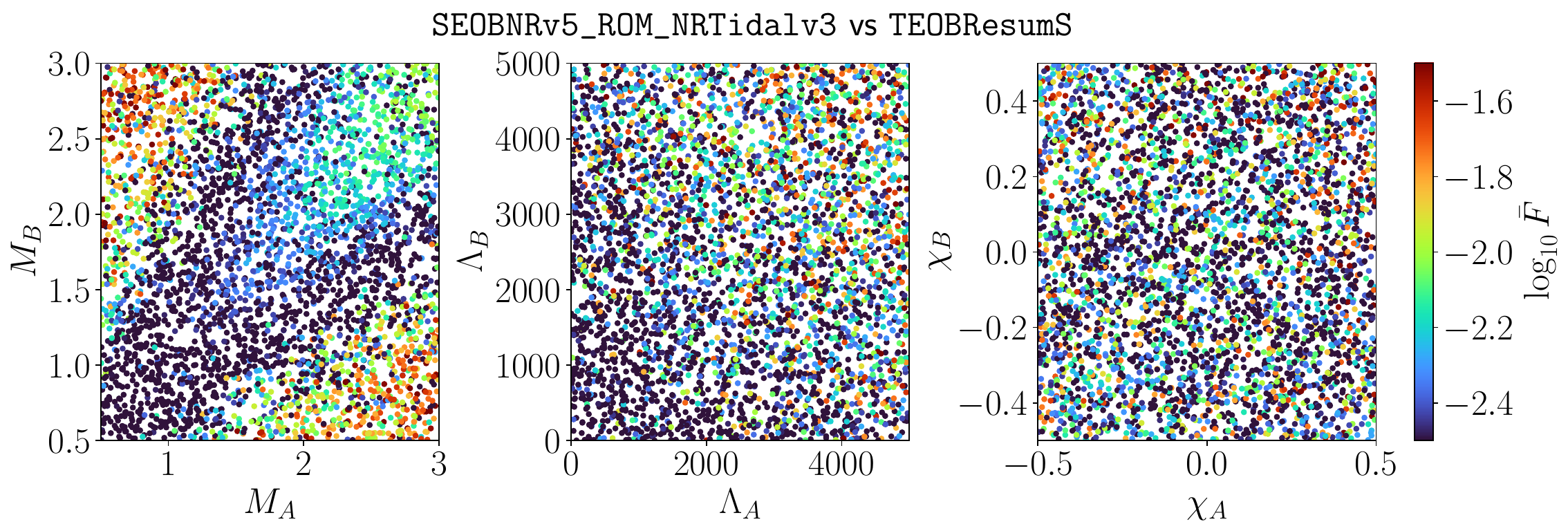}
  \includegraphics[width=0.77\linewidth]{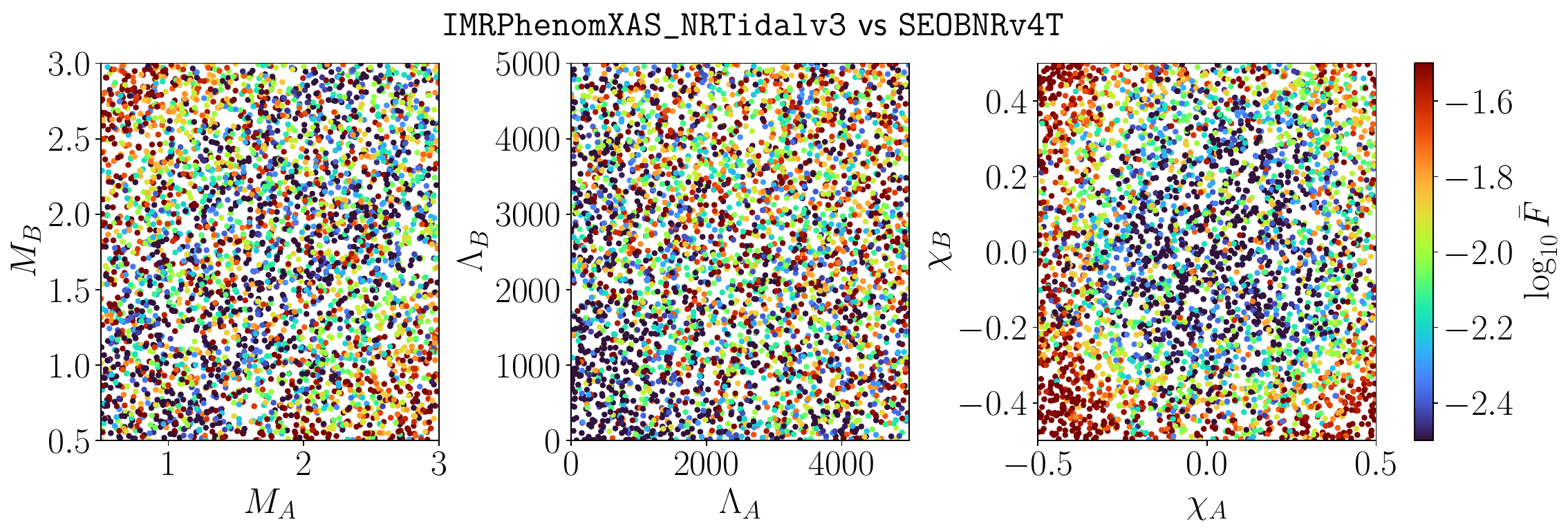}
  \includegraphics[width=0.77\linewidth]{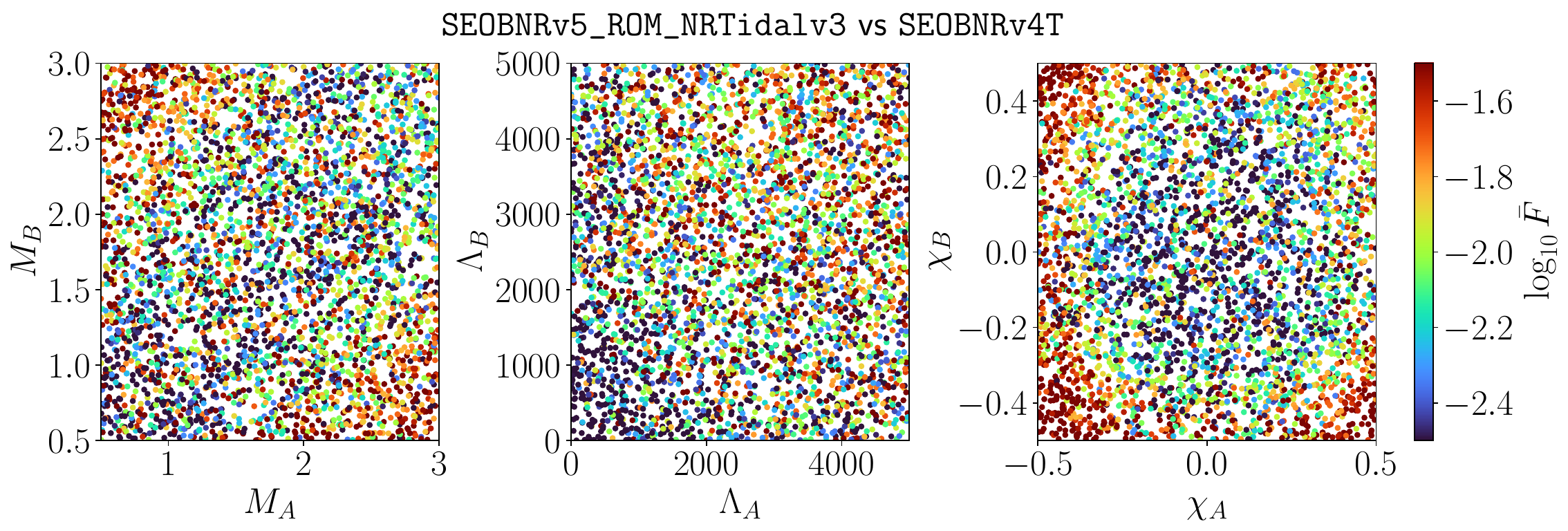}
\caption{Mismatch Comparisons of \nrtidalvthree\ with other tidal waveform models for aligned-spin configurations. Each subfigure has 4000 random configurations of mass $M_{A,B} = [1.0, 3.0]$, tidal deformability $\Lambda_{A,B} = [0, 5000]$ and spin $\chi_{A,B} = [-0.5,0.5]$. We also put the mismatches with \nrtidalvtwo\ for comparison. For each subfigure we include a color bar indicating the values of the mismatches.}
\label{fig: mm aligned spin}
\end{figure*}

\begin{figure}
    \centering
    \includegraphics[width = 0.9\linewidth]{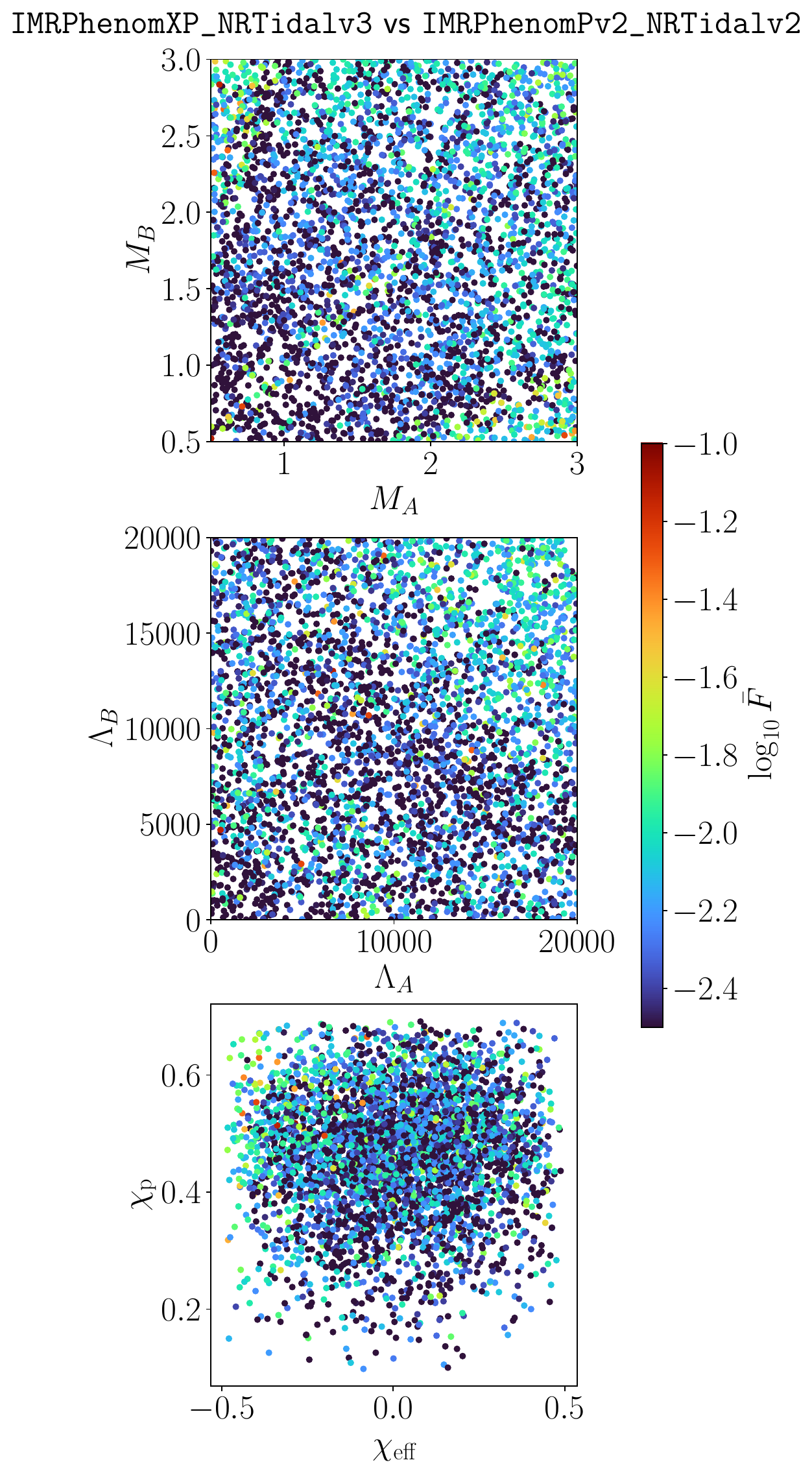}
    \caption{Mismatches for \imrphenomxpnrtidalthree\ vs \imrphenompvtwonrtidaltwo\ for precessing configurations. For this comparison, we generate 4000 random samples with mass $M_{A,B} = [0.5, 3.0]$, tidal deformabilities $\Lambda_{A,B} = [0, 20000]$, and spins $\chi_{A,B,C}^{x,y,z} = [-0.5,0.5]$. For the spins, we plot the effective spin precession $\chi_p$ against the effective aligned spin $\chi_{\rm eff}$. We include a colorbar indicating the values of the mismatches.} 
    \label{fig: prec spin}
\end{figure}

In this section, we compare the waveform models utilizing \nrtidalvthree\ with other waveforms, particularly their corresponding \nrtidalvtwo\ versions, as well as the \seobnrvfourt\ model using NR waveforms from BAM \cite{Dietrich:2018phi, Gonzalez:2022mgo}, SACRA~\cite{Kiuchi:2017pte, Kawaguchi:2018gvj, Kiuchi:2019kzt}, and SpEC~\cite{Foucart:2018lhe}. The configurations of the NR waveforms used for comparison are found in Table~\ref{table: bns nr configs}. We note that the waveforms named BAM:0001, BAM:0037, SACRA:15H\_135\_135\_00155\_182\_135, and SACRA:HB\_121\_151\_00155\_182\_135 are also used in the calibration of \nrtidalvthree\ (see Table~\ref{table: 55nr} in Appendix~\ref{subsection: appA}). For \nrtidalvthree\ we employ \imrphenomxasnrtidalthree\  and \seobnrvfiveromnrtidalvthree, and the other tidal waveforms for the comparison are \teobresums~\cite{Damour:2014sva, Nagar:2015xqa, Nagar:2018zoe, Nagar:2019wds, Nagar:2020pcj, Riemenschneider:2021ppj}, \seobnrvfourt~\cite{Lackey:2018zvw}, \seobnrvfiveromnrtidalvtwo\ (also implemented by this work), and \imrphenomxasnrtidaltwo.

For the time-domain comparison, we are first aligning the waveforms using the same procedure as for the construction of the EOB-NR hybrids described in Sec.~\ref{section: nr data and extraction}. 
Fig.~\ref{fig: timedomaincomparisons} shows our comparison for the BAM waveforms, where we also indicate the numerical uncertainty as shaded bands. In particular, we use a blue band for setups for which we just calculate the phase difference between the two highest resolutions~\footnote{The exception to this would be the two SACRA waveforms, whose errors were computed using the highest and third-highest resolutions. This is due to the SACRA waveforms having higher resolutions (than BAM) in general but not showing a clear convergence order, so the errors may be underestimated.}. For NR simulations where we find a clear convergence order and employ the Richardson-extrapolated waveform, we are using a green band as an error measure given by the phase difference between the Richardson-extrapolated waveform and the highest resolution.

We observe from Fig.~\ref{fig: timedomaincomparisons} that for the BAM NR data, both \nrtidalvthree\ models perform well in terms of the dephasing with respect to the NR waveforms, and the \nrtidalvthree\ models are consistent with the other models such that they generally fall within the estimated numerical uncertainties. The exception to this would be for BAM:0081 and BAM:0094 (the same can be said for the other waveform models), which are characterized by large mass ratios, and large tidal deformabilities which are far outside the space of calibration for \nrtidalvthree, though we still observe \nrtidalvthree\ to perform a bit better than \nrtidalvtwo. We also note that the employed EoS for these setups (MS1b) is already disfavored by the observation of GW170817~\cite{LIGOScientific:2017vwq, LIGOScientific:2017zic, LIGOScientific:2017ync}, i.e., this could be considered as a reasonable upper bound for a realistic BNS setup. We also present the comparison between the models and the SACRA and SpEC waveforms. 
Generally, due to the higher resolution, the computed uncertainties are generally smaller than for the BAM waveforms, however, we do not see a clear convergence order, which means that we cannot estimate the error based on Richardson extrapolation. 
For SACRA waveforms, we see that the \nrtidalvthree\ models perform better than the other waveform models. However, we note that this might also simply be due to the fact that the same waveforms were also used in the calibration of \nrtidalvthree. Meanwhile, no model was able to capture the merger part of SXS:NSNS:0001\footnote{An alternative time-domain dephasing comparison was done with \seobnrvfourt, but not using a quasi-universal relation to compute the waveform parameters for the polytropic EoS (used for SXS:NSNS:0001), and the result stays the same as in Fig.~\ref{fig: timedomaincomparisons}.}, while \nrtidalvthree\ performed better than the others in terms of the dephasing for SXS:NSNS:002.

\subsection{Mismatch against EOB-NR hybrid data}
\label{subsection: mismatch against eob-nr}
We further test the accuracy of our fits by comparing mismatches of the \nrtidalvtwo\ and \nrtidalvthree\ models with the hybrid EOB-NR waveforms. The mismatch (or unfaithfulness) between two frequency-domain complex waveforms $\tilde{h}_1$ and $\tilde{h}_2$ is given by
\begin{equation}\label{eq: mismatch}
    \bar{F} = 1 - \text{max}_{\phi_c, t_c}\frac{(h_1(\phi_c, t_c)|h_2)}{\sqrt{(h_1|h_1)(h_2|h_2)}},
\end{equation}
where the overlap is
\begin{equation}\label{eq: overlap}
    (h_1|h_2) = 4 {\rm Re} \int_{f_\text{min}}^{f_{\text{max}}}\frac{\tilde{h}_1^{*}(f)\tilde{h}_2(f)}{S_n(f)}df,
\end{equation}
and the maximization of the overlap for some arbitrary phase $\phi_c$ and time shift $t_c$ (in the time domain, corresponding to a phase and frequency shift in the frequency domain) ensures the alignment between the two waveforms. Here, we assume the spectral density of the detector as $S_n(f) = 1.0$ so that the computation is detector-agnostic. We set $f_{\text{max}} = f_{\text{mrg}}$. For the waveforms, we choose a sampling rate of $2^{13} = 8192$ Hz. For a full comparison, we construct full waveforms with corresponding BBH baselines (from the model we want to compare with) and add to its phase the EOB-NR hybrid tidal phase, i.e., we have \texttt{IMRPhenomD\_Data}, \texttt{IMRPhenomXAS\_Data}, and \texttt{SEOBNRv5\_ROM\_Data}. The minimum frequency $f_{\text{min}}$ is chosen to be either 20~Hz or 350~Hz. The former takes into account almost the entire hybrid waveform, which is dominated in the very early inspiral by the EOB model used for constructing the EOB-NR hybrid (\seobnrvfourt), while the latter value considers mainly the NR contribution (typically where the hybridization starts).

We show results as mismatch histograms for  \imrphenomdnrtidalthree, \imrphenomxasnrtidalthree, and \seobnrvfiveromnrtidalvthree\ in Fig.~\ref{fig: mismatchhistograms}. Taking a look, for example, at the histograms for \imrphenomdnrtidalthree\ with $f_{\rm min} = 20\, {\rm Hz}$ (the uppermost left panel), we compute the mismatches between  \imrphenomdnrtidalthree\ and \texttt{IMRPhenomD\_Data}, between  \imrphenomdnrtidalthree\ and \texttt{IMRPhenomD\_NRTidalv2}, and between \texttt{IMRPhenomD\_NRTidalv2} and \texttt{IMRPhenomD\_Data}. For each histogram, the mean mismatch $\mu_{\bar{F}}$  is indicated by a vertical dashed line. The mismatches are larger with larger $f_{\rm min}$. We note the improvement in the accuracy of the \nrtidalvthree\ waveforms with respect to \nrtidalvtwo, as indicated by their lower mismatches with respect to the hybrid data. In general, \nrtidalvthree\ has a mismatch of about half of an order of magnitude smaller than that of \nrtidalvtwo. We also note that for $f_{\rm min} = 350 \, {\rm Hz}$, \nrtidalvtwo\ has a tail in its distribution for $\bar{F} \gtrsim \mathcal{O}(10^{-2})$, while this is not present for \nrtidalvthree. While the reduction of the mismatch is to some extent also caused by the fact that all of the employed waveforms have been used in the calibration of \nrtidalvthree, it still shows the overall robustness and performance of the model. 

\begin{table*}[t]
\caption{\label{table: meanmax_mm} The mean mismatch $\mu_{\bar{F}}$ and maximum mismatch ${\rm max}(\bar{F})$ for various waveform model comparisons for the non-spinning, aligned-spin, and precessing configurations. A `$-$' symbol indicates that the mismatches were not calculated between the two waveform models.} 
\begin{tabular}{p{3.6cm}|l|c|c}
\hline
\multicolumn{4}{c}{\multirow{2}{*}{\textbf{Non-Spinning   Case}}} \\
\multicolumn{4}{c}{} \\
\hline
\hline
  &   & \multirow{2}{*}{\parbox{3.6cm}{\centering \texttt{IMRPhenomXAS\_NRTidalv3}}} &  \multirow{2}{*}{\parbox{3.6cm}{\centering \texttt{SEOBNRv5\_ROM\_NRTidalv3}}} \\
   &   &    &   \\
\hline
\multirow{2}{*}{\parbox{2cm}{\centering \texttt{IMRPhenomXAS\_NRTidalv3}}}  & $\mu_{\bar{F}}$ & $-$ & $3.608\times 10^{-4}$ \\
              & ${\rm max}(\bar{F})$ & $-$ & 0.005032  \\
\hline
\multirow{2}{*}{\parbox{2cm}{\centering \texttt{SEOBNRv5\_ROM\_NRTidalv3}}}  & $\mu_{\bar{F}}$ & $3.608\times 10^{-4}$ & $-$ \\
              & ${\rm max}(\bar{F})$ & 0.005032 & $-$ \\
\hline
\multirow{2}{*}{\parbox{2cm}{\centering \teobresums}} & $\mu_{\bar{F}}$      & 0.004238  & 0.004736  \\ 
           & ${\rm max}(\bar{F})$ & 0.02527  & 0.02648  \\
\hline
\multirow{2}{*}{\parbox{2cm}{\centering \texttt{IMRPhenomXAS\_NRTidalv2}}}  & $\mu_{\bar{F}}$ &  0.004279  &$-$ \\
              & ${\rm max}(\bar{F})$ & 0.02804 & $-$ \\
\hline
\multirow{2}{*}{\parbox{2cm}{\centering \texttt{SEOBNRv5\_ROM\_NRTidalv2}}}  & $\mu_{\bar{F}}$ & $-$ & 0.006038 \\
              & ${\rm max}(\bar{F})$ & $-$ & 0.03243 \\
\hline
\multirow{2}{*}{\parbox{2cm}{\centering \seobnrvfourt}} & $\mu_{\bar{F}}$      & 0.005874  &   0.006729 \\ 
           & ${\rm max}(\bar{F})$ & 0.03282 &  0.03415 \\
\hline

\multicolumn{4}{c}{\multirow{2}{*}{\textbf{Aligned Spin Case}}}                       \\
\multicolumn{4}{c}{} \\
\hline
\hline
            &                      & \teobresums & \seobnrvfourt    \\
\hline
\multirow{2}{*}{\parbox{2cm}{\centering \texttt{IMRPhenomXAS\_NRTidalv3}}}  & $\mu_{\bar{F}}$ & 0.005577 & 0.01456 \\
              & ${\rm max}(\bar{F})$ & 0.06703 & 0.1713\\
\hline
\multirow{2}{*}{\parbox{2cm}{\centering \texttt{SEOBNRv5\_ROM\_NRTidalv3}}}  & $\mu_{\bar{F}}$ & 0.006278 & 0.01454 \\
              & ${\rm max}(\bar{F})$ &  0.06221 & 0.1708 \\
\hline

\multicolumn{4}{c}{\multirow{2}{*}{\textbf{Precessing Case}}}                        \\
\multicolumn{4}{c}{} \\
\hline
\hline
            &                      & \imrphenomxpnrtidalthree     &   $-$        \\
            \hline
\multirow{2}{*}{\parbox{2cm}{\centering \texttt{IMRPhenomPv2\_NRTidalv2}}}  & $\mu_{\bar{F}}$ & 0.0056560 & $-$ \\
              & ${\rm max}(\bar{F})$ &  0.08585 & $-$ \\
\hline
\end{tabular}
\end{table*}

\subsection{Mismatches against other tidal waveform models}
\label{subsection: mismatch against other waveforms}
Finally, we want to compare our model with other tidal models in three cases: (a) non-spinning configurations, (b) aligned spins, and (c) precessing systems. For the non-spinning configurations, we compute the mismatch between \nrtidalvthree\ (specifically, \imrphenomxasnrtidalthree\ and \seobnrvfiveromnrtidalvthree) and \teobresums, \seobnrvfourt, and \nrtidalvtwo.  The mismatches computed in this section assume $f_{\rm min} = 40\, {\rm Hz}$, $f_{\rm max} = 2048\, {\rm Hz}$ and a waveform sampling rate of $2^{13} = 8192 \, {\rm Hz}$, except for when comparisons are done with \seobnrvfourt, where a sampling rate of $2^{13} = 4096 \, {\rm Hz}$ was used for the sake of efficiency.

\paragraph{Non-spinning setups:}
In the non-spinning case, the mismatches are computed for 4000 random configurations of masses $M_{A,B} = [0.5, 3.0] M_{\odot}$ and tidal deformabilities $\Lambda_{A,B} = [0, 5000]$, for comparisons done with \teobresums\ and \seobnrvfourt. For $\Lambda_{A,B} \ge 5000$, and especially when combined with a high-mass-ratio or high-spin configuration, we find both \teobresums\ and \seobnrvfourt\ to fail to produce a waveform. For comparisons which do not involve these two waveform models, we use $\Lambda_{A,B} = [0, 20000]$. These mismatches are then plotted in a 2D scatter plot (in the x-y plane) with the masses and tidal deformabilites (Fig.~\ref{fig: mm nospin}). In addition, we also directly compare the mismatches between the \nrtidalvthree\ models. We note the very small mismatches, with $\mu_{\tilde{F}} = 3.608 \times 10^{-4}$, indicating that the waveforms behave very similarly to each other. In general, we find very small mismatches for relatively low masses and low-mass ratios, with the largest mismatches occurring at higher masses ($M_{A,B} \gtrsim 2.0M_{\odot}$) and high mass-ratios ($q \gtrsim 1.5$). Large mismatches also occur for large tidal deformabilities ($\Lambda_{A,B} \gtrsim \mathcal{O}(10^3)$). For all comparisons in the non-spinning case (except \imrphenomxasnrtidalthree\ vs \seobnrvfiveromnrtidalvthree) $\mu_{\bar{F}} = \mathcal{O}(10^{-3})$ and ${\rm max}(\bar{F}) = \mathcal{O}(10^{-2})$ (see Table~\ref{table: meanmax_mm}
).\\

\paragraph{Aligned-spin setups:}For the aligned-spin configurations, we compute the mismatches between two \nrtidalvthree\ variants, \imrphenomxasnrtidalthree\ and \seobnrvfiveromnrtidalvthree\, and these \nrtidalvthree\ variants against \teobresums\ and \seobnrvfourt. In addition to the 4000 random configurations of the masses and tidal deformabilities used in the non-spinning case, that we generated for the non-spinning case, we also generate 4000 random aligned spins $\chi_{A,B} = [-0.5,0.5]$. In the mismatch scatter plots, we plot the masses, the tidal deformabilities, and the aligned-spin components in the $x-y$ plane, and the mismatches in the color bar (see Fig.~\ref{fig: mm aligned spin}). When the \nrtidalvthree\ variants are compared against \teobresums, large tidal deformabilities ($\Lambda = \mathcal{O}(10^3)$), mass-ratios ($q \gtrsim 1.5$), and masses ($M_{A,B} \gtrsim 2.0$) result to large mismatches. In general, comparisons with \seobnrvfourt\ yield on average larger mismatches than that with \teobresums\ (see Table~\ref{table: meanmax_mm}). The mismatches are largest ($\mathcal{O}(10^{-1})$) for the configurations with very high spin magnitudes $|\chi_{A,B}| \gtrsim 0.4$.\\

\begin{figure*}
    \centering
    \includegraphics[width = 0.7\linewidth]{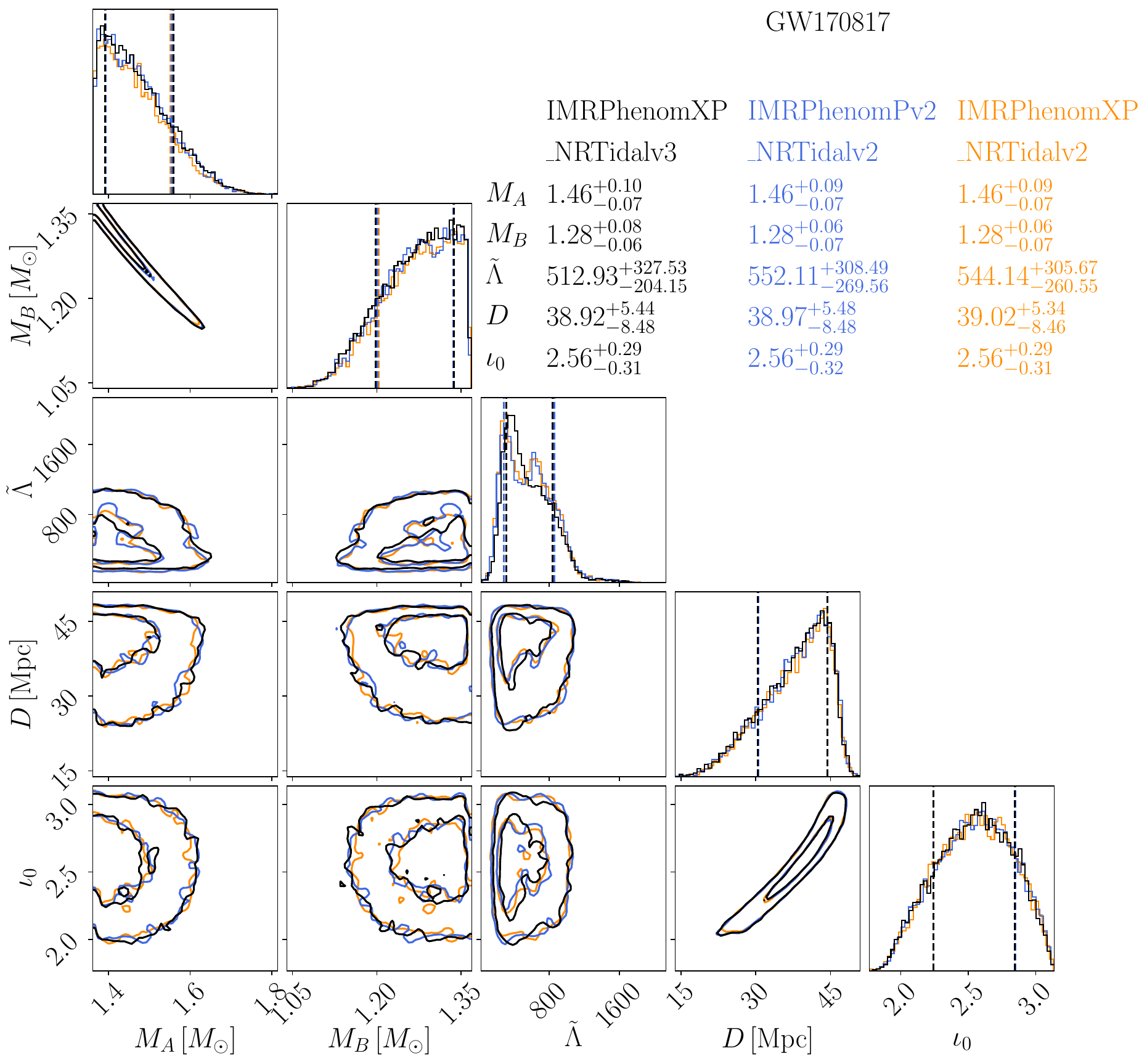}
    \caption{The marginalized 1D and 2D posterior probability distributions for selected parameters of GW170817, obtained with \imrphenomxpnrtidalthree\ (black), \imrphenompvtwonrtidaltwo\ (blue), and \imrphenomxpnrtidaltwo\ (orange). The parameters shown here are the individual star masses $M_{A,B}$, binary tidal deformability $\tilde{\Lambda}$, luminosity distance $D$, and inclination angle $\iota_0$. The 68\% and 90\% confidence intervals are indicated by contours for the 2D posterior plots, while vertical lines in the 1D plots indicate 90\% confidence interval. We note a narrow constraint on the tidal deformability for \imrphenomxpnrtidalthree\ compared to the other models, due to the updated tidal information that was used.}  
    \label{fig: PE_GW170817}
\end{figure*}

\begin{figure}[h]
    \centering
    \includegraphics[width = \linewidth]{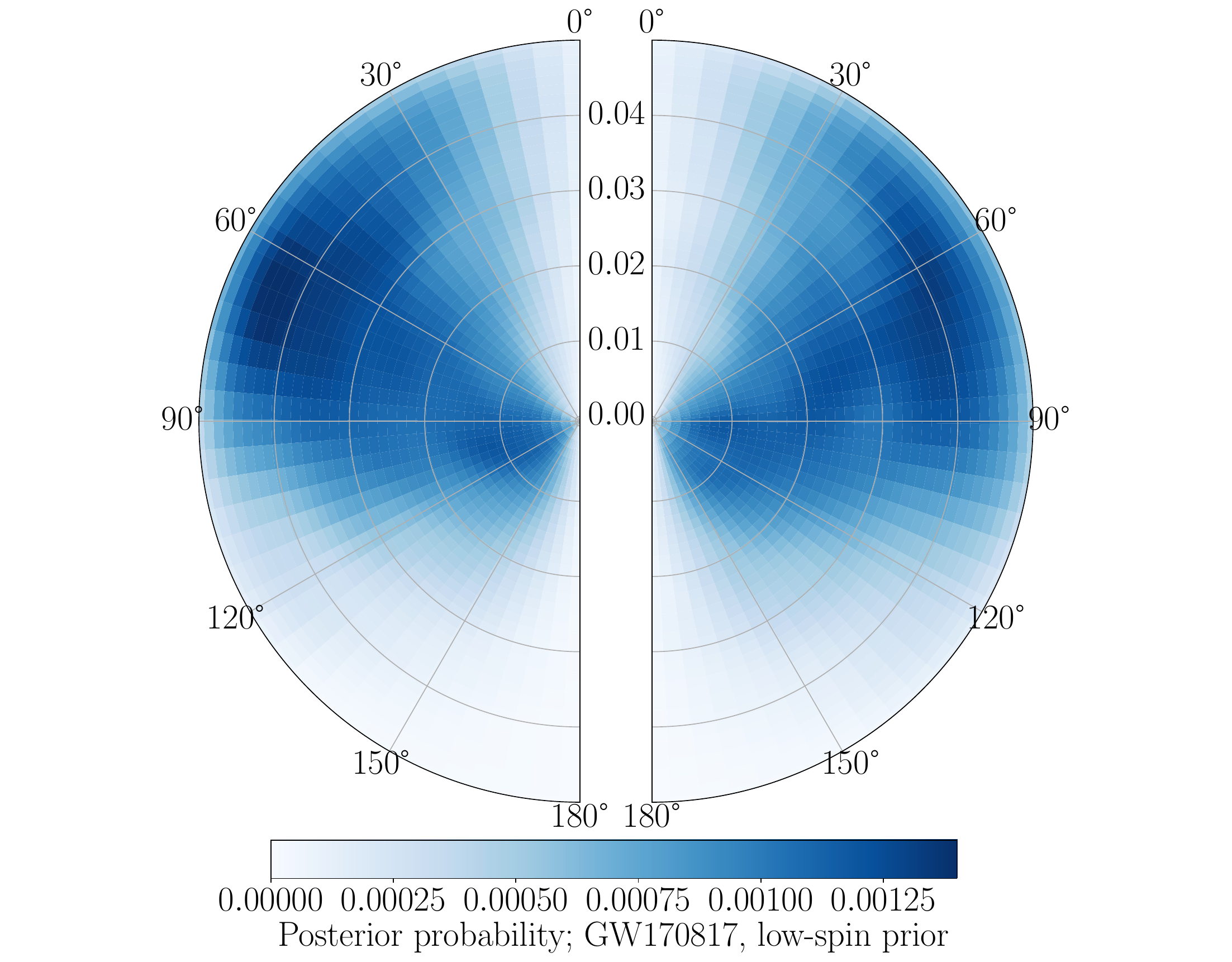}
    \caption{Inferred spin parameters for GW170817 from a low-spin prior ($\chi \le 0.05$) using \imrphenomxpnrtidalthree. Plotted here are the probability densities for the dimensionless spin components $\chi_1$ (left hemisphere) and $\chi_2$ (right hemisphere) relative to the orbital angular momentum $\mathbf{L}$ and tilt angles (a tilt angle $0^{\circ}$ means that the spin is aligned with the $\mathbf{L}$). The plot was done using a reference frequency of 20 Hz.} 
    \label{fig: lowspin_GW170817}
\end{figure}

\paragraph{Precessing setups:}
For the precessing configurations, we compare \imrphenomxpnrtidalthree\ and \imrphenompvtwonrtidaltwo~\cite{Husa:2015iqa, Dietrich:2019kaq}. Aside from the randomly generated masses and tidal deformabilities used in the non-spinning and aligned-spin cases, we also generate 4000 random values of the  $x, y, z$ components of the spin, $\chi_{A,B}^{x,y,z} = [-0.5,0.5]$. Here, for the spins, we plot in Fig.~\ref{fig: prec spin} on the $x-y$ plane the effective spin precession $\chi_p$ against the effective aligned spin $\chi_{\rm eff}$, where
\begin{equation}\label{eq: chip}
    \chi_p = \frac{S_p}{K_A M_A^2},
\end{equation}
where $S_p$ is the average magnitude of the spins:
\begin{equation}
    S_p = {\rm max}(K_A S_{A, \perp}, K_B S_{B, \perp}),
\end{equation}
with $K_A = 2 + 3/(2q)$ (with $M_A > M_B$), and $S_{A, \perp}$ is the magnitude of the in-plane spin for body $A$~\cite{Schmidt:2014iyl}. Meanwhile, the effective aligned spin is given by 
\begin{equation}
    \chi_{\rm eff} = \mathbf{m}_A\chi_{A} + \mathbf{m}_B\chi_B,
\end{equation}
where $\mathbf{m}_{A,B} = (1/2)(1\pm \sqrt{1 -4 \nu})$. From Fig.~\ref{fig: prec spin}, we observe large mismatches ($\mathcal{O}(10^{-2})$) for configurations with large masses ($M_{A,B} \gtrsim 2.5M_{\odot}$), large tidal deformabilities ($\Lambda_{A,B} \gtrsim \mathcal{O}(10^3)$), or large $\chi_p$ and $|\chi_{\rm eff}|$ ($\gtrsim 0.4$).

For all the configurations discussed above, we calculate the mean mismatch $\mu_{\bar{F}}$ and maximum mismatch ${\rm max}(\bar{F}$) for all waveform comparisons done. The results are shown in Table~\ref{table: meanmax_mm}.

\section{\label{section: pe analysis}Parameter Estimation}

Finally, to test the applicability of the developed \nrtidalvthree\ models, we will reanalyze the two BNS detections GW170817~\cite{LIGOScientific:2017vwq} and GW190425~\cite{LIGOScientific:2020aai}. 
For this purpose, we utilize parallel \texttt{bilby}~\cite{Ashton:2018jfp, Smith:2019ucc}, which performs GW parameter estimation using a parallelized nested sampler named \texttt{dynesty}~\cite{Speagle:2019ivv}. 
Parallel \texttt{bilby} uses Bayes' theorem:
\begin{equation}
    P(\theta|d, \mathcal{H}) = \frac{\mathcal{L}(d|\theta, \mathcal{H})p(\theta|\mathcal{H})}{E(d|\mathcal{H})},
\end{equation}
where $P(\theta|d, \mathcal{H})$ is the posterior probability distribution of the parameters $\theta$ given some data $d$ (which consists the waveform) and hypothesis $\mathcal{H}$, $p(\theta|\mathcal{H})$ is the prior probability distribution, and $E(d|\mathcal{H})$ is the evidence, serving as normalization constant to $P(\theta|d, \mathcal{H})$. Meanwhile, $\mathcal{L}(d|\theta, \mathcal{H})$ is the likelihood of obtaining the data $d$ given that the parameters $\theta$ are under the hypothesis $\mathcal{H}$. For further details, we refer to Refs.~\cite{Ashton:2018jfp, Romero-Shaw:2020owr, Ashton:2021cub}. 
In our study, we use GW170817 and GW190425 as our GW events, and we use the following waveform models: \imrphenomxpnrtidalthree, \imrphenompvtwonrtidaltwo, and \imrphenomxpnrtidaltwo, and also compare their results with each other. We also use low-spin and high-spin priors for both GW events following the standard LIGO Scientific-, Virgo, KAGRA Collaboration (LVK) analyses~\cite{LIGOScientific:2018hze, LIGOScientific:2017vwq, Romero-Shaw:2020owr, Dietrich:2020efo, LIGOScientific:2020aai, Ashton:2021cub}.

\subsection{GW170817}
We show the results of the inferred marginalized 1D and 2D posterior probability distribution of a selection of source parameters for GW170817 in Fig.~\ref{fig: PE_GW170817}, for a low-spin prior with $\chi \le 0.05$.  In this figure, we show the individual masses of the stars $M_{A,B}$, the luminosity distance $D$ in Mpc, and the inclination angle $\iota_0$. We also include here dimensionless tidal deformability defined in terms of the individual masses and tidal deformabilities of the stars~\cite{Dietrich:2020efo}:
\begin{equation}
    \tilde{\Lambda} = \frac{16}{13}\frac{(M_A + 12M_B)M_A^4\Lambda_A + (M_B + 12M_A)M_B^4\Lambda_B}{(M_A + M_B)^5}.
\end{equation}
From Fig.~\ref{fig: PE_GW170817}, we note that the performance of \imrphenomxpnrtidalthree\ in terms of the inferred parameters is consistent with the results of  \imrphenompvtwonrtidaltwo~\cite{LIGOScientific:2018hze}. The main difference is the slightly tighter constraint of \imrphenomxpnrtidalthree\ for $\tilde{\Lambda}$ compared to the other models (which are based on \nrtidalvtwo). This is due to the reduction of the secondary peak that \nrtidalvtwo\ models show at higher $\tilde{\Lambda}$.

Finally, we investigate the spin constraints of \imrphenomxpnrtidalthree\ on GW170817 by plotting the inferred spin component magnitudes and orientation (in terms of tilt angles). A tilt angle of $0^{\circ}$ means that the spin components are aligned with the orbital angular momentum $\mathbf{L}$. The results are shown in Fig.~\ref{fig: lowspin_GW170817}, where the left hemisphere is for $\chi_1$ and the right hemisphere corresponds to $\chi_2$. We note that large negative components and anti-aligned spins are ruled out, which is consistent with previous studies using~\imrphenompvtwonrtidaltwo~\cite{LIGOScientific:2018hze}.

For the corresponding parameter estimation for the high-spin prior ($\chi \le 0.5$), we refer to Appendix~\ref{section: PE_high_spin}.

\begin{figure*}
    \centering
    \includegraphics[width = 0.7\linewidth]{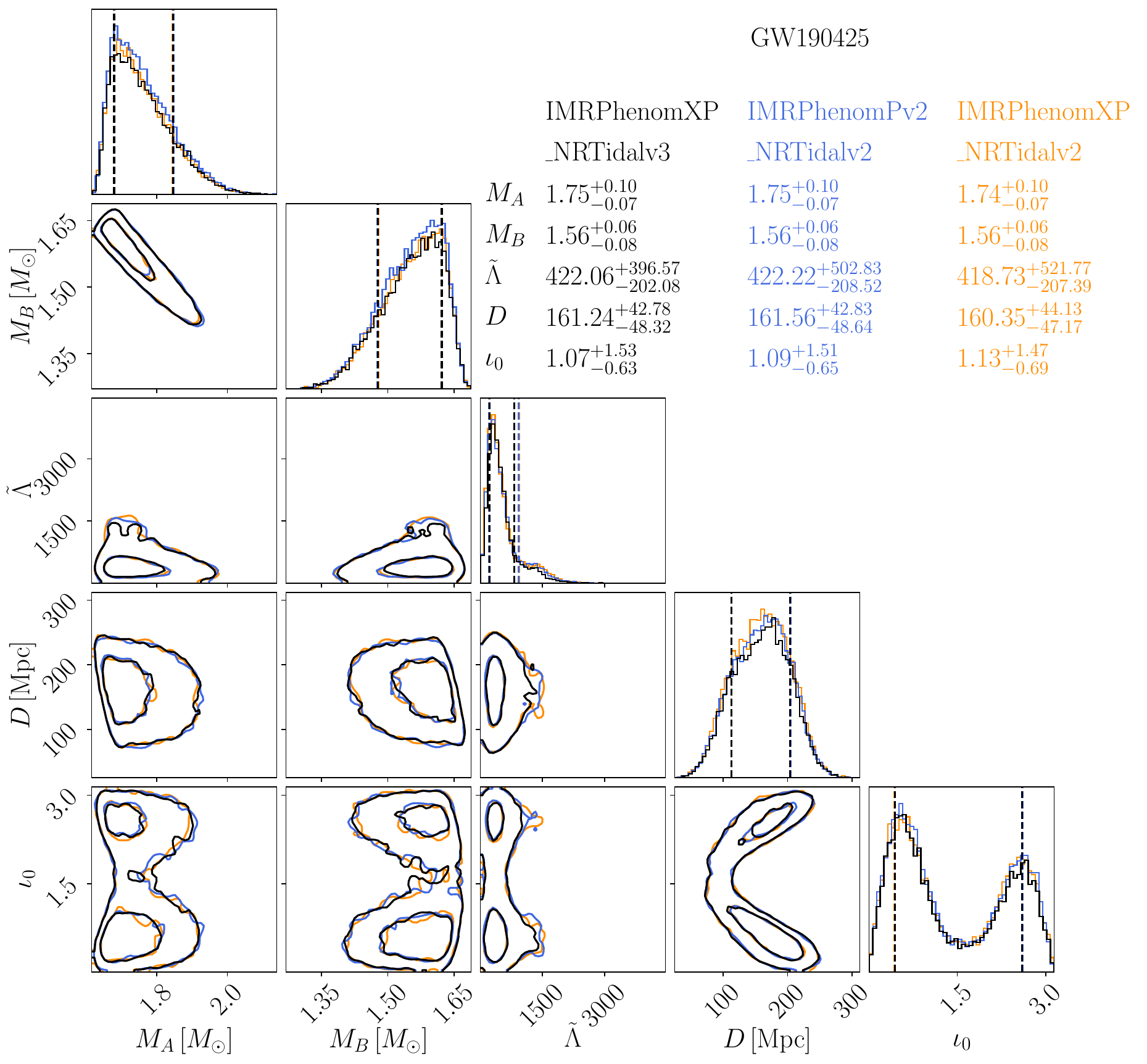}
    \caption{The marginalized 1D and 2D posterior probability distributions for selected parameters of GW190425, obtained with \imrphenomxpnrtidalthree\ (black), \imrphenompvtwonrtidaltwo\ (blue), and \imrphenomxpnrtidaltwo\ (orange). The parameters shown here are the individual star masses $M_{A,B}$, binary tidal deformability $\tilde{\Lambda}$, luminosity distance $D$, and inclination angle $\iota_0$. The 68\% and 90\% confidence intervals are indicated by contours for the 2D posterior plots, while vertical lines in the 1D plots indicate 90\% confidence interval. As with GW170817, note a narrow constraint on the tidal deformability for \imrphenomxpnrtidalthree\ compared to the other models, due to the updated tidal information that was used.} 
    \label{fig: PE_GW190425}
\end{figure*}

\begin{figure}[hbt!]
    \centering
    \includegraphics[width = \linewidth]{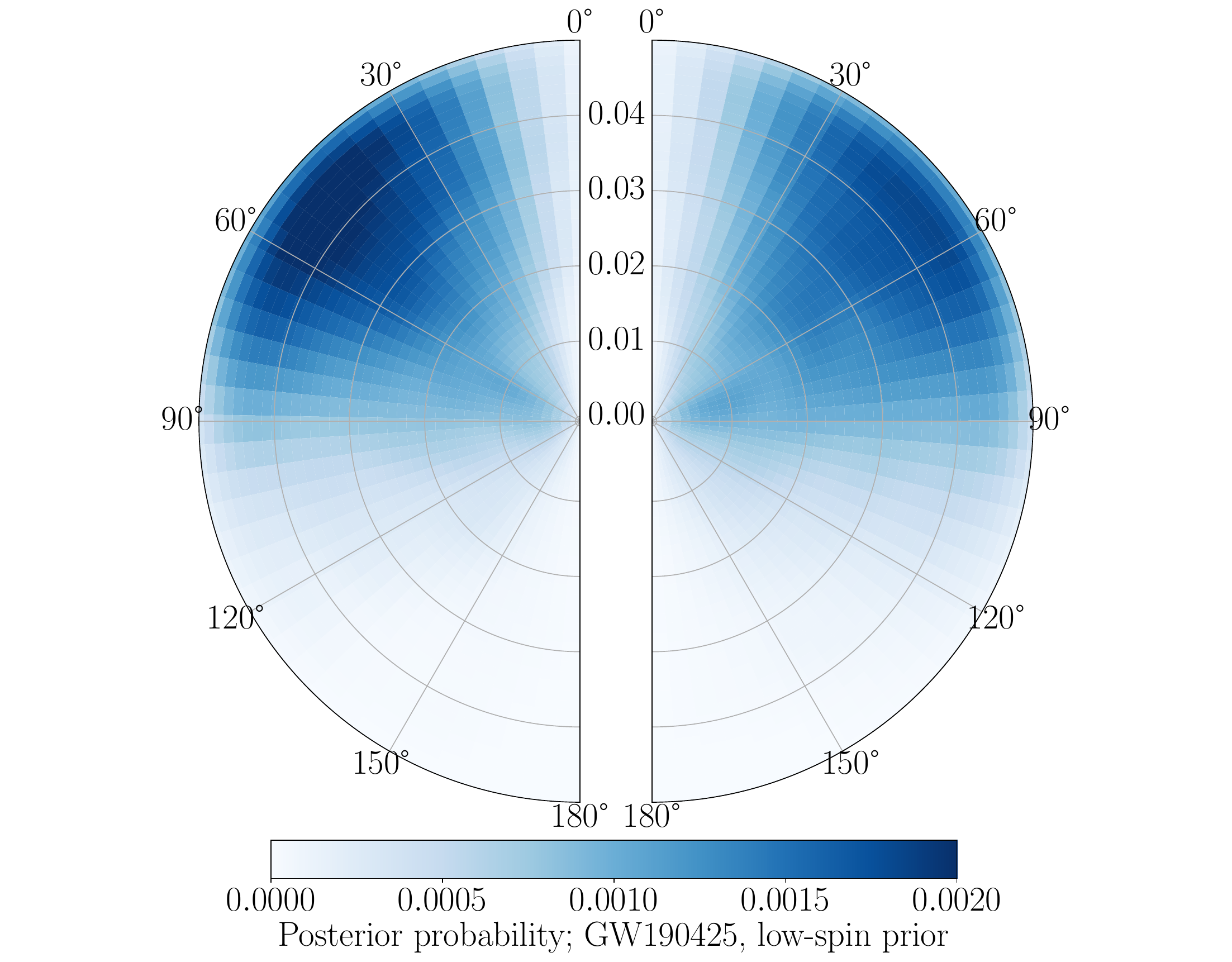}
    \caption{Inferred spin parameters for GW190425 from a low-spin prior ($\chi \le 0.05$) using \imrphenomxpnrtidalthree. Plotted here are the probability densities for the dimensionless spin components $\chi_1$ (left hemisphere) and $\chi_2$ (right hemisphere) relative to the orbital angular momentum $\mathbf{L}$ and tilt angles (i.e. a tilt angle $0^{\circ}$ means that the spin is aligned with the $\mathbf{L}$). The plot was done at a reference frequency of 20 Hz.} 
    \label{fig: lowspin_GW190425}
\end{figure}

\subsection{GW190425}
The results for the marginalized posterior distributions for GW190425 can be found in Fig.~\ref{fig: PE_GW190425}, where (as in the case for GW170817), we see a slightly tighter constraint on the tidal deformability $\tilde{\Lambda}$ from \imrphenomxpnrtidalthree, than the other waveform models. However, all results are consistent between the different GW models. 

Finally, the spin magnitudes and orientation inferred for GW190425 using \imrphenomxpnrtidalthree\ in Fig.~\ref{fig: lowspin_GW190425} shows that negative values of the spin component magnitudes are heavily disfavored, as well as orientation greater than $90^{\circ}$. The corresponding parameter estimation for the high-spin prior ($\chi \le 0.5$) is found in Appendix~\ref{section: PE_high_spin}.

\subsection{Performance of \nrtidalvthree}
\begin{figure}[hbt!]
    \centering
    \includegraphics[width = \linewidth]{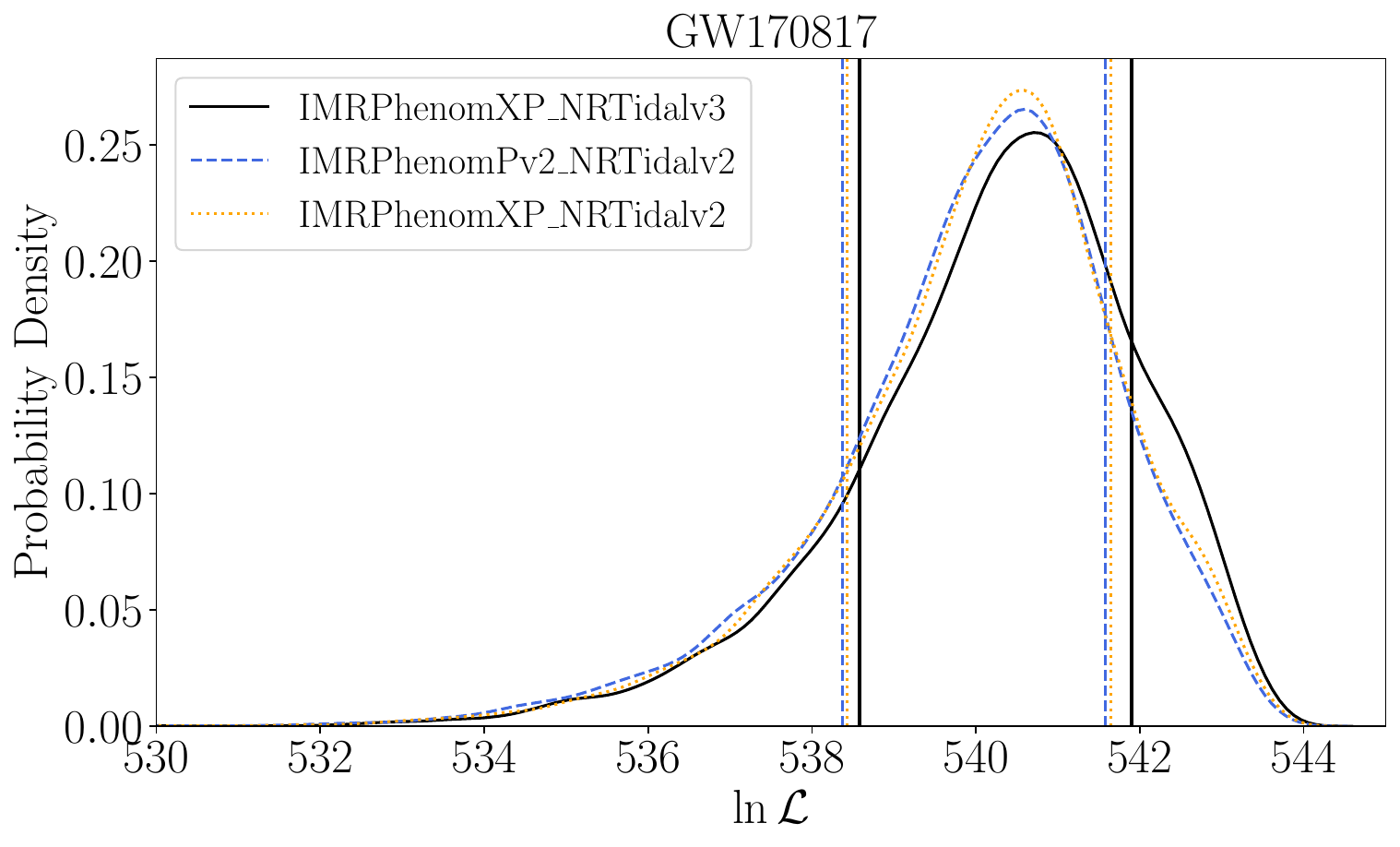}
    \includegraphics[width = \linewidth]{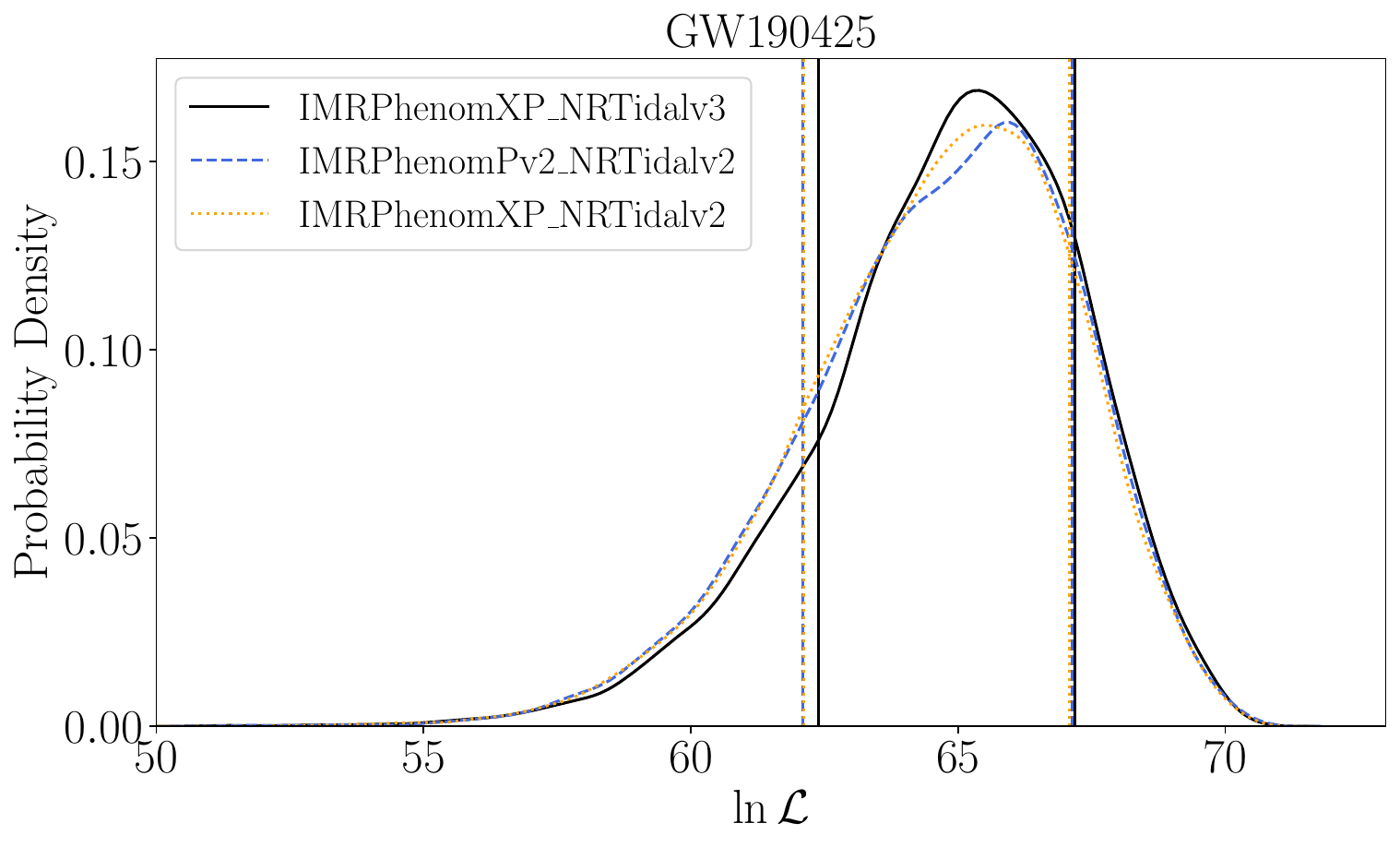}
    \caption{The distribution of the log-likelihood $\ln \mathcal{L}$ for GW170817 (top) and GW190425 (bottom). The vertical lines indicate 90\% confidence intervals for the models. We note the shifting of the distribution using \imrphenomxpnrtidalthree\ towards larger $\ln \mathcal{L}$.} 
    \label{fig: likelihood}
\end{figure}
To further investigate the performance of \imrphenomxpnrtidalthree\, we plot the posterior probability distribution of the natural logarithm of its likelihood (also known as the log-likelihood), $\ln \mathcal{L}$, and compare it with the other two waveform models, as shown in the top panel of Fig.~\ref{fig: likelihood}. We note that for GW170817, the $\ln \mathcal{L}$ of \imrphenomxpnrtidalthree\ is generally shifted towards slightly larger values when compared to that of \imrphenompvtwonrtidaltwo\ and \imrphenomxpnrtidaltwo. In addition, we compute the Bayes factors (which is the ratio of the evidences or posterior probabilities) with respect to the null hypothesis of a non-detection and find \imrphenomxpnrtidalthree\ has $\ln \mathcal{B} = 526.726 \pm 0.085$ which is larger than that of \imrphenompvtwonrtidaltwo\ ($\ln \mathcal{B} = 526.647 \pm 0.084$) and \imrphenomxpnrtidaltwo\ ($\ln \mathcal{B} = 526.668 \pm 0.079$). Although these numbers might indicate a slight preference for \imrphenomxpnrtidalthree, they are not statistically significant. 

Similar to GW170817, we also present the log-likelihood for GW190425 in the bottom panel of Fig.~\ref{fig: likelihood}. As before, we find slightly larger $\ln \mathcal{L}$  for \imrphenomxpnrtidalthree, and at the same time Bayes Factors of \imrphenomxpnrtidalthree\, $\ln \mathcal{B} = 53.849 \pm 0.079$, which is larger than that of \imrphenompvtwonrtidaltwo\ ($\ln \mathcal{B} = 53.756 \pm 0.079$) and \imrphenomxpnrtidaltwo\ ($\ln \mathcal{B} = 53.766 \pm 0.079$), though with overlapping uncertainties and not being statistically significant.

\section{Conclusions and Outlook}
\label{section: conclusions}

\paragraph*{Summary:} In this work, we introduced \nrtidalvthree\ as a new description of the tidal phase contribution to the total GW phase of BNS systems. This model improves upon the previous version (\nrtidalvtwo) by employing a larger set of NR waveforms across a wide range of tidal deformabilities with mass ratios up to $q = 2.0$. 
To construct the model, we employed a representation of the frequency-domain, dynamical enhancement factor for the Love number, as well as a Padé approximant, which imposes additional constraints pertaining to the masses and tidal deformabilities of the component stars.

The model was then augmented onto existing BBH baseline models in \lalsuite\ (\texttt{IMRPhenomD}, \texttt{IMRPhenomXAS}, \texttt{IMRPhenomXP}, and \texttt{SEOBNRv5\_ROM}). To test the performance of \nrtidalvthree\, we calculated its dephasing relative to existing NR waveforms and found overall consistency between \nrtidalvthree\ with respect to the uncertainties in the NR waveforms and with respect to other waveform models. 
The computed mismatches between \nrtidalvthree\ and NR waveforms, as well as NR hybrids, were smaller than for the previously constructed model \nrtidalvtwo~\cite{Dietrich:2019kaq}. With respect to other tidal waveform models, we observe the largest mismatches for high masses, mass ratios, tidal deformabilities, and spin magnitudes. However, we find that overall \nrtidalvthree\ can be employed for masses $M_{A,B}\in [0.5,3.0]M_\odot$, dimensionless spins with magnitude below $|\chi_{A,B}| \leq 0.5$, and for dimensionless tidal deformalities $\Lambda_{A,B}\in [0.,20000]$.

To finalize our investigations, we performed parameter estimation analyses for the GW events GW170817 and GW190425 using \imrphenomxpnrtidalthree, \imrphenompvtwonrtidaltwo, and \imrphenomxpnrtidaltwo. 
We find a slightly tighter constraint on the tidal deformability $\tilde{\Lambda}$ for \nrtidalvthree\ than \nrtidalvtwo. For both events, the \nrtidalvthree\ model displays a slightly higher log Bayes factor, but the difference is too small to be of statistical significance. In general, the performance of \nrtidalvthree\ is consistent with previous analyses done by LVK on these GW events~\cite{LIGOScientific:2018hze, Ashton:2021cub}.\\

\paragraph*{Outlook:} A suitable extension to \nrtidalvthree\ would be the inclusion of higher-order modes since \nrtidalvthree\ only includes the (2,2)-mode in its tidal phase description. It has recently been shown that the higher-order modes for BNS can be rescaled with respect to the (2,2)-mode, in the same manner as found on the BBH waveform phase contributions of higher-modes~\cite{Ujevic:2022qle}. Moreover, the inclusion of higher-order spin contributions to the BBH baseline phase can also be investigated. 
Aligned with the inclusion of higher modes in \nrtidalvthree\ models would be the extension of existing BHNS models such as \texttt{IMRPhenomNSBH}~\cite{Thompson:2020nei} or \texttt{SEOBNRv4\_ROM\_NRTidalv2\_NSBH}~\cite{Matas:2020wab} with \nrtidalvthree\ phase contributions. 

\newtext{Another possible extension of the model would be a generalization of the frequency dependence of the Love numbers to be able to set the fundamental f-mode frequency $f_{02}$ as a free parameter. In the current implementation, this cannot be straightforwardly done, as the time-domain effective enhancement factor $k_2^{\rm eff}$ (and subsequently the frequency-domain $\bar{k}_2^{\rm eff}$) depends on the quasi-universal relation Eq.~\eqref{eq: univrel} between $f_{02}$ and $\Lambda_2$.}

Finally, other NR waveforms may also be added in a future version of the model, such as BNS and/or NSBH waveforms with~\cite{Markin:2023fxx} and without~\cite{LIGOScientific:2021qlt} sub-solar-mass components, as well as waveforms whose NS have more exotic EoS. This will allow the model to accommodate an even wider range of neutron star properties and to constrain these EoSs with GW observations. 

\section*{\label{section: Acknowledgments}Acknowledgments}

This material is based upon work supported by the NSF's LIGO Laboratory which is a major facility fully funded by the National Science Foundation.

The authors would like to thank Marta Colleoni, Nathan Johnson-McDaniel, Antoni Ramos Buades, Hector Estelles, Hector Silva, Lorenzo Pompili, Elise Sänger, Steffen Grunewald, Guglielmo Faggioli, Marcus Haberland, Peter James Nee, Henrik Rose, Anna Neuweiler, Hao-Jui Kuan, Ivan Markin, Thibeau Wouters and Nina Kunert for fruitful discussions that have led to the improvement of this work. In particular, the authors acknowledge the work of Marta Colleoni and Nathan Johnson-McDaniel in the implementation of \imrphenomxpnrtidaltwo\, which was used as the basis for the implementation of \imrphenomxpnrtidalthree.

TD acknowledges funding from the EU Horizon under ERC Starting Grant, no.\ SMArt-101076369, support from the Deutsche Forschungsgemeinschaft, DFG, project number DI 2553/7, and from the Daimler and Benz Foundation for the project `NUMANJI'.

The parameter estimation runs were done using the hypatia cluster at the Max Planck Institute for Gravitational Physics.
\appendix

\begin{table*} 
    \caption{\label{table: 55nr} Properties of the 55 NR waveforms used in the calibration of \nrtidalvthree. The first 46 waveforms are from SACRA \cite{Kiuchi:2017pte, Kawaguchi:2018gvj, Kiuchi:2019kzt}, while the last nine are from the CoRe database simulated with the BAM code \cite{Dietrich:2018phi, Ujevic:2022qle, Gonzalez:2022mgo}. For each waveform, we indicate its equation of state (EoS), the masses $M_{A,B}(M_{\odot})$ of the individual bodies, the total mass $M(M_{\odot})$, the mass ratio $q = M_A/M_B \, (M_A > M_B)$, the dimensionless tidal deformabilities $\Lambda_{A,B}$, effective tidal deformability $\tilde{\Lambda}$, and the radii $R_{A,B}$ [km].}
    \begin{ruledtabular}
        \begin{tabular}{l || cccccccccc}
\textbf{NR Waveform Name} & EoS  & $M_A$ & $M_B$ & $M$  & $q$  & $\Lambda_A$ & $\Lambda_B$ & $\tilde{\Lambda}$ & $R_A$ & $R_B$\\
\hline
\hline
SACRA:15H\_135\_135\_00155\_182\_135  & 15H  & 1.35 & 1.35 & 2.70 & 1.00 & 1211 & 1211 & 1211 & 13.69 & 13.69 \\
SACRA:125H\_135\_135\_00155\_182\_135 & 125H & 1.35 & 1.35 & 2.70 & 1.00 & 863  & 863  & 863  & 12.97 & 12.97 \\
SACRA:H\_135\_135\_00155\_182\_135    & H    & 1.35 & 1.35 & 2.70 & 1.00 & 607  & 607  & 607  & 12.27 & 12.27 \\
SACRA:HB\_135\_135\_00155\_182\_135   & HB   & 1.35 & 1.35 & 2.70 & 1.00 & 422  & 422  & 422  & 11.61 & 11.61 \\
SACRA:B\_135\_135\_00155\_182\_135    & B    & 1.35 & 1.35 & 2.70 & 1.00 & 289  & 289  & 289  & 10.96 & 10.96 \\
SACRA:15H\_125\_146\_00155\_182\_135  & 15H  & 1.46 & 1.25 & 2.71 & 1.17 & 760  & 1871 & 1201 & 13.72 & 13.65 \\
SACRA:125H\_125\_146\_00155\_182\_135 & 125H & 1.46 & 1.25 & 2.71 & 1.17 & 535  & 1351 & 858  & 12.99 & 12.94 \\
SACRA:H\_125\_146\_00155\_182\_135    & H    & 1.46 & 1.25 & 2.71 & 1.17 & 369  & 966  & 605  & 12.18 & 12.26 \\
SACRA:HB\_125\_146\_00155\_182\_135   & HB   & 1.46 & 1.25 & 2.71 & 1.17 & 252  & 684  & 423  & 11.59 & 11.61 \\
SACRA:B\_125\_146\_00155\_182\_135    & B    & 1.46 & 1.25 & 2.71 & 1.17 & 168  & 477  & 290  & 10.92 & 10.98 \\
SACRA:15H\_121\_151\_00155\_182\_135  & 15H  & 1.51 & 1.21 & 2.72 & 1.25 & 625  & 2238 & 1197 & 13.73 & 13.63 \\
SACRA:125H\_121\_151\_00155\_182\_135 & 125H & 1.51 & 1.21 & 2.72 & 1.25 & 435  & 1621 & 855  & 12.98 & 12.93 \\
SACRA:H\_121\_151\_00155\_182\_135    & H    & 1.51 & 1.21 & 2.72 & 1.25 & 298  & 1163 & 604  & 12.26 & 12.25 \\
SACRA:HB\_121\_151\_00155\_182\_135   & HB   & 1.51 & 1.21 & 2.72 & 1.25 & 200  & 827  & 421  & 11.57 & 11.60 \\
SACRA:B\_121\_151\_00155\_182\_135    & B    & 1.51 & 1.21 & 2.72 & 1.25 & 131  & 581  & 289  & 10.89 & 10.98 \\
SACRA:15H\_118\_155\_00155\_182\_135  & 15H  & 1.55 & 1.18 & 2.73 & 1.31 & 530  & 2575 & 1192 & 13.74 & 13.62 \\
SACRA:125H\_118\_155\_00155\_182\_135 & 125H & 1.55 & 1.18 & 2.73 & 1.31 & 366  & 1875 & 853  & 12.98 & 12.92 \\
SACRA:H\_118\_155\_00155\_182\_135    & H    & 1.55 & 1.18 & 2.73 & 1.31 & 249  & 1354 & 605  & 12.26 & 12.24 \\
SACRA:HB\_118\_155\_00155\_182\_135   & HB   & 1.55 & 1.18 & 2.73 & 1.31 & 165  & 966  & 422  & 11.55 & 11.60 \\
SACRA:B\_118\_155\_00155\_182\_135    & B    & 1.55 & 1.18 & 2.73 & 1.31 & 107  & 681  & 291  & 10.87 & 10.98 \\
SACRA:15H\_117\_156\_00155\_182\_135  & 15H  & 1.56 & 1.17 & 2.73 & 1.33 & 509  & 2692 & 1196 & 13.74 & 13.61 \\
SACRA:125H\_117\_156\_00155\_182\_135 & 125H & 1.56 & 1.17 & 2.73 & 1.33 & 350  & 1963 & 856  & 12.98 & 12.91 \\
SACRA:H\_117\_156\_00155\_182\_135    & H    & 1.56 & 1.17 & 2.73 & 1.33 & 238  & 1415 & 607  & 12.25 & 12.24 \\
SACRA:HB\_117\_156\_00155\_182\_135   & HB   & 1.56 & 1.17 & 2.73 & 1.33 & 157  & 1013 & 424  & 11.55 & 11.60 \\
SACRA:B\_117\_156\_00155\_182\_135    & B    & 1.56 & 1.17 & 2.73 & 1.33 & 101  & 719  & 293  & 10.86 & 10.98 \\
SACRA:15H\_116\_158\_00155\_182\_135  & 15H  & 1.58 & 1.16 & 2.74 & 1.36 & 465  & 2863 & 1189 & 13.73 & 13.60 \\
SACRA:125H\_116\_158\_00155\_182\_135 & 125H & 1.58 & 1.16 & 2.74 & 1.36 & 319  & 2085 & 851  & 12.98 & 12.90 \\
SACRA:H\_116\_158\_00155\_182\_135    & H    & 1.58 & 1.16 & 2.74 & 1.36 & 215  & 1506 & 603  & 12.25 & 12.23 \\
SACRA:HB\_116\_158\_00155\_182\_135   & HB   & 1.58 & 1.16 & 2.74 & 1.36 & 142  & 1079 & 423  & 11.53 & 11.59 \\
SACRA:B\_116\_158\_00155\_182\_135    & B    & 1.58 & 1.16 & 2.74 & 1.36 & 91   & 765  & 292  & 10.84 & 11.98 \\
SACRA:15H\_125\_125\_0015\_182\_135   & 15H  & 1.25 & 1.25 & 2.50 & 1.00 & 1875 & 1875 & 1875 & 13.65 & 13.65 \\
SACRA:125H\_125\_125\_0015\_182\_135  & 125H & 1.25 & 1.25 & 2.50 & 1.00 & 1352 & 1352 & 1352 & 12.94 & 12.94 \\
SACRA:H\_125\_125\_0015\_182\_135     & H    & 1.25 & 1.25 & 2.50 & 1.00 & 966  & 966  & 966  & 12.26 & 12.26 \\
SACRA:HB\_125\_125\_0015\_182\_135    & HB   & 1.25 & 1.25 & 2.50 & 1.00 & 683  & 683  & 683  & 11.61 & 11.61 \\
SACRA:B\_125\_125\_0015\_182\_135     & B    & 1.25 & 1.25 & 2.50 & 1.00 & 476  & 476  & 476  & 10.98 & 10.98 \\
SACRA:15H\_112\_140\_0015\_182\_135   & 15H  & 1.40 & 1.12 & 2.52 & 1.25 & 975  & 3411 & 1838 & 13.71 & 13.58 \\
SACRA:125H\_112\_140\_0015\_182\_135  & 125H & 1.40 & 1.12 & 2.52 & 1.25 & 693  & 2490 & 1329 & 12.98 & 12.89 \\
SACRA:H\_112\_140\_0015\_182\_135     & H    & 1.40 & 1.12 & 2.52 & 1.25 & 484  & 1812 & 953  & 12.28 & 12.23 \\
SACRA:HB\_112\_140\_0015\_182\_135    & HB   & 1.40 & 1.12 & 2.52 & 1.25 & 333  & 1304 & 675  & 11.60 & 11.59 \\
SACRA:B\_112\_140\_0015\_182\_135     & B    & 1.40 & 1.12 & 2.52 & 1.25 & 225  & 933  & 474  & 10.95 & 10.97 \\
SACRA:15H\_107\_146\_0015\_182\_135   & 15H  & 1.46 & 1.07 & 2.53 & 1.36 & 760  & 4361 & 1848 & 13.72 & 13.54 \\
SACRA:125H\_107\_146\_0015\_182\_135  & 125H & 1.46 & 1.07 & 2.53 & 1.36 & 535  & 3196 & 1337 & 12.99 & 12.86 \\
SACRA:H\_107\_146\_0015\_182\_135     & H    & 1.46 & 1.07 & 2.53 & 1.36 & 369  & 2329 & 959  & 12.18 & 12.22 \\
SACRA:HB\_107\_146\_0015\_182\_135    & HB   & 1.46 & 1.07 & 2.53 & 1.36 & 252  & 1695 & 685  & 11.59 & 11.60 \\
SACRA:B\_107\_146\_0015\_182\_135     & B    & 1.46 & 1.07 & 2.53 & 1.36 & 168  & 1216 & 481  & 10.92 & 10.97 \\
SACRA:SFHo\_135\_135\_00155\_182\_135 & SFHo & 1.35 & 1.35 & 2.70 & 1.00 & 460  & 460  & 460  & 11.91 & 11.91 \\
CoRe:BAM:0137                         & SLy  & 1.50 & 1.20 & 2.70 & 1.25 & 191  & 812  & 409  & 11.42 & 11.46 \\
CoRe:BAM:0136                         & SLy  & 1.62 & 1.08 & 2.70 & 1.50 & 108  & 1503 & 453  & 11.34 & 11.43 \\
CoRe:BAM:0131                         & SLy  & 1.72 & 0.98 & 2.70 & 1.75 & 66   & 2557 & 504  & 11.27 & 11.40 \\
CoRe:BAM:0130                         & SLy  & 1.80 & 0.90 & 2.70 & 2.00 & 43   & 4088 & 566  & 11.14 & 11.36 \\
CoRe:BAM:0095                         & SLy  & 1.35 & 1.35 & 2.70 & 1.00 & 390  & 390  & 390  & 11.46 & 11.46 \\
CoRe:BAM:0097                         & SLy  & 1.35 & 1.35 & 2.70 & 1.00 & 390  & 390  & 390  & 11.46 & 11.46 \\
CoRe:BAM:0037                         & H4   & 1.37 & 1.37 & 2.74 & 1.00 & 1006 & 1006 & 1006 & 13.53 & 13.53 \\
CoRe:BAM:0001                         & 2B   & 1.35 & 1.35 & 2.70 & 1.00 & 127  & 127  & 127  & 9.72  & 9.72  \\
CoRe:BAM:0064                         & MS1b & 1.35 & 1.35 & 2.70 & 1.00 & 1531 & 1532 & 1532 & 14.02 & 14.02
        \end{tabular}
    \end{ruledtabular}
\end{table*}

\section{\label{subsection: appA}Configurations of the NR waveforms used in \nrtidalvthree}

We present in Table~\ref{table: 55nr} the configurations of all the 55 NR waveforms from SACRA~\cite{Kiuchi:2017pte, Kawaguchi:2018gvj, Kiuchi:2019kzt} and CoRe (using the BAM code)~\cite{Dietrich:2018phi, Ujevic:2022qle, Gonzalez:2022mgo} that were used in the calibration of \nrtidalvthree. We include in the table the individual masses of the stars $M_{A,B}(M_{\odot})$, the total mass $M(M_{\odot})$, mass ratio $q$, dimensionless tidal deformabilities $\Lambda_{A,B}$, and radii $R_{A,B}$ in km. The radii data for the SACRA waveforms were adapted from Ref.~\cite{Kiuchi:2017pte, Kawaguchi:2018gvj, Kiuchi:2019kzt}, while the radii for the CORE (BAM) waveforms were computed using the adiabatic Love number and tidal deformability, i.e. from $\Lambda = (2/3)k_2R^5/M^5$.

\begin{figure}[hbt!]
    \centering
    \includegraphics[width = \linewidth]{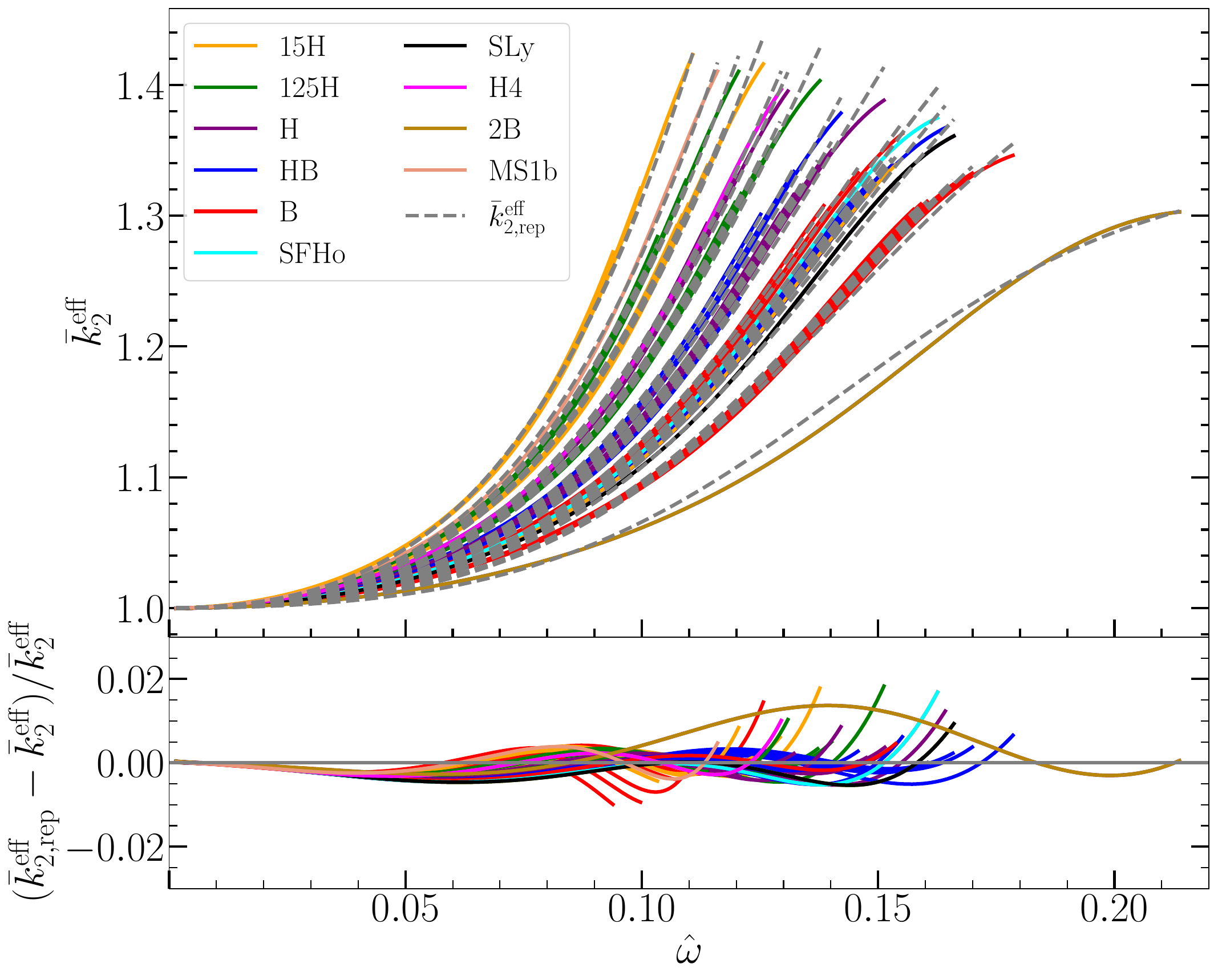}
    \caption{\newtext{\textit{Top}: The $\bar{k}_{2, \rm{rep}}^{\rm eff}$ plotted on top of the numerically calculated $\bar{k}_2^{\rm eff}$, color-coded according to EoS. \textit{Bottom}: Fractional differences between the numerical calculation and the fit.}} 
    \label{fig: k2effcomp}
\end{figure}

\begin{figure*}[hbt!]
    \centering
    \includegraphics[width = 0.7\linewidth]{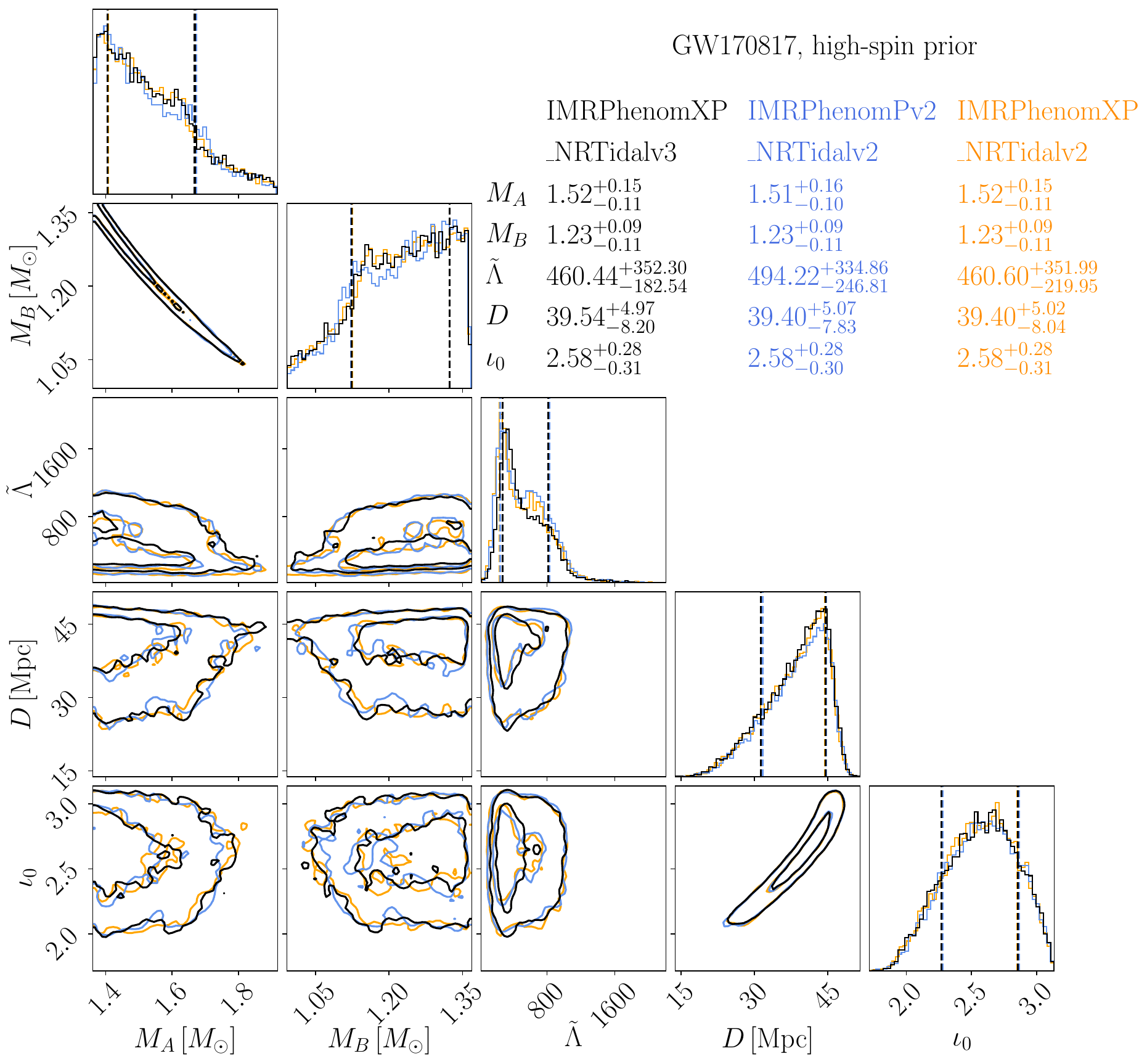}
    \caption{The marginalized 1D and 2D posterior probability distributions for selected parameters of GW170817, with a high-spin prior ($\chi \le 0.5)$ obtained with \imrphenomxpnrtidalthree\ (black), \imrphenompvtwonrtidaltwo\ (blue), and \imrphenomxpnrtidaltwo\ (orange). The parameters shown here are the individual star masses $M_{A,B}$, binary tidal deformability $\tilde{\Lambda}$, luminosity distance $D$, and inclination angle $\iota_0$. The 68\% and 90\% confidence intervals are indicated by contours for the 2D posterior plots, while vertical lines in the 1D plots indicate 90\% confidence interval. We note a narrow constraint on the tidal deformability for \imrphenomxpnrtidalthree\ compared to the other models, due to the updated tidal information that was used.}  
    \label{fig: PE_GW170817_highspin}
\end{figure*}

\begin{figure}[hbt!]
    \centering
    \includegraphics[width = \linewidth]{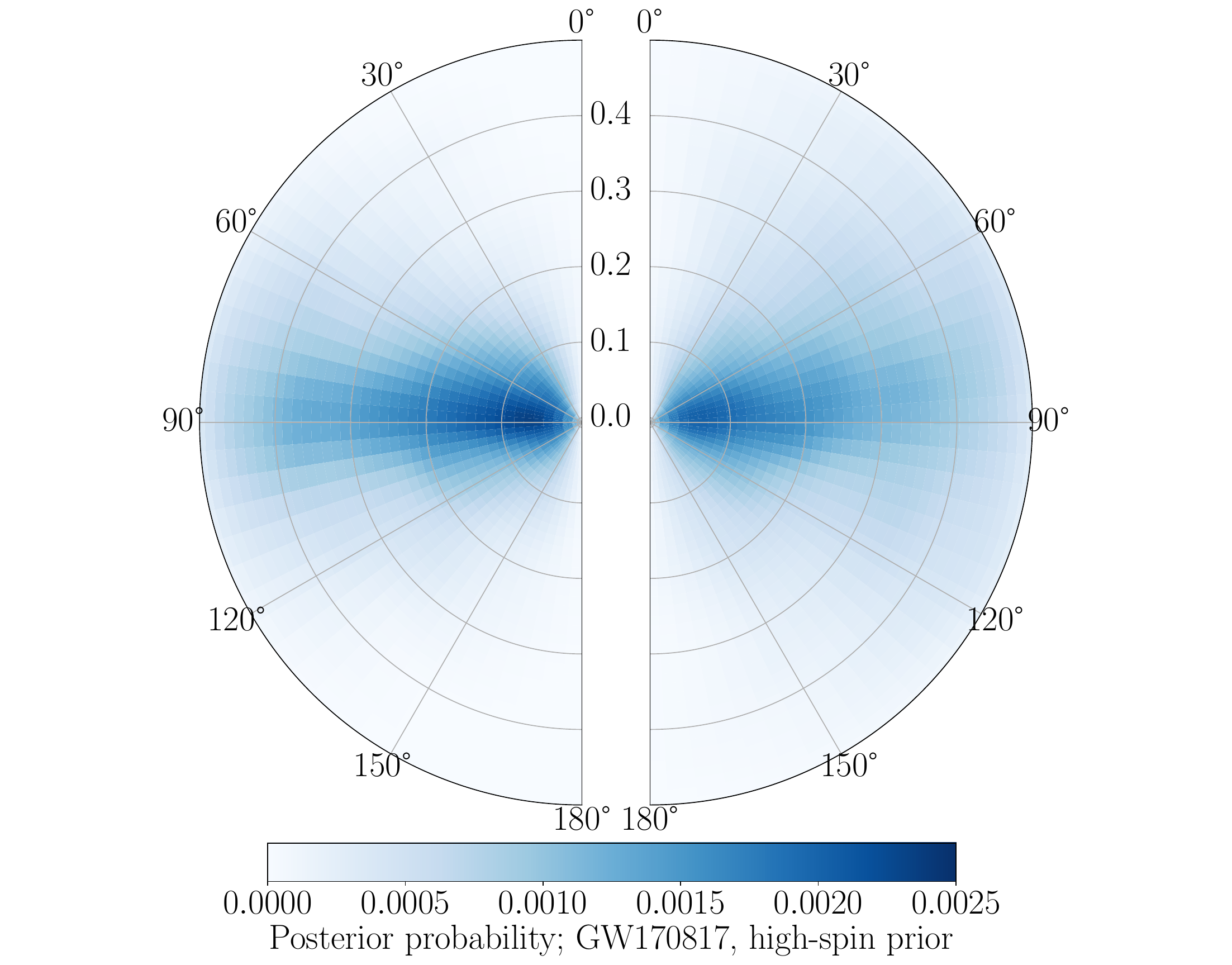}
    \caption{Inferred spin parameters for GW170817 from a high-spin prior ($\chi \le 0.5$) using \imrphenomxpnrtidalthree. Plotted here are the probability densities for the dimensionless spin components $\chi_1$ (left hemisphere) and $\chi_2$ (right hemisphere) relative to the orbital angular momentum $\mathbf{L}$ and tilt angles (i.e. a tilt angle $0^{\circ}$ means that the spin is aligned with the $\mathbf{L}$). The plot was done at a reference frequency of 20 Hz.} 
    \label{fig: highspin_GW170817}
\end{figure}

\section{\label{section: k2eff} Comparison between the numerically computed $\bar{k}_2^{\rm eff}$ and $\bar{k}_{2, \rm{rep}}^{\rm eff}$}

\newtext{We show here a comparison between the frequency-domain effective enhancement factor $\bar{k}_2^{\rm eff}$ numerically calculated from Eq.~\eqref{eq: dek2} and the effective representation $\bar{k}_{2, \rm{rep}}^{\rm eff}$ in Eq.~\eqref{eq: k2bareff}. The results are shown in Fig.~\ref{fig: k2effcomp}. We also show the relative error between $\bar{k}_2^{\rm eff}$ and $\bar{k}_{2, \rm{rep}}^{\rm eff}$. The fitting formula satisfies the numerically calculated $\bar{k}_2^{\rm eff}$, with a maximum (deviation) error of $1.8 \%$ and mean error of $0.2 \%$.}

\section{\label{section: PE_high_spin} Parameter Estimation for high-spin priors}

We present here the results for the parameter estimation runs for GW170817 and GW190425, as was done in Sec.~\ref{section: pe analysis}, but this time, using high-spin priors. For both events, we use a spin prior of up to $\chi \le 0.50$. We do not employ spins up to 0.89 as done in some LVK studies, given that this spin is well above the breakup spin of neutron stars and is unrealistically large for BNS systems.

\paragraph{GW170817.}The inferred properties of GW170817 with a high-spin prior, together with the marginalized 1D and 2D posterior distributions are shown in Fig.~\ref{fig: PE_GW170817_highspin}. As with the low-spin prior case, the result is consistent with the other waveform models and with previous results~\cite{LIGOScientific:2018hze}, and we observe a slightly narrow constraint for \imrphenomxpnrtidalthree\ in the tidal deformability $\tilde{\Lambda}$. As for the inferred spin parameters, Fig.~\ref{fig: highspin_GW170817} shows that relatively large components of the spin which are aligned or anti-aligned with the orbital angular momentum $\mathbf{L}$ are heavily disfavored, and constraints are placed on the in-plane spin components.

\begin{figure*}[hbt!]
    \centering
    \includegraphics[width = 0.7\linewidth]{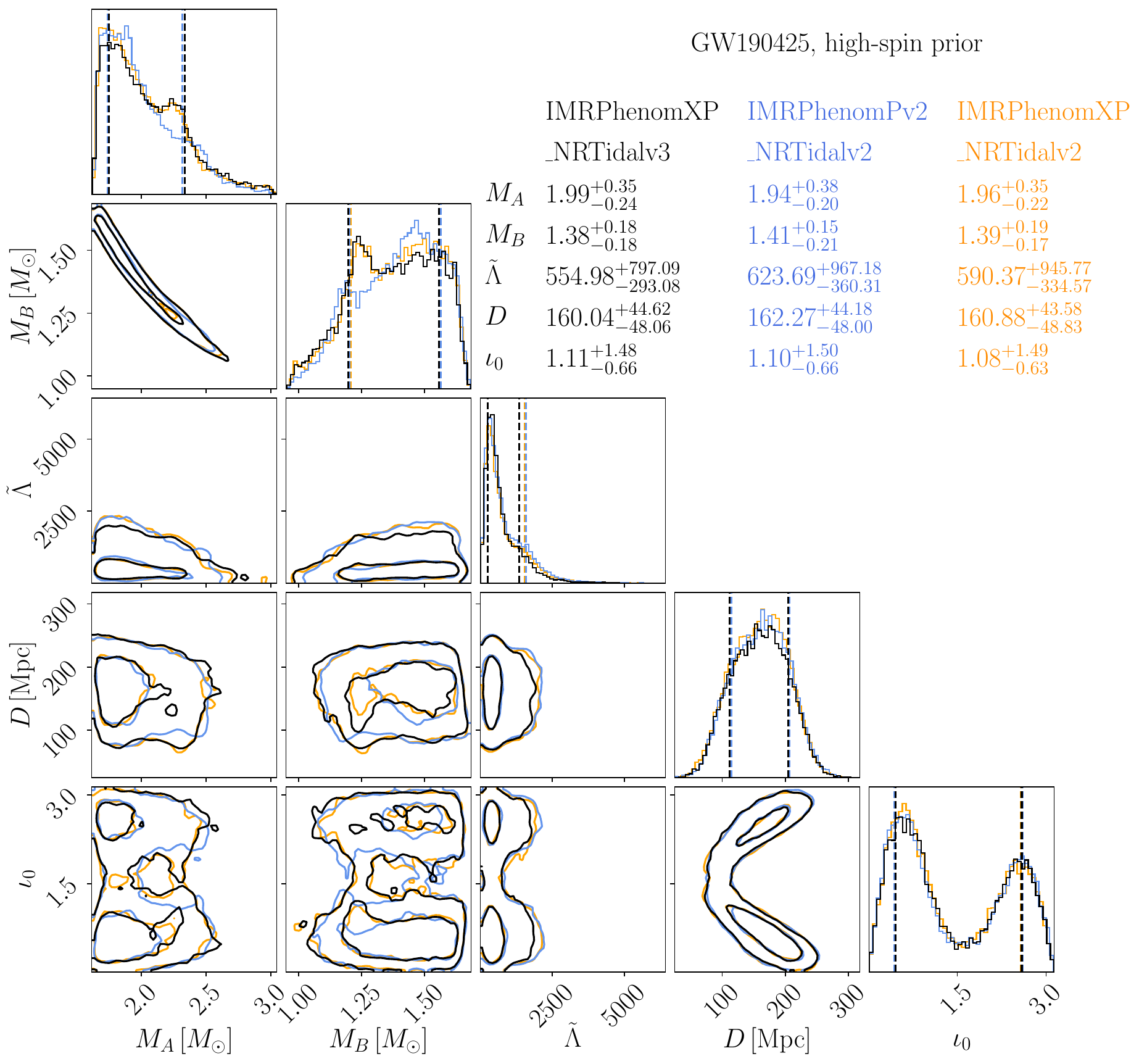}
    \caption{The marginalized 1D and 2D posterior probability distributions for selected parameters of GW190425, obtained with \imrphenomxpnrtidalthree\ (black), \imrphenompvtwonrtidaltwo\ (blue), and \imrphenomxpnrtidaltwo\ (orange), for a high-spin prior ($\chi \le 0.50$). The parameters shown here are the individual star masses $M_{A,B}$, binary tidal deformability $\tilde{\Lambda}$, luminosity distance $D$, and inclination angle $\iota_0$. The 68\% and 90\% confidence intervals are indicated by contours for the 2D posterior plots, while vertical lines in the 1D plots indicate 90\% confidence interval. We note a narrow constraint on the tidal deformability for \imrphenomxpnrtidalthree\ compared to the other models, due to the updated tidal information that was used.} 
    \label{fig: PE_GW190425_highspin}
\end{figure*}

\begin{figure}[hbt!]
    \centering
    \includegraphics[width = \linewidth]{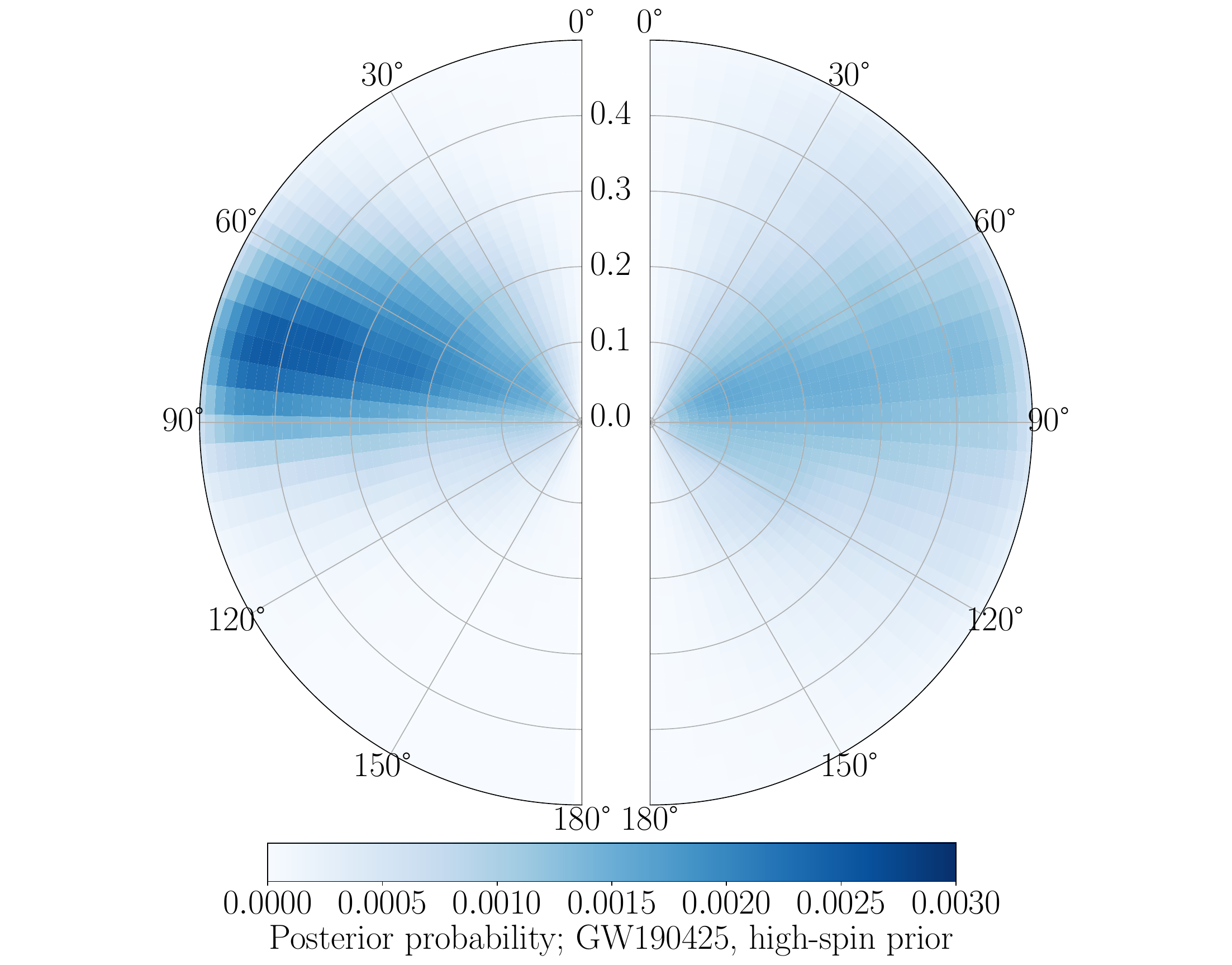}
    \caption{Inferred spin parameters for GW190425 from a high-spin prior ($\chi \le 0.50$) using \imrphenomxpnrtidalthree. Plotted here are the probability densities for the dimensionless spin components $\chi_1$ (left hemisphere) and $\chi_2$ (right hemisphere) relative to the orbital angular momentum $\mathbf{L}$ and tilt angles (i.e. a tilt angle $0^{\circ}$ means that the spin is aligned with the $\mathbf{L}$). The plot was done at a reference frequency of 20 Hz.} 
    \label{fig: highspin_GW190425}
\end{figure}

\paragraph{GW190425.}We show inferred properties of GW190425 with a high-spin prior, together with the marginalized 1D and 2D posterior distributions in Fig.~\ref{fig: PE_GW190425_highspin}. As with the low-spin prior case, the result is consistent with the other waveform models, and we observe a narrower constraint for \imrphenomxpnrtidalthree\ in the tidal deformability $\tilde{\Lambda}$. Fig.~\ref{fig: highspin_GW190425} shows similar behavior as in Fig.~\ref{fig: highspin_GW170817} such that relatively large components of the spin which are aligned or anti-aligned with the orbital angular momentum $\mathbf{L}$ are heavily disfavored, and constraints are placed on the in-plane spin components~\cite{LIGOScientific:2020aai}.

\begin{figure}[hbt!]
    \centering
    \includegraphics[width = \linewidth]{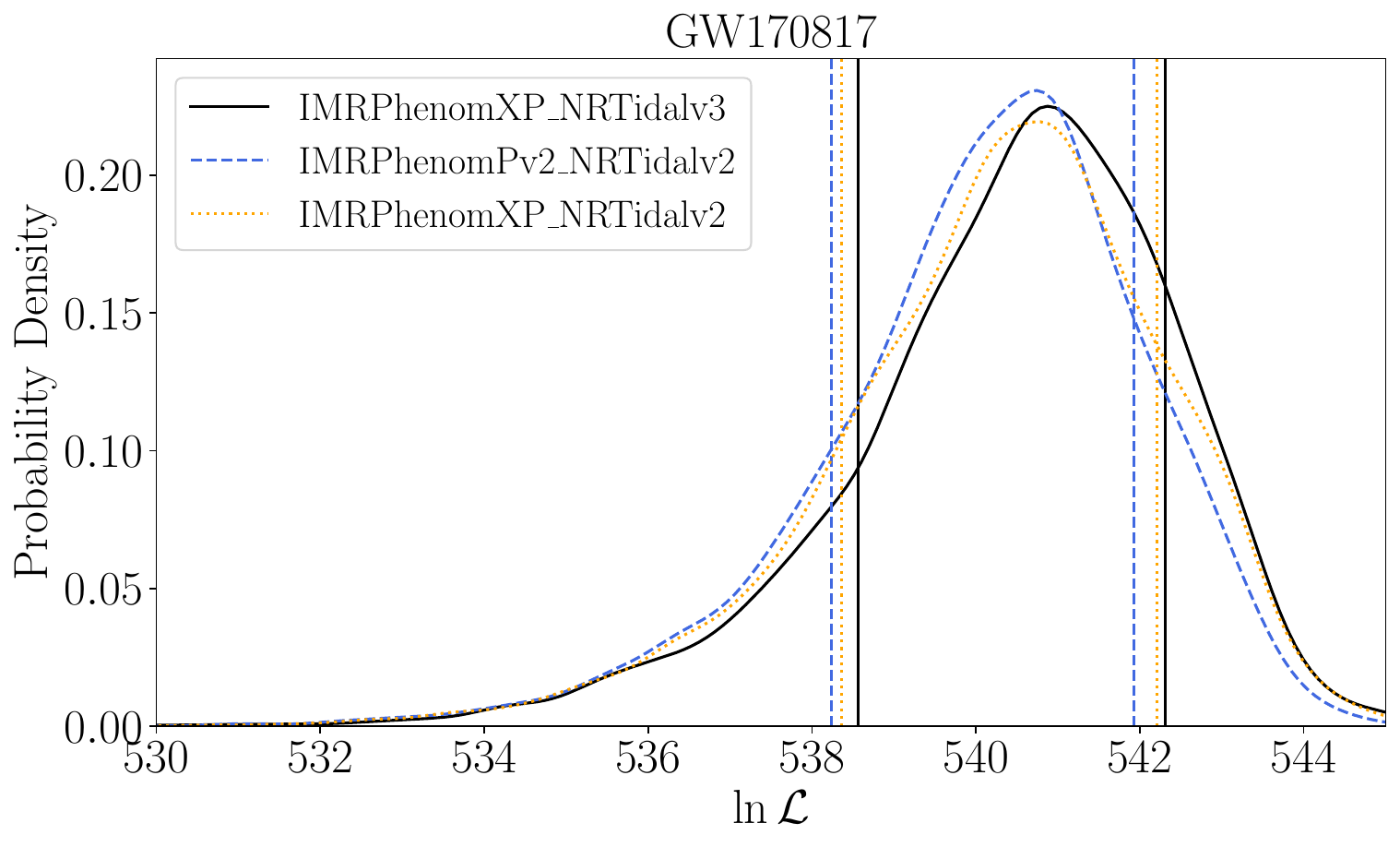}
    \includegraphics[width = \linewidth]{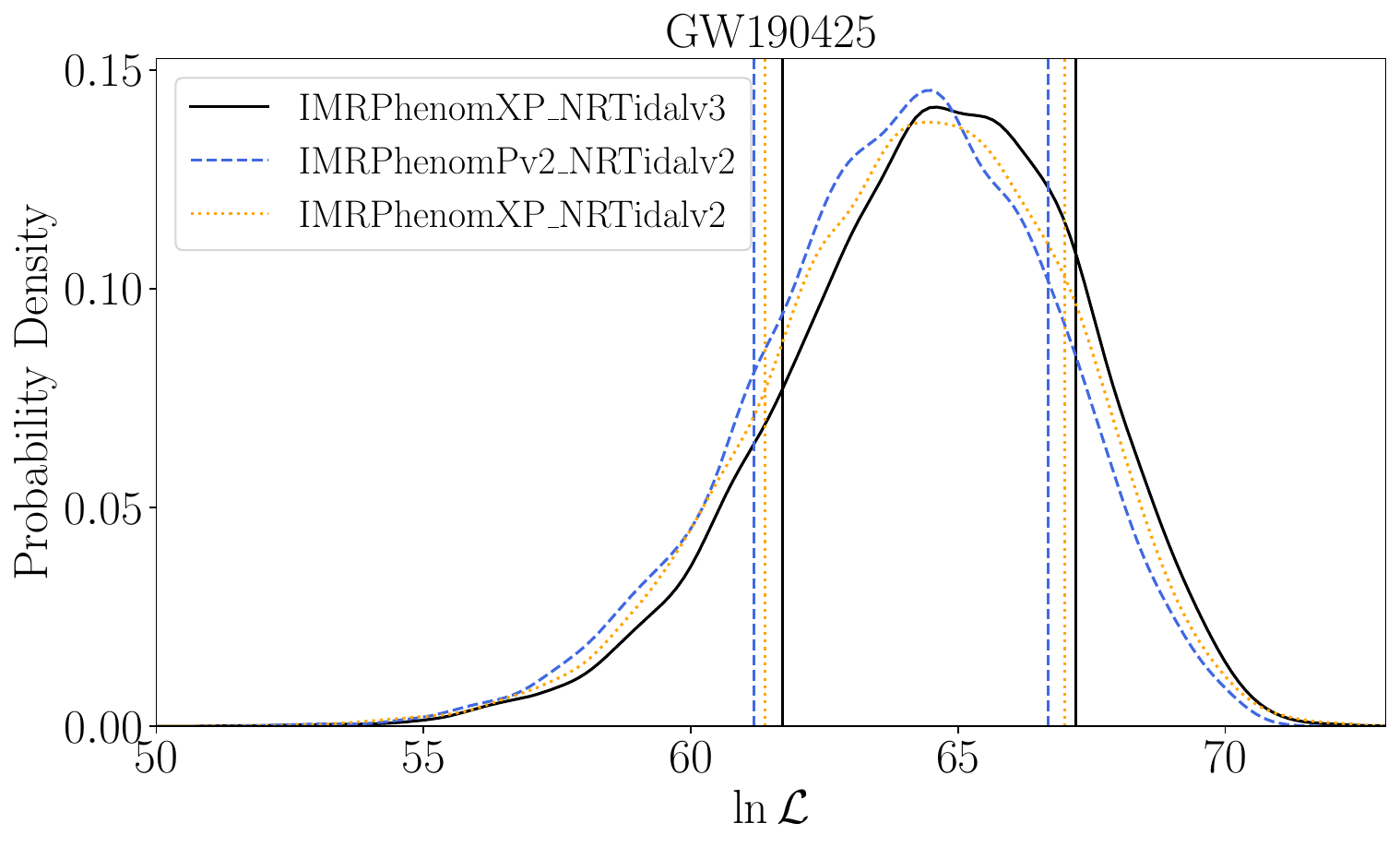}
    \caption{The distribution of the log-likelihood $\ln \mathcal{L}$ for GW170817 (top) and GW190425 (bottom) with high-spin priors ($\chi \le 0.5$). The vertical lines indicate 90\% confidence intervals for the models. We note the shifting of the distribution using \imrphenomxpnrtidalthree\ towards larger $\ln \mathcal{L}$.} 
    \label{fig: likelihoodhighspin}
\end{figure}

\paragraph{Performance.} Finally, the posterior probability distribution of the log likelihood $\ln \mathcal{L}$ of the three models used with GW170817 and GW190425 and shown in  Fig.~\ref{fig: likelihoodhighspin}. We note that for both GW events, the $\ln \mathcal{L}$ values for \imrphenomxpnrtidalthree\ are shifted towards larger values relative to the other two models. For GW170817, we find $\ln \mathcal{B} = 525.114 \pm 0.091$ with \imrphenomxpnrtidalthree, $\ln \mathcal{B} = 524.926 \pm 0.090$ with \imrphenompvtwonrtidaltwo, and $\ln \mathcal{B} = 525.131 \pm 0.090$ with \imrphenomxpnrtidaltwo. Meanwhile, for GW190425, we obtain $\ln \mathcal{B} = 53.831 \pm 0.078$ with \imrphenomxpnrtidalthree, $\ln \mathcal{B} = 53.630 \pm 0.077$ with \imrphenompvtwonrtidaltwo, and $\ln \mathcal{B} = 53.632 \pm 0.077$ with \imrphenomxpnrtidaltwo.

\nocite{*}

\bibliography{Abac_NRTidalv3Notes}

\end{document}